\documentclass[aps,superscriptaddress,groupedaddress,nofootinbib,twocolumn,floatfix]{revtex4}  % for double-spaced preprint
\usepackage{graphicx}  % needed for figures
\usepackage{dcolumn}   % needed for some tables
\usepackage{bm}        % for math
\usepackage{amssymb}   % for math
\usepackage{amsmath,amsfonts} %for math
\usepackage{slashed}  %for math
\usepackage{rotating}
\usepackage{overpic}
\usepackage[absolute,overlay]{textpos}
\usepackage{xcolor}
\newcommand{\unit}{1\!\!1}
\DeclareMathOperator{\e}{e}
\DeclareMathOperator{\img}{i}

\DeclareMathOperator{\su}{SU}
\DeclareMathOperator{\dd}{d}
\begin{document}

% The following information is for internal review, please remove them for submission
%\widetext
% \leftline{Version xx as of \today}
% \leftline{Primary authors: Joe E. Physics}
% \leftline{To be submitted to (PRL, PRD-RC, PRD, PLB; choose one.)}
% \leftline{Comment to {\tt d0-run2eb-nnn@fnal.gov} by xxx, yyy}
% \centerline{\em D\O\ INTERNAL DOCUMENT -- NOT FOR PUBLIC DISTRIBUTION}

% the following line is for submission, including submission to the arXiv!!
%\hspace{5.2in} \mbox{Fermilab-Pub-04/xxx-E}

\title{Influence of broken flavor and C and P symmetry on the quark propagator}
%\input author_list.tex       % D0 authors (remove the first 3 lines
                             % of this file prior to submission, they
                             % contain a time stamp for the authorlist)
                             % (includes institutions and visitors)
\date{\today}

\author{Axel Maas}
 \email{axel.maas@uni-graz.at}
\author{Walid Ahmed Mian}%
 \email{walid.mian@uni-graz.at}
\affiliation{
Institute of Physics, NAWI Graz, University of Graz,
Universit\"atsplatz 5, A-8010 Graz, Austria
}%

%%%%%%%%%%%%%%%%%%%%%%%%%%%%%%%%%%%%%%%%%%%%%%%%%%%%%%%%%%%%%%%%%%%%%%%%%%%%%%%%%%%%%%%%%555
%%%%%%%%%%%%%%%%%%%%%%%%%%%%%%%%%%%%%%%%%%%%%%%%%%%%%%%%%%%%%%%%%%%%%%%%%%%%%%%%%%%%%%%%%555
\begin{abstract}
Embedding QCD into the standard model breaks various symmetries of QCD explicitly, especially C and P.
While these effects are usually perturbatively small, they can be amplified in extreme 
environments like merging neutron stars or by the interplay with new physics. 
To correctly treat these cases requires fully backcoupled calculations. 
To pave the way for later investigations of hadronic physics, we study the QCD quark 
propagator coupled to an explicit breaking. 
This substantially increases the tensor structure even for this simplest correlation function. 
To cope with the symmetry structure, and covering all possible quark masses, 
from the top quark mass to the chiral limit, we employ Dyson-Schwinger equations. 
While at weak breaking the qualitative effects have similar trends as in perturbation theory, 
even moderately strong breakings lead to qualitatively different effects, non-linearly amplified by the strong interactions.
\end{abstract}

%\pacs{}
\maketitle

%%%%%%%%%%%%%%%%%%%%%%%%%%%%%%%%%%%%%%%%%%%%%%%%%%%%%%%%%%%%%%%%%%%%%%%%%%%%%%%%%%%%%%%%%555
%%%%%%%%%%%%%%%%%%%%%%%%%%%%%%%%%%%%%%%%%%%%%%%%%%%%%%%%%%%%%%%%%%%%%%%%%%%%%%%%%%%%%%%%%555
\section{Introduction}

With the detection of gravitational waves \cite{Abbott:2016blz} a whole new era of astronomy has begun. 
Eventually, this will allow to investigate neutron star mergers. 
In such dense environments, weak interactions become so prevalent that the dynamical 
backcoupling between the weak and the strong interaction becomes relevant, see e.\ g.\ 
\cite{Rosswog:2003rv,Sekiguchi:2011zd,Faber:2012rw,Neilsen:2014hha,Palenzuela:2015dqa,
Sekiguchi:2015dma,Foucart:2015gaa,Caballero:2016lof}. 
This requires therefore a fully coupled, and necessarily non-perturbative, description. 
This is so far only possible at the level of comparatively simple effective models, 
but not yet in an ab-initio calculation.

This is not the only reason to consider this problem. 
Beyond the known standard model (hidden) sectors may exist in which strongly 
interacting parity-conserving and parity-breaking interactions both exist. 

Both these considerations motivate to understand such backcouplings better. 
The main hallmark of both scenarios is the appearance of explicit C and P symmetry breaking, 
as well as, to a lesser extent, flavor breaking. The presence of additional, or non-negligible, 
symmetry breaking effects implies always a more involved tensor structure of correlation functions. 
Therefore, we focus here on the simplest object, which exhibits 
the full additional complexity, the quark propagator (QP). 
Since we are mainly interested in how the strong interaction may amplify, 
or modify, the symmetry breaking effects, we will consider only an explicit source of the symmetry breaking, 
as will be discussed in greater detail in section \ref{Basics}. 
For the weak interactions, which, due to explicit masses of the $W$ and $Z$ bosons do not 
show a strong momentum dependence at low energies, this is a sufficient approximation.

However, the inclusion of such a breaking, and the large differences in relevant energies, 
already limits the possible choices of methods. Especially, lattice gauge theory is not suitable. 
This is on the one hand due to the immense computational costs for the vastly different energy levels involved. 
On the other hand, there is no fully proven way yet to upgrade the static breaking considered 
here to the full weak interactions using lattice methods \cite{Hasenfratz:2007dp}.

As an alternative, here functional methods in the form of Dyson-Schwinger equations (DSEs) will be employed. 
These have been used very successfully to determine the quark propagator in QCD in various levels of sophistication 
\cite{Alkofer:2000wg,Fischer:2006ub,Alkofer:2008tt,Roberts:2015lja,Williams:2015cvx}. 
In the more exploratory investigations here, the most important features are the dynamical mass generation 
as well as the correct implementation of chiral symmetry. 
These features are particularly well implemented in the so-called rainbow-ladder truncation 
\cite{Alkofer:2000wg,Fischer:2006ub,Roberts:2015lja}. 
This completes our setup, which we describe in much more detail in section \ref{Basics} as well as 
in the appendices \ref{Structure of QP} and \ref{DSE for QP}. Especially in appendix \ref{Structure of QP}, 
we will discuss the tensor structure of the quark propagator, which has now four matrix-valued, 
rather than two real-valued, dressing functions, demonstrating the much higher complexity compared to QCD alone.

The most important results will be discussed in section \ref{Result}. 
Probably the most relevant insight gained, aside from the necessary technology to deal with the increase of complexity, 
is that the fully coupled system behaves as expected from perturbation theory only at very weak breaking. 
Already at moderately small breaking, the amplification by the strong interaction can lead to qualitatively 
different behaviors for various dressing functions of the quark propagator than perturbatively expected. 
Of course, only future investigations of observable quantities will tell what the implications for physics are, 
but the present results mandates caution with extrapolation of perturbative notions. 
In this section, we will also present information on the analytic structure of the quark propagator using 
its Schwinger function, an issue which is already in QCD alone highly non-trivial 
\cite{Alkofer:2000wg,Fischer:2006ub,Roberts:2015lja}.

Finally, we list many further results in appendix \ref{Numerical Results}, providing a complete 
picture of the quark propagator in this setup for a wide range of parameters and quark masses. 
All of the insights and results are finally summarized in section \ref{Conclusion}.

Some preliminary results have already been reported in \cite{Mian:2016eel}.

%%%%%%%%%%%%%%%%%%%%%%%%%%%%%%%%%%%%%%%%%%%%%%%%%%%%%%%%%%%%%%%%%%%%%%%%%%%%%%%%%%%%%%%%%555
%%%%%%%%%%%%%%%%%%%%%%%%%%%%%%%%%%%%%%%%%%%%%%%%%%%%%%%%%%%%%%%%%%%%%%%%%%%%%%%%%%%%%%%%%555
\section{Basics and Methods}
\label{Basics}

%%%%%%%%%%%%%%%%%%%%%%%%%%%%%%%%%%%%%%%%%%%%%%%%%%%%%%%%%%%%%%%%%%%%%%%%%%%%%%%%%%%%%%%%%555
\subsection{Ansatz}

The breaking of the symmetries will be realized by including an explicitly symmetry-violating term in the Lagrangian.
The breaking is thus generated already at tree-level. 
Since the weak interactions motivate our study, the term will break C, P, and flavor within generations.
Thus, our QP becomes matrix-valued in flavor space, with the off-diagonal elements mediating flavor changes.

Flavor violation within a generation is actually not possible without involving further particles, 
due to electric charge conservation. 
In the standard model, this is ensured by the emission of a lepton and a (anti)neutrino. 
To avoid this additional complexity, here the leptons are modeled as an external background field. 
Given that our ultimate interest is in neutron star mergers, where such a reservoir is readily available,
this appears like a reasonable approximation.

In the following the subscript $u$ and $d$ denotes up-like quarks and down-like quarks
and the superscript $L$ and $R$ left-handed and right-handed quarks, respectively. 
The strength of the weak interaction, or more precisely the coupling to the symmetry breaking external field, 
will be denoted by $g_{\text{w}}$, the effective weak strength. 
We will vary the value of $g_{\text{w}}$ from small values to large values to turn on the effects smoothly. 
Finally, the quark fields are denoted by $\psi$.

All together leads to the Lagrangian
\begin{align}
  \mathcal{L} =&  \mathcal{L}_{\text{QCD}} +  \mathcal{L}_{\text{Effective}}, 
 \nonumber   \\
  \mathcal{L}_{\text{QCD}}  =&
  \overline{\psi}_{u} \left[ - \slashed{\partial} +m_u \right] \psi_u +
  \overline{\psi}_{d} \left[ - \slashed{\partial} +m_d \right] \psi_d\nonumber\\
  &+  
  g_s \overline{\psi} \slashed{A^{i}} T^{i} \psi +
  \mathcal{L}_{\text{Rest}},
 \nonumber  \\ 
  \mathcal{L}_{\text{Effective}} =&
  - 2 g_{\text{w}} \left( \overline{\psi}^{L}_{u} \slashed{\partial} \psi^{L}_{d}   +  
  \overline{\psi}^{L}_{d} \slashed{\partial} \psi^{L}_{u} \right),
 \label{Eqn:Ansatz} \\
  \psi^{L} =& \frac{1}{2}(\unit - \gamma^{5})\psi,
 \nonumber
\end{align}
where $A^{i}$ are the gluon fields and $T^{i}$ are the generators of $\su(3)$.
$m_{u}$ and $m_{d}$ are the masses for up-like and down-like quarks. 
$\mathcal{L}_{\text{Rest}}$ is the remainder of the QCD Lagrangian, which includes the gluon self 
interaction and the gluon-ghost part, and does not play an explicit role in the following. 
It also contains the gauge-fixing terms of our choice of (minimal) Landau gauge. 
For brevity, we also suppressed the renormalization constants.

%%%%%%%%%%%%%%%%%%%%%%%%%%%%%%%%%%%%%%%%%%%%%%%%%%%%%%%%%%%%%%%%%%%%%%%%%%%%%%%%%%%%%%%%%555
\subsection{Quark Propagator}
\label{QP}

In the following, $P_{AB}$ is the propagator from flavor $A$ to $B$  with $A,B \in \left\{ u, d\right\}$ for up-like and
down-like quarks. Its inverse will be denoted by $P^{-1}_{AB}$.
Because of parity violation the QPs have in addition to the usual vector- and
scalar channels also non-vanishing axial- and pseudo-scalar channels.
The standard notation for the vector- and scalar channel dressing functions 
of the inverse Propagator in the literature is $A$ and $B$.
We will keep this and denote the dressing functions for the
axial- and pseudo-scalar channels of the inverse propagator with $C$ and $D$.
The corresponding dressing functions of the propagator are denoted by a tilde.

This leads to the form
\begin{align}
 P_{AB} (p^2) &= \tilde{A}_{AB}(p^2) \img \slashed{p} + \tilde{B}_{AB}(p^2) \unit\nonumber\\
 &+ 
 \tilde{C}_{AB}(p^2) \img \slashed{p} \gamma^5 + \tilde{D}_{AB}(p^2) \gamma^5,
 \nonumber \\
 P^{-1}_{AB} (p^2) &= - A_{AB}(p^2) \img \slashed{p} + B_{AB}(p^2) \unit\nonumber\\
 &+ 
 C_{AB}(p^2) \img \slashed{p} \gamma^5 + D_{AB}(p^2) \gamma^5.
 \label{Eqn:SQP}
\end{align}
The dressing functions of the propagator and its inverse are related with each other.
In general the dressing functions of the propagator depends on all the dressing functions
of the inverse propagator in a complicated way, see for details appendix \ref{Structure of QP}, and generically denoted as
\begin{align}
 \tilde{A}_{AB} &= \tilde{A}_{AB}(A_{CD},B_{CD},C_{CD},D_{CD}).
 \label{Eqn:RDF}
\end{align}
Instead of splitting the Lorenz channels into the vector- and axial channels, 
it can also be split into left-handed  $\tilde{L}(p^2)$ and right-handed $\tilde{R}(p^2)$ components, leading to
\begin{align}
 P_{AB} (p^2) &= \frac{\tilde{L}_{AB}(p^2)}{\sqrt{2}} \img \slashed{p}(\unit - \gamma^5) +
 \frac{\tilde{R}_{AB}(p^2)}{\sqrt{2}} \img \slashed{p}(\unit + \gamma^5)\nonumber\\
 &+ \tilde{B}_{AB}(p^2) \unit + 
 \tilde{D}_{AB}(p^2) \gamma^5,
 \label{Eqn:SQPLR}
\end{align}
with the relations
\begin{align}
 \tilde{L}_{AB}(p^2)&= \frac{1}{\sqrt{2}} \left( \tilde{A}_{AB}(p^2)-\tilde{C}_{AB}(p^2) \right),
 \nonumber \\
 \tilde{R}_{AB}(p^2)&= \frac{1}{\sqrt{2}} \left( \tilde{A}_{AB}(p^2)+\tilde{C}_{AB}(p^2) \right).
 \label{Eqn:RLC}
\end{align}
\noindent At tree-level the propagator reads
\begin{align}
P_{0,uu}(p^2)=& \frac{1}{N(p^{2})} \left[ (m_{d}^{2}+(1-2g_{\text{w}}^{2})p^{2}) \img \slashed{p}\right.\nonumber\\ 
&\left.+ m_{u}(m_{d}^{2}+p^{2}) \unit + 2g_{\text{w}}^{2}p^{2} \img \slashed{p} \gamma^{5} \right],
\nonumber 
\\
P_{0,dd}(p^2)=& \frac{1}{N(p^{2})} \left[ (m_{u}^{2} + (1-2g_{\text{w}}^{2})p^{2}) \img \slashed{p}\right.\nonumber\\
&\left.+ m_{d} (m_{u}^{2}+p^{2}) \unit +2g_{\text{w}}^{2}p^{2} \img \slashed{p} \gamma^{5} \right],
\label{Eqn:TLprop} \\
P_{0,ab,a\neq b}(p^2)&= \frac{g_{\text{w}}}{N(p^{2})}\left[ (m_{u}m_{d}-p^{2}) \img \slashed{p} \right.\nonumber\\
&\left.- (m_{u}+m_{d})p^{2} \unit -  (m_{u}m_{d}+p^{2}) \img \slashed{p} \gamma^{5}\right.\nonumber\\
&\left.-\sigma_{ab}(m_{u}-m_{d})p^{2} \gamma^{5}\right],
\nonumber
\end{align}
where the common denominator is given by
\begin{align}
N(p^{2})=& m_{d}^{2} m_{u}^{2} + (m_{u}^{2} + m_{d}^{2}) p^{2}+ (1-4g_{\text{w}}^{2}) p^{4}
\nonumber 
\\
=& (1-4g_{\text{w}}^{2})(p^{2}+M_{l}^{2})(p^{2}+M_{h}^2),
\nonumber 
\\
M_{l}=& \sqrt{ \frac{m_{u}^{2}+m_{d}^{2} - \sqrt{ (m_{u}^{2}-m_{d}^{2})^{2}
 +16g_{\text{w}}^2m_{u}^{2}m_{d}^{2}}} {2(1-4g_{\text{w}}^{2})}},
\label{Eqn:DNQP}
\\
M_{h}=& \sqrt{ \frac{m_{u}^{2}+m_{d}^{2} + \sqrt{ (m_{u}^{2}-m_{d}^{2})^{2}
 +16g_{\text{w}}^2m_{u}^{2}m_{d}^{2}}} {2(1-4g_{\text{w}}^{2})}}.
\nonumber
\end{align}
The quantity $\sigma_{ab}$ is 1 for $a=u$ and $b=d$ and $-1$ in the other case. 
The tree-level propagator already reveals the major contributions for the QPs. 
They are separated in the different channels at tree-level, but will mix in the full case.

Consider the denominator.
In the second line of equation (\ref{Eqn:DNQP}) we have factorized
the denominator to see both poles of the tree-level propagator.
For $g_{\text{w}} \to 0$ $M_{l}$ goes to the mass of the lighter quark  and 
$M_{h}$ to the mass of the heavier quark.
By increasing $g_\text{w}$, the value of $M_{l}$ is decreased
and $M_{h}$ is increased and thus increases the effect
of the mass splitting.

Note that $M_{l}$ and $M_{h}$ diverge at $g_w=0.5$. 
At this point the poles turn imaginary, indicating a breakdown of the trivial vacuum 
around which the perturbative expansion is performed. 
This feature will actually not be lifted in the full non-perturbative treatment, 
and we are only able to find solutions as long as $g_{\text{w}}\lesssim 0.4$. 
Since this is a very large breaking, probably far too strong for the setting of neutron star mergers 
guiding this work, we did not endeavor to find out what happens beyond this point, 
and restrict ourselves to breaking strengths below this value.

The explicit chiral symmetry breaking by the tree-level masses manifests itself in the scalar channel of the pure
and mixed flavors.
In the chiral limit the scalar channel vanishes at tree-level, but due to 
dynamical chiral breaking the scalar channel does not vanish for the full propagator.

The vector channel of the mixed propagator is proportional to $g_{\text{w}}(m_u m_d -p^2)$,
which changes its sign for $p^2>m_u m_d$.
The influence of this is a contribution in different direction for large and low momenta.
For heavier bare quark masses this contribution is shifted to higher momenta.

The most remarkable contribution, a difference of both quark masses, 
appears in the pseudo-scalar channel of the mixed propagator. 
This will have a significant impact on the full propagator. 
It is remarkable that this contribution appears with opposite sign for the propagator from up-like quarks
to down-like quarks and the other way around: 
Although we have taken the same strength for the propagation of both mixed QP in our ansatz,
we get a difference, if the quarks have different masses.

%%%%%%%%%%%%%%%%%%%%%%%%%%%%%%%%%%%%%%%%%%%%%%%%%%%%%%%%%%%%%%%%%%%%%%%%%%%%%%%%%%%%%%%%%%%%5
\subsection{DSEs}
\label{DSE}

\begin{figure*}
 \includegraphics[scale=0.41]{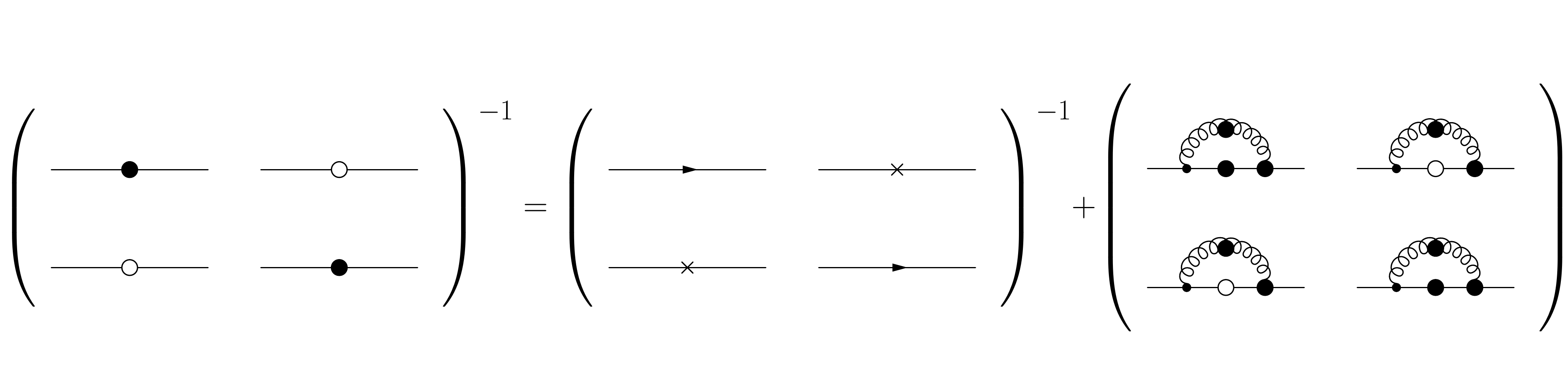}
  \caption{Diagrammatic representation of the DSEs for the QPs.
  The solid lines with an arrow and x represents tree-level QPs
  for pure and mixed flavors, respectively.
  Analog the solid lines with filled and empty blob represents the full QPs
  for pure and mixed flavors, respectively.
  The wiggly lines represents the propagation of gluons.
  A small and big filled blob at a vertex indicates a bare and a full vertex,
  respectively.}
  \label{Pic:DSE_QP}
\end{figure*}

We can write the QP in a matrix form, where the diagonal elements are the QPs for pure flavor
and the off-diagonal elements are the QPs for mixed flavors, 
see for details appendix \ref{Structure of QP}.
A graphical representation of the DSEs is given in figure \ref{Pic:DSE_QP}.
The equations look  similar to those of QCD, where the QP can also be given in a matrix form, 
but with vanishing off-diagonal elements. 
It is only the appearance of the off-diagonal tree-level
elements, which gives rise to all differences.

The full quark-gluon-vertex appears in the self energy graph, 
which is determined by a separate DSEs involving even higher-order correlation functions.

To avoid this complication, the DSEs are truncated at this level, 
which is known as the rainbow truncation \cite{Alkofer:2000wg,Fischer:2006ub,Roberts:2015lja}. 
The required inputs of the full gluon propagator and the full quark-gluon-vertex 
are then replaced by a bare gluon propagator and a quark-gluon vertex given by the tree-level tensor structure, 
but dressed with an effective running coupling $\alpha$. 
Denoting the tree-level inverse propagator with $P^{-1}_{0,AB}$, the final DSE reads
\begin{align}
 &&P^{-1}_{AB}(p^2,\mu^2) = \sqrt{Z_{2,A}(\mu^2,\Lambda^2) Z_{2,B}(\mu^2,\Lambda^2) } P^{-1}_{0,AB}\nonumber\\
 &&+ \frac{Z_{2,A}(\mu^2,\Lambda^2) Z_{2,B}(\mu^2,\Lambda^2)}{3 \pi^3} \times \nonumber \\
 &&\times \int^{\Lambda}{\dd^{4} q \frac{\alpha (k^2)}{k^2}\left(\delta_{\nu \rho} - \frac{k_{\nu}k_{\rho}}{k^2}\right)
 \gamma_{\nu}P_{BA}(q^2,\mu^2) \gamma_{\rho}},
 \label{Eqn:DSE_QP}
\end{align}
where $Z_{2,A}$ and $Z_{2,B}$ are the quark wave function-renormalization constants for flavor A and B. 
$\int^{\Lambda}$ represents a translationally-invariant regularization with the UV-cutoff $\Lambda$. 
$\mu$ is the renormalization point and $k=q-p$.

For the dressing function $\alpha$ we choose the Maris-Tandy coupling \cite{Maris:1999nt}
\begin{align}
  \alpha(q^2) =& \frac{\pi}{\omega^6} D q^4 \e^{-\frac{q^2}{\omega^2}}
  + \frac{2\pi \gamma_m [1-\exp{(-\frac{q^2}{m_t^2})}]}{\ln[\e^2-1+(1+\frac{q^2}{\Lambda^2_{\text{QCD}}})^2]},
 \label{Eqn:MTC}.
\end{align}
Here the parameters are adapted to describe pions in the vacuum adequately. 
In the literature these parameters are fitted for degenerate masses of up 
and down quarks \cite{Goecke:2011pe}. 
For a detailed analyses of different parameter sets see, e.\ g., \cite{Sanchis-Alepuz:2014wea,Hilger:2015ora}. 
To allow for comparison, we choose here one such set, namely
$\Lambda_{\text{QCD}} = 0.234$ GeV, $m_t = 1.0$ GeV, $\omega = 0.4 $ GeV, 
and $D=0.93$ GeV. $\gamma_{m}=12/(11 N_{c}-2 N_{f})$ is the anomalous dimension of the quark propagator. 
Because we consider each quark generation on its own, we choose $N_f=2$ and $N_c=3$.

For the bare quark masses we take $m_{\text{up}}=2.3$ MeV, $m_{\text{down}}=4.8$ MeV, 
$m_{\text{strange}}=95$ MeV, $m_{\text{charm}}=1.275$ GeV, $m_{\text{bottom}}=4.18$ GeV and $m_{\text{top}}=160$ GeV, 
always at the renormalization point of $\mu=10^{6}$ GeV. In addition, we will also consider the chiral limit as well degenerate cases.

Further details can be found in appendix \ref{DSE for QP}.

%%%%%%%%%%%%%%%%%%%%%%%%%%%%%%%%%%%%%%%%%%%%%%%%%%%%%%%%%%%%%%%%%%%%%%%%%%%%%%%%%%%%%%%%%
\subsection{Schwinger function and Masses}
\label{Schwinger}

To obtain information on the analytic structure, we also determine the Schwinger function \cite{Alkofer:2003jj, Maas:2011se}. 
It is defined as
\begin{align}
 \Delta_{AB}(t) =& \frac{1}{\pi} \int_{0}^{\infty}{\dd p_{4} \cos(t p_{4}) \sigma_{AB}(p_4^2)}.
 \label{Eqn:SWF}
\end{align}
$\sigma_{AB}(p_4^2)$ is one of the dressing functions from the propagator evaluated at zero spatial momenta ($\vec{p}=0$).

The actual analytic form of the propagator is yet unknown, 
but poles and/or cuts appear likely \cite{Alkofer:2003jj,Alkofer:2000wg,Roberts:2015lja}. 
If there would be only an ordinary mass pole, the Schwinger function would show an exponential decay \cite{Maas:2011se}
\begin{align}
 \Delta(t) \sim \e^{-m t}
 \label{Eqn:RPS}
\end{align}
\noindent and $m$ would be the mass.

However, investigation of pure QCD in the rainbow truncation yielded rather a structure with complex conjugated poles 
\cite{Alkofer:2003jj,Alkofer:2000wg,Roberts:2015lja}, which is e.\ g.\ expected in the Gribov-Stingl scenario 
\cite{Habel:1989aq,Habel:1990tw,Stingl:1994nk}. 
Note, however, that this may be a truncation artifact. 
In this case the Schwinger function is roughly given by
\begin{align}
 \Delta(t) \sim \e^{-a t} \cos(bt+\delta).
 \label{Eqn:CPS}
\end{align}
The decay rate is given by the real part and the oscillation frequency by the imaginary part of the mass pole.

%%%%%%%%%%%%%%%%%%%%%%%%%%%%%%%%%%%%%%%%%%%%%%%%%%%%%%%%%%%%%%%%%%%%%%%%%%%%%%%%
\section{Results}
\label{Result}
For each quark generation, we have 2 dressing functions for pure flavor and 2 dressing functions for mixed flavor 
and each has 4 Lorentz channels, resulting in 16 dressing functions. 
We will consider 6 cases in total, the chiral limit and three physical quark generations and two cases of degenerate masses. 
Therefore we have numerical results for $192$ dressing functions and each of them as a function of the weak strength. 
In addition to that, we also have the Schwinger function for each dressing function. 
To avoid cluttering up the main text, most of these results are relegated to appendix \ref{Numerical Results}. 
Here, only the qualitatively most remarkable results will be analyzed. 
The results in the appendix do not add any conceptual new to this section.

The result for the QP in QCD are usually \cite{Alkofer:2003jj,Alkofer:2000wg,Roberts:2015lja} given in 
terms of the wave function renormalization $Z(p^2)=1/A(p^2)$ and mass function $M(p^2)=B(p^2)/A(p^2)$. 
For ease of comparison, the case with $g_{\text{w}}=0$ will serve as reference. 
Therefore in section \ref{WFR} the results for $Z$ and $\tilde{A}_{AA}$ will be explored. 
Afterwards, the mass function and the related $\tilde{B}_{AA}$ will be discussed in section \ref{MF}. 
In section \ref{Parity Violation} we will analyze the results for the axial channel $\tilde{C}_{AA}$, 
and study the impact of parity violation. 
The results for the Schwinger function will be discussed in section \ref{Masspole}. 
But first, it is necessary to discuss the involved scales, as the problem is now a multi-scale one. 
Note that in the chiral limit there is no difference between the up-like quark and down-like quark, 
and thus the flavor-diagonal elements coincide. In these cases, always the up-type one will be shown.

\subsection{Relative scales}

\begin{figure*}
 \includegraphics[width=0.5\linewidth]{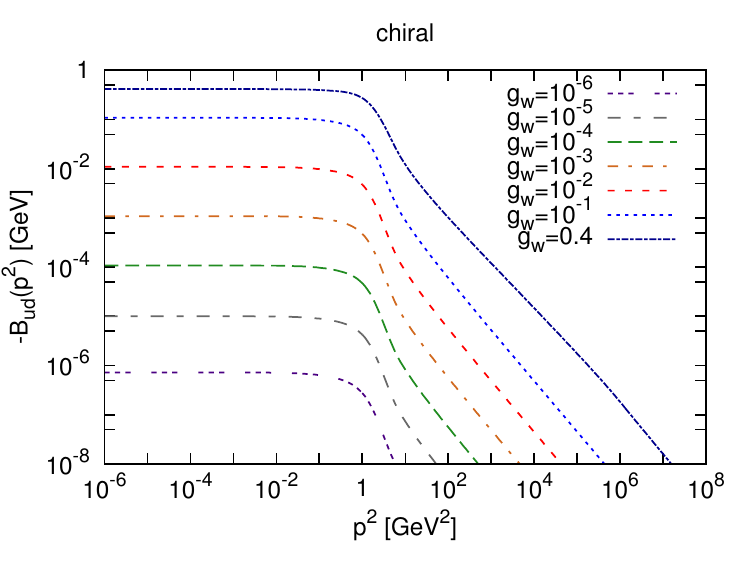}\includegraphics[width=0.5\linewidth]{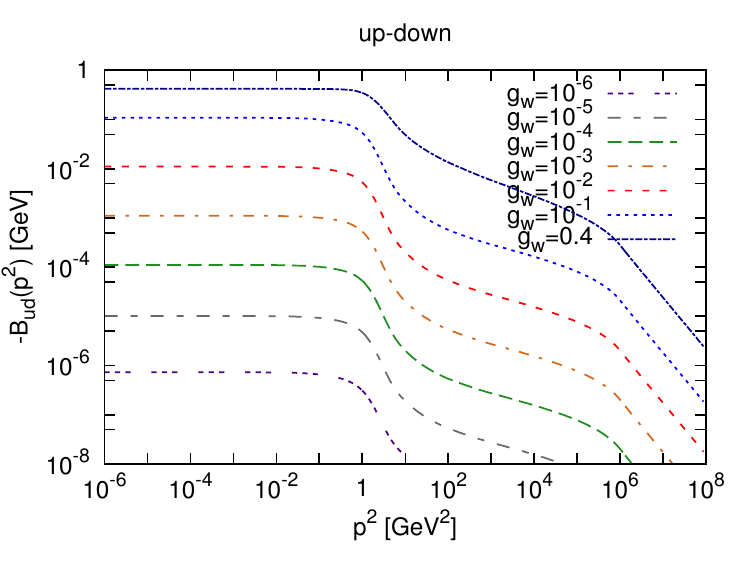}\\
 \includegraphics[width=0.5\linewidth]{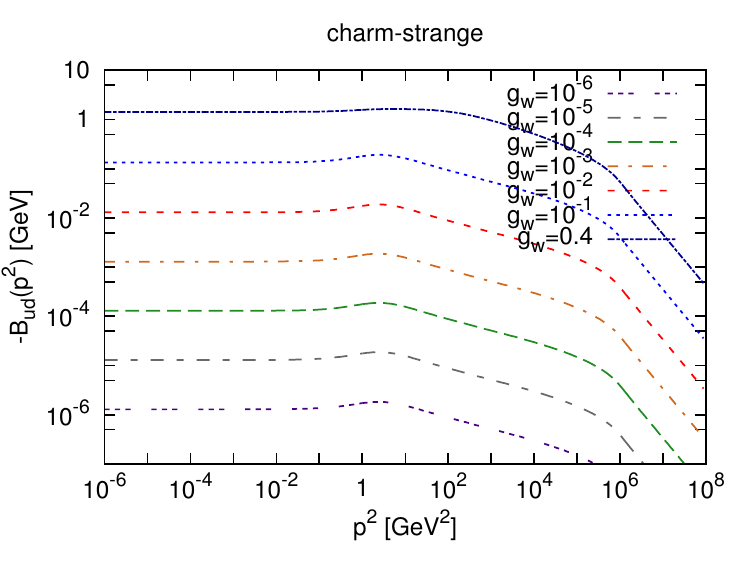}\includegraphics[width=0.5\linewidth]{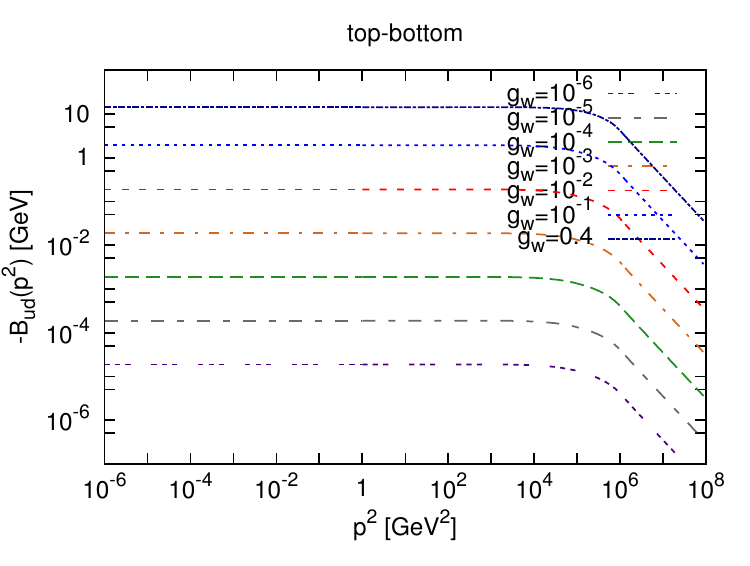}
  \caption{The scalar channel for the inverse mixed propagator in the chiral limit and for all three quark generations. 
  For $g_{\text{w}} \leq 0.1$ the mass in the IR is approximately proportional to $g_{\text{w}}$.}
  \label{Pic:IBM}
\end{figure*}

$g_{\text{w}}$ is dimensionless. 
Thus, a comparison of the strength of breaking with the strong interaction scale $\Lambda_{\text{QCD}}$ is not directly possible. 
However, in the scalar channel, the interaction is found to be transmuted into a momentum scale. 
This is particularly true for the flavor off-diagonal-element, which is zero without breaking. 
It is shown for various quark masses in figure \ref{Pic:IBM}

It is seen that in the IR this dressing function is approximately proportional to the weak strength for $g_{\text{w}} \leq 0.1$. 
Also, for small values of the weak strength, the scale generated is small compared to $\Lambda_{\text{QCD}}$.
At the largest values of $g_{\text{w}}$ the generated scale becomes of the same order as $\Lambda_{\text{QCD}}$, 
which therefore substantially deviates from nature. 
This also justifies our choice to restrict to not too large values of $g_{\text{w}}$. 
However, such effects may play a role in theories with strongly-interacting chiral sectors.

Note that the generated scale depends on the quark masses. 
For both light generations the difference to the chiral limit is small. 
For top and bottom quark, the generated scale is one order of magnitude bigger at the same $g_{\text{w}}$. 
Thus, there is a linking of the different involved scales.

%%%%%%%%%%%%%%%%%%%%%%%%%%%%%%%%%%%%%%%%%%%%%%%%%%%%%%%%%%%%%%%%%%%%%%%%%%%%%%%%%%%%%%%%%%%%5
\subsection{Wave function renormalization}
\label{WFR}

\begin{figure}
 \includegraphics[width=\linewidth]{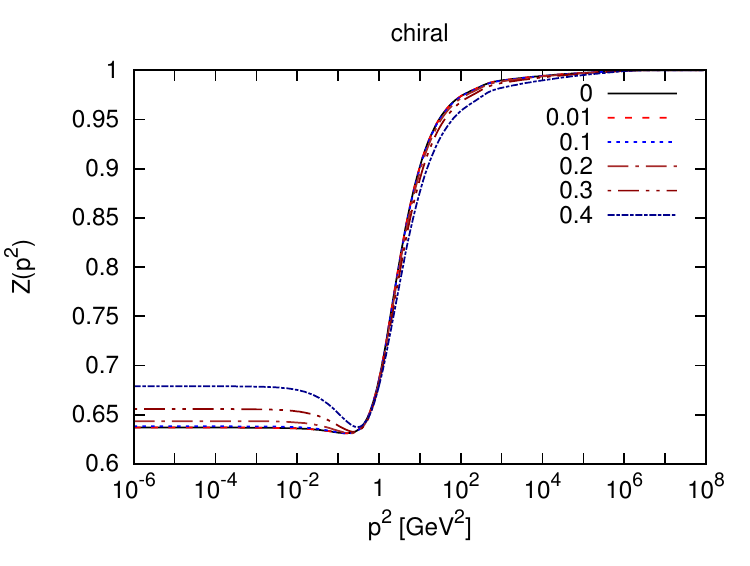}
 \includegraphics[width=\linewidth]{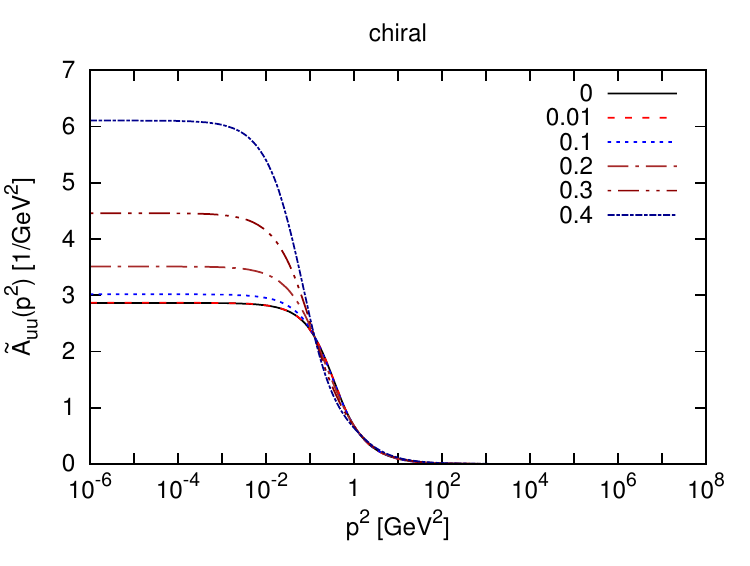}
  \caption{The wave function renormalization $Z(p^2)$ (top panel) and the vector channel (bottom panel) 
  for different values of $g_{\text{w}}$ in the chiral limit.}
  \label{Pic:WF01}
\end{figure}

From equation (\ref{Eqn:DSE_DF1}) follows that $A_{uu}$ and $\tilde{A}_{uu}$ are directly linked with each other, 
and thus $Z=1/A_{uu}$ is also directly related to $\tilde{A}_{uu}$. 
These dressing functions are shown in figure \ref{Pic:WF01} for different values of $g_{\text{w}}$ in the chiral limit. 
For values of $g_{\text{w}} \lesssim 0.01$ no appreciable effect is seen. 
At larger values $\tilde{A}_{uu}$ slightly increases in the UV when increasing $g_{\text{w}}$. 
In the mid momenta regime it is slightly decreased and in the IR it is significantly increased. 
A consequence of this is that $Z$ also increases in the IR. 
This can be understood from (\ref{Eqn:DSE_DF1}), as $A_{uu}$ is obtained from integrating
$\tilde{A}_{uu}$ multiplied with a kernel over all momenta.
Because $\tilde{A}_{uu}$ is increased in the UV very little and more decreased in the mid range, 
$Z$ is slightly decreased in the UV range due to the integration. 
For the same reason $Z$ is not increased as much as $\tilde{A}_{uu}$ is increased in the IR.

\begin{figure*}
 \includegraphics[width=0.5\linewidth]{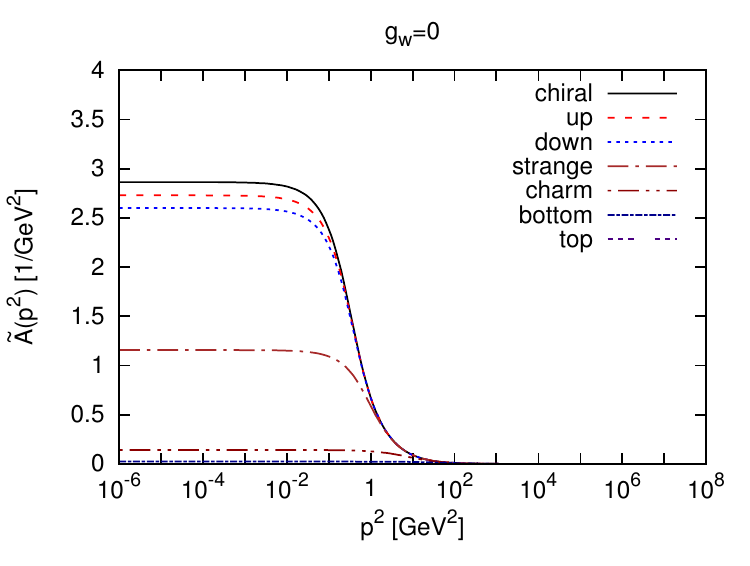}\includegraphics[width=0.5\linewidth]{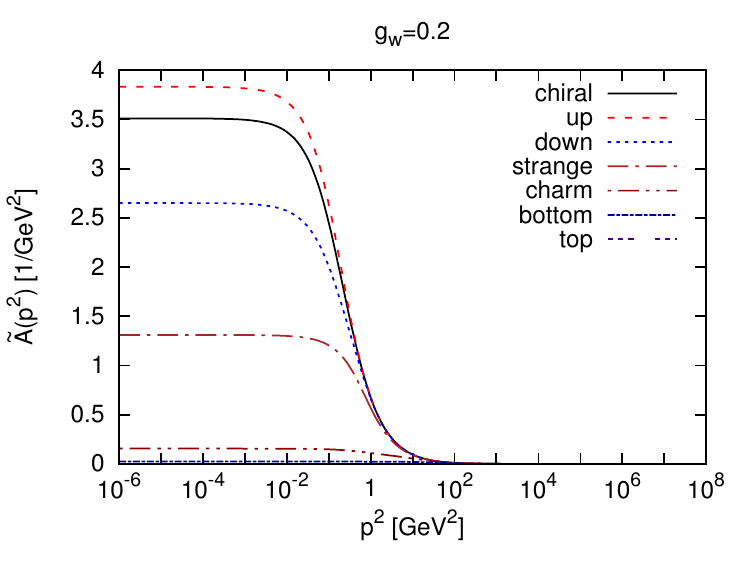}\\
 \includegraphics[width=0.5\linewidth]{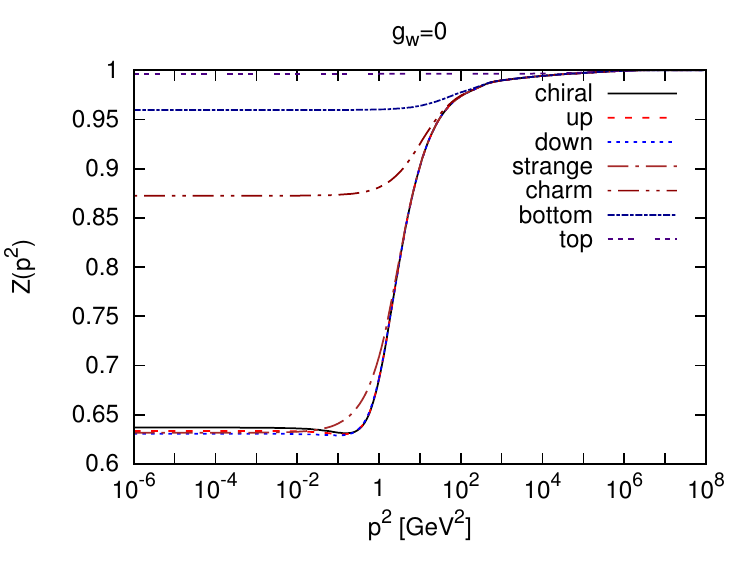}\includegraphics[width=0.5\linewidth]{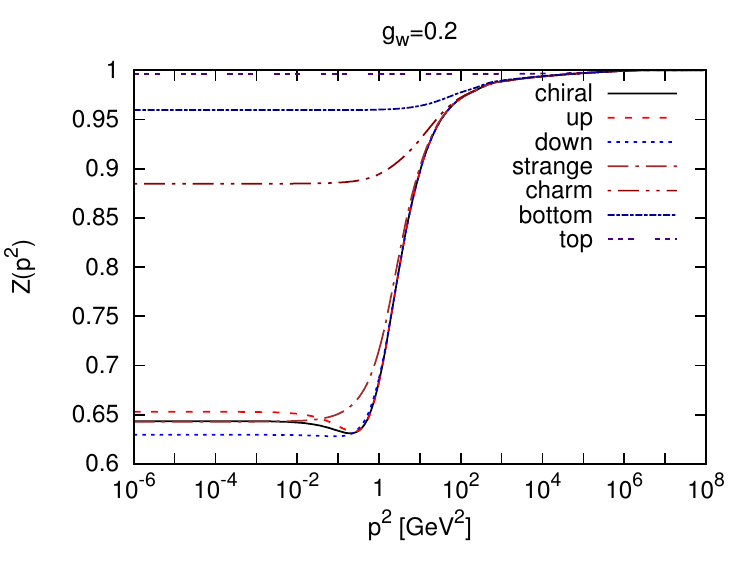}
  \caption{The flavor-diagonal vector channel without (left panels) and with (right panel) explicit breaking. 
  The lower panels show the same for $Z(p^2)$.}
  \label{Pic:WF02}
\end{figure*}

For other quark masses the same behavior is seen, as shown in figure \ref{Pic:WF02}. 
The graph shows that $\tilde{A}$ is increased in the IR for all quark flavors and thus $Z$ is also increased.

The effect comes from different sources. 
One is from $g_{\text{w}}$ and the other from a combination of $g_{\text{w}}$ and the bare quark masses. 
Especially, $A$ and $Z$ are increased for the up quarks more than in the chiral limit. 
Also, in general the value for up-like quarks is increased more than for down-like quarks. 
This is also seen in figure \ref{Pic:APAF01} in appendix \ref{VC} in more detail.

This can be understood from equation (\ref{Eqn:TLprop}) for the tree-level case. 
One of the contributions arises from the bare quark masses and another from mass splitting with different signs. 
This creates the cross-talks leading to the observed effects. 
We will return to this later in section \ref{Parity Violation}. 
In addition, the absolute value is decreased for higher bare quark masses, as anticipated because 
the masses of the quarks enter in the denominator of the QP. 
This can bee seen already for the tree-level propagator in equation (\ref{Eqn:DNQP}).

%%%%%%%%%%%%%%%%%%%%%%%%%%%%%%%%%%%%%%%%%%%%%%%%%%%%%%%%%%%%%%%%%%%%%%%%%%%%%%%%%%%%%%%%%%%%%%
\subsection{Mass function}
\label{MF}

The relations between the dressing functions of the QP and its inverse are more involved as in QCD, 
see also appendix \ref{Structure of QP}. 
In QCD, the relation is given by
\begin{align}
 \tilde{A}_{AA}(p^2) =& \frac{A_{AA}(p^2)}{A^2_{AA}(p^2)p^2+B^2_{AA}(p^2)}
 = \frac{Z_{AA}(p^2)}{p^2+M^2_{AA}(p^2)},
 \nonumber \\
 \tilde{B}_{AA}(p^2) =& \frac{B_{AA}(p^2)}{A^2_{AA}(p^2)p^2+B^2_{AA}(p^2)}
 = \frac{Z_{AA}(p^2) M_{AA}(p^2)}{p^2+M^2_{AA}(p^2)}.
 \label{Eqn:RPQCD}
\end{align}
To be able to compare, we therefore choose to define a (pseudo) mass function as  
\begin{align}
 M_{AA}(p^2) =& \frac{B_{AA}(p^2)}{A_{AA}(p^2)},
 \label{Eqn:MFQCD}
\end{align}
\noindent which by construction coincides with the usual one in the QCD case. 
Of course, neither in QCD nor here this function needs to coincide with the actual mass. 
Any such statement requires the Schwinger function in section \ref{Masspole}. 
Nonetheless, we will stick here with the usual convention and call this quantity mass function.

\begin{figure}
 \includegraphics[width=\linewidth]{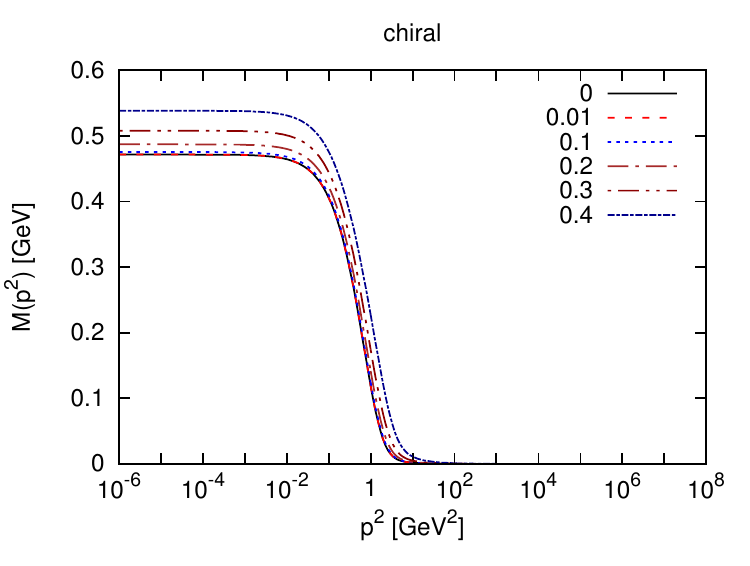}
 \includegraphics[width=\linewidth]{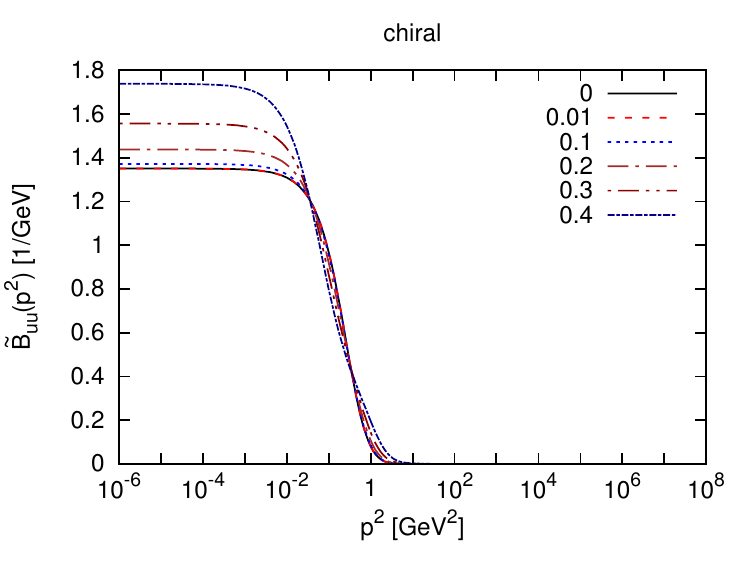}
  \caption{The mass function $M(p^2)$ (top panel) and the flavor-diagonal scalar channel (bottom panel) 
  for different values of $g_{\text{w}}$ in the chiral limit.}
  \label{Pic:MF01}
\end{figure}

The dependence on $g_{\text{w}}$ of this mass function is shown in figure \ref{Pic:MF01}. 
As in QCD, the mass function is non-zero, indicative of chiral symmetry breaking.
The mass function starts to change appreciably for $g_{\text{w}} \gtrsim 0.01$, 
like the wave function renormalization. 
The same is true for $\tilde{B}_{uu}$, which is also shown in figure \ref{Pic:MF01}. 
Since the connection between  $B_{uu}$ and $\tilde{B}_{uu}$, due to equation (\ref{Eqn:DSE_DF1}), 
is similar as for $A_{uu}$ and $\tilde{A}_{uu}$, the same analysis as before applies, 
and the response to $g_{\text{w}}$ follows the same pattern.

\begin{figure}
 \includegraphics[width=\linewidth]{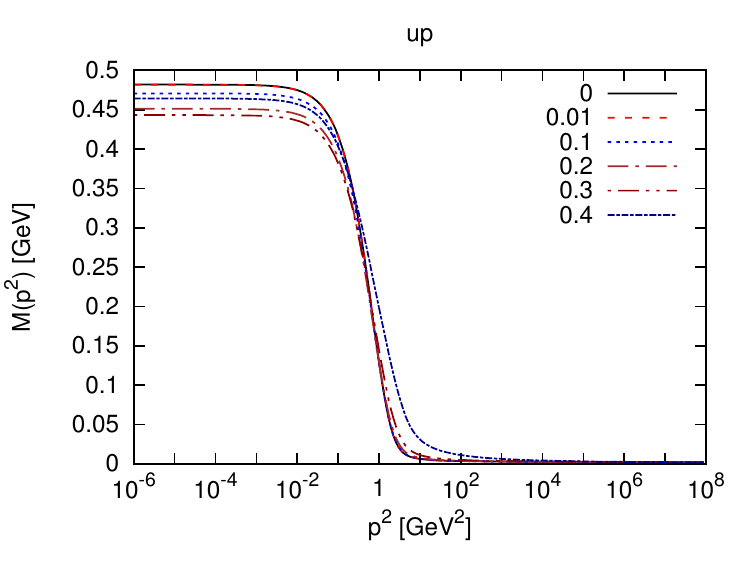}
 \includegraphics[width=\linewidth]{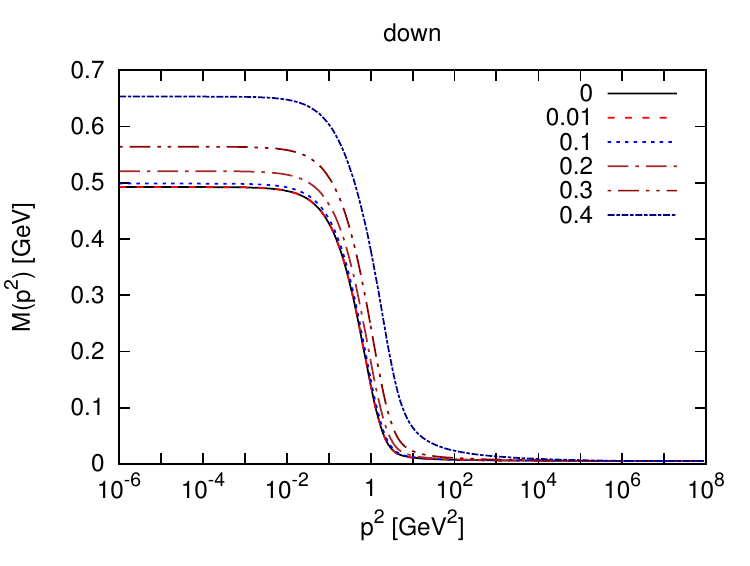}
  \caption{The mass function $M(p^2)$ for different values of 
  $g_{\text{w}}$ for the up quark (top panel) and down quark (bottom panel).}
  \label{Pic:MF02}
\end{figure}

At non-zero masses, the picture changes. This is shown in figure \ref{Pic:MF02} for up and down quarks. 
The mass function $M$ of the up quark is decreased by increasing $g_{\text{w}}$ for $g_{\text{w}}\lesssim 0.3$. 
For larger values it increases again, but here our approximations start to break down. 
This replicates the result of the tree-level propagator in section \ref{QP}: 
The mass of the heavier quark in a generation is increased and the mass of 
the lighter quark is decreased.

\begin{figure*}
 \includegraphics[width=0.5\linewidth]{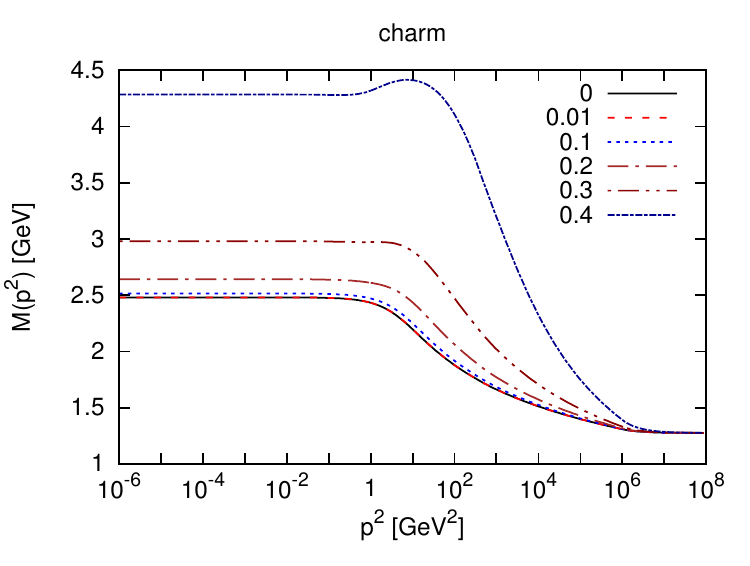}\includegraphics[width=0.5\linewidth]{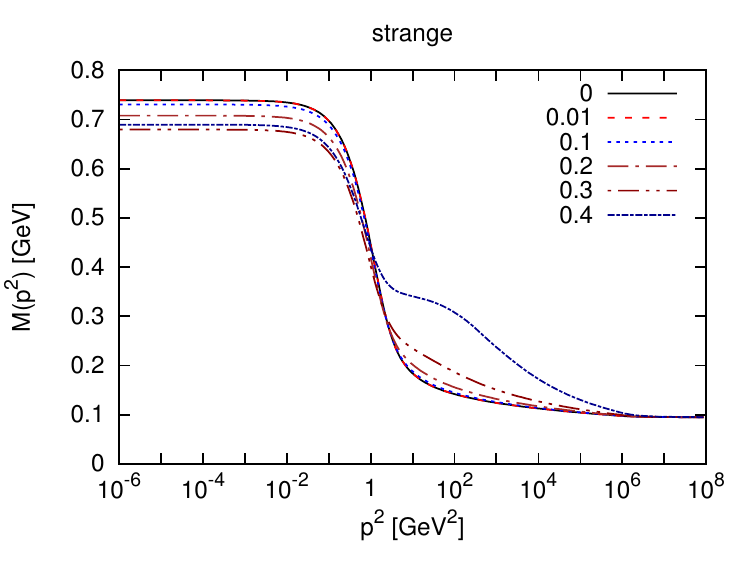}\\
 \includegraphics[width=0.5\linewidth]{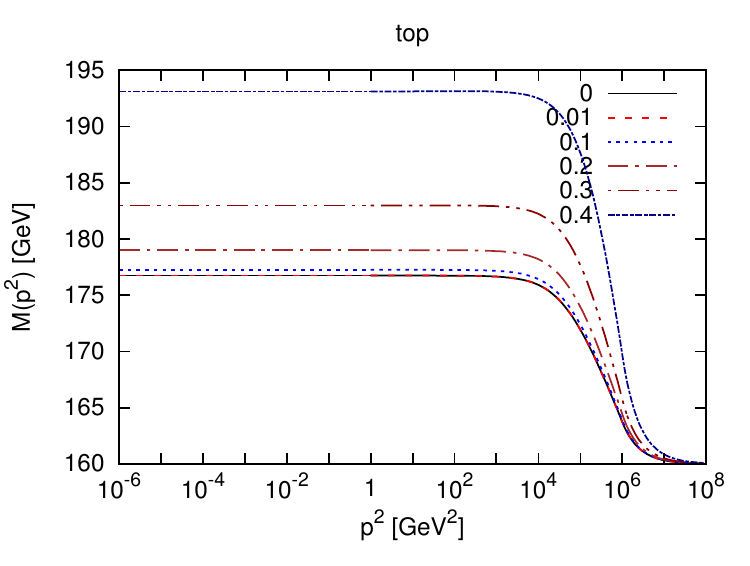}\includegraphics[width=0.5\linewidth]{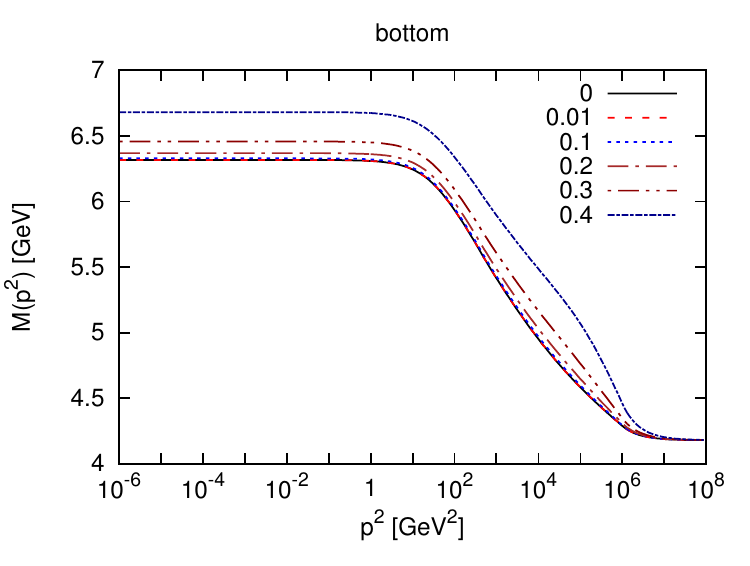}
 \caption{The mass function for different values of $g_{\text{w}}$ and different quarks masses.}
  \label{Pic:MPAF01}
\end{figure*}

The other flavors are shown in figure \ref{Pic:MPAF01}. 
The second generation follows the pattern of the first, but not the third. 
In the latter case the mass function of both quarks increases. 
This implies different contributions to the mass functions. 
One increases the mass function of the heavier quark and decreases the one of the lighter quark. 
The other contribution increases with the mass of both quarks. 
An indication of this is already seen in the second generation, albeit not creating a qualitative change.

%%%%%%%%%%%%%%%%%%%%%%%%%%%%%%%%%%%%%%%%%%%%%%%%%%%%%%%%%%%%%%%%%%%%%%%%%%%%%%%%%%%%%%%%%%%%%%
\subsection{Parity Violation}
\label{Parity Violation}

\begin{figure}
 \includegraphics[width=\linewidth]{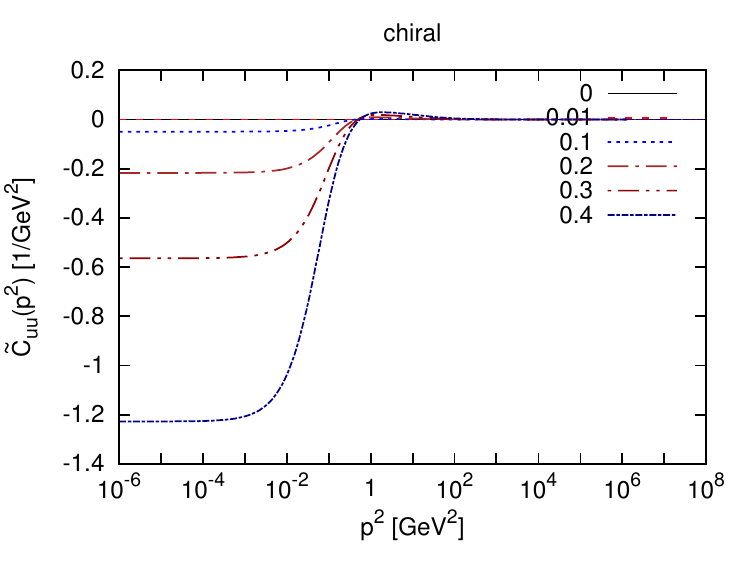}
 \includegraphics[width=\linewidth]{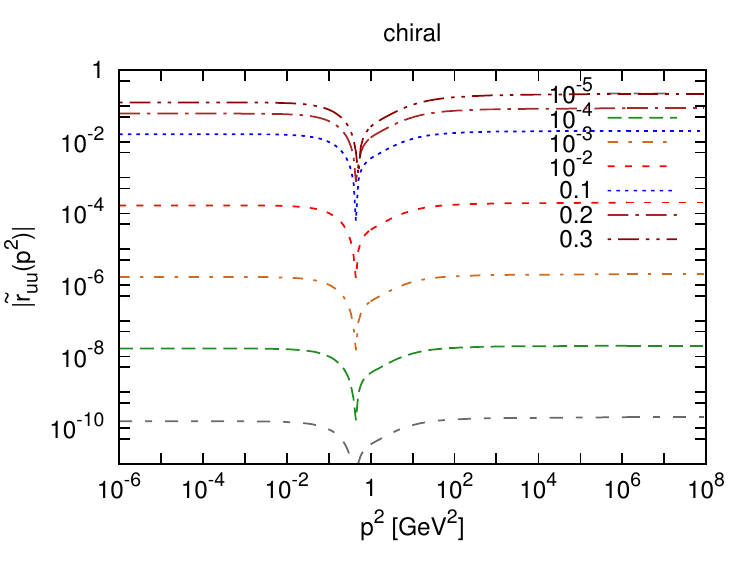}
  \caption{Flavor-diagonal axial channel (top panel) and the ratio (\ref{Eqn:RLR}) 
  (bottom panel) in the chiral limit for different values of $g_{\text{w}}$.}
  \label{Pic:CPRPch}
\end{figure}

In the following the handiness of the quarks is investigated, 
using the definition (\ref{Eqn:RLC}). 
This requires the axial channel, shown in figure \ref{Pic:CPRPch} for the chiral limit. 
The corresponding dressing function  $\tilde{C}$ is for the flavor-diagonal elements 
found to be positive for higher momenta and negative in the IR. 
At the same time, for increasing $g_{\text{w}}$ the absolute value of $\tilde{C}$ also increases. 
Of course, at large momenta the dressing function goes to its tree-level part, 
which from equation (\ref{Eqn:TLprop}) is
\begin{align}
 \tilde{C}_{0,AA} &= \frac{2g_{\text{w}}^{2}p^{2}}{N(p^{2})},
 \label{Eqn:TLC}
\end{align}
which is positive, actually for all momenta. 
Therefore the backcoupling to QCD forces it to be negative in the IR. 
This happens at a transition scale of approximately 1 GeV$^2$, which is the typical QCD scale.

To assess the consequences of this for the left-handed and right-handed contributions, 
it is useful to define their relative ratio as
\begin{align}
 \tilde{r}_{AB}(p^2) &= \frac{\tilde{L}_{AB}(p^2)-\tilde{R}_{AB}(p^2)}{\tilde{L}_{AB}(p^2)+\tilde{R}_{AB}(p^2)}
 = - \frac{\tilde{C}_{AB}(p^2)}{\tilde{A}_{AB}(p^2)}.
 \label{Eqn:RLR}
\end{align}
Since $\tilde{A}$ is always positive, the sign of $\tilde{r}$ is given by the sign of $\tilde{C}$. 
This already entails a change of sign, and that the left-handed part is larger in the infrared. 
This is also shown in figure \ref{Pic:CPRPch}. 
The effect increases non-linearly with $g_{\text{w}}$: 
For $g_{\text{w}} \lesssim 0.1 $ the absolute value $|\tilde{r}|$ in the UV and IR is 
increased by two order of magnitudes, when $g_{\text{w}}$ is increased by one order of magnitude.

\begin{figure}
 \includegraphics[width=\linewidth]{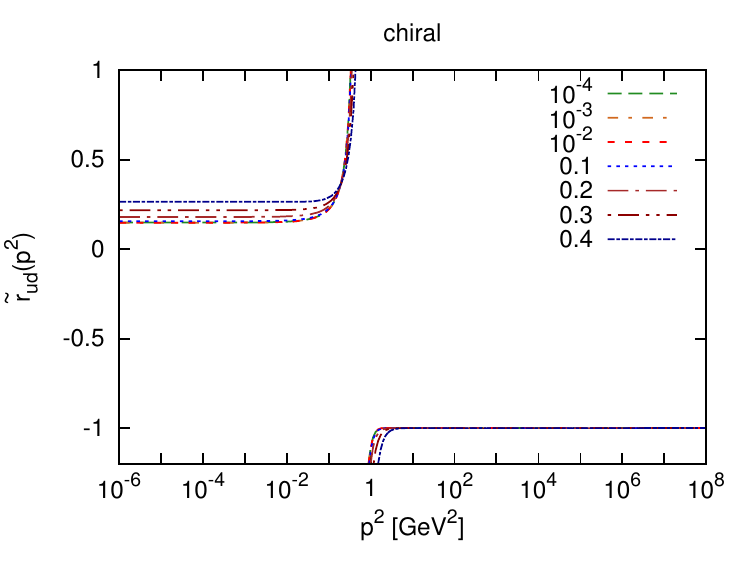}
 \includegraphics[width=\linewidth]{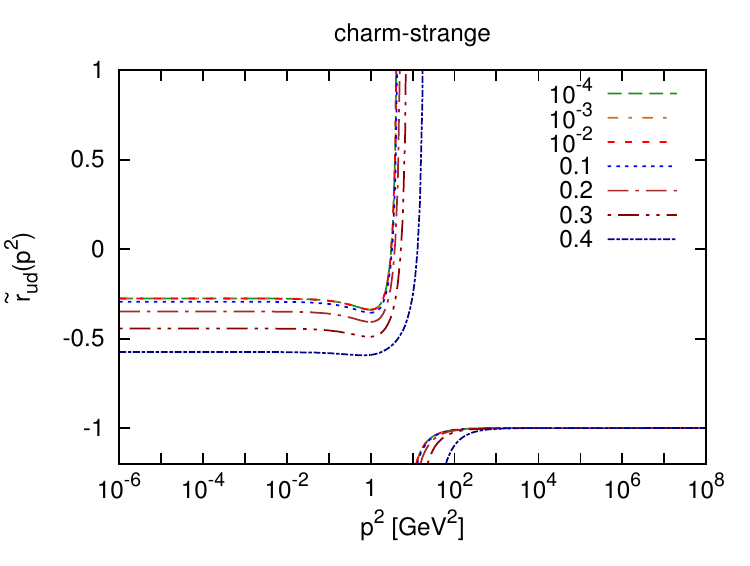}
  \caption{The flavor-off-diagonal ratio (\ref{Eqn:TLC}) in the chiral case (top panel)
  and for the second generation (bottom panel) for different values of $g_{\text{w}}$.}
  \label{Pic:Rch01}
\end{figure}

The flavor-off-diagonal $\tilde{C}$ is always negative in the chiral limit, 
but $\tilde{A}$ changes its sign, see figures \ref{Pic:AMAF01} 
and \ref{Pic:CMAF01} in appendix \ref{Numerical Results}. 
At the same transition scale of approximately 1 GeV$^2$ as for the flavor-diagonal case, 
and in the chiral limit, $\tilde{A}$ has a zero crossing and $\tilde{C}$ not. 
This leads to a diverging $\tilde{r}_{ud}$ at this scale. 
Thus also in this case there is a transition from right-handed in the UV to left-handed in the IR. 
This is shown in figure \ref{Pic:Rch01}.

\begin{figure*}
 \includegraphics[width=0.5\linewidth]{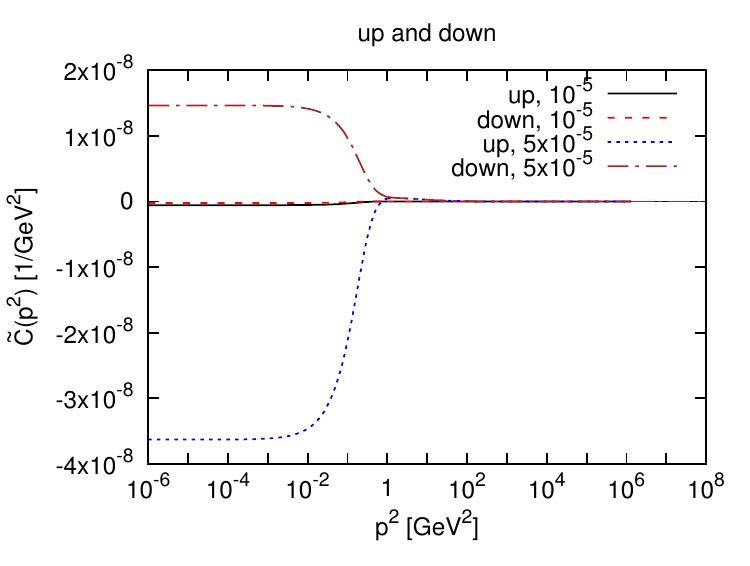}\includegraphics[width=0.5\linewidth]{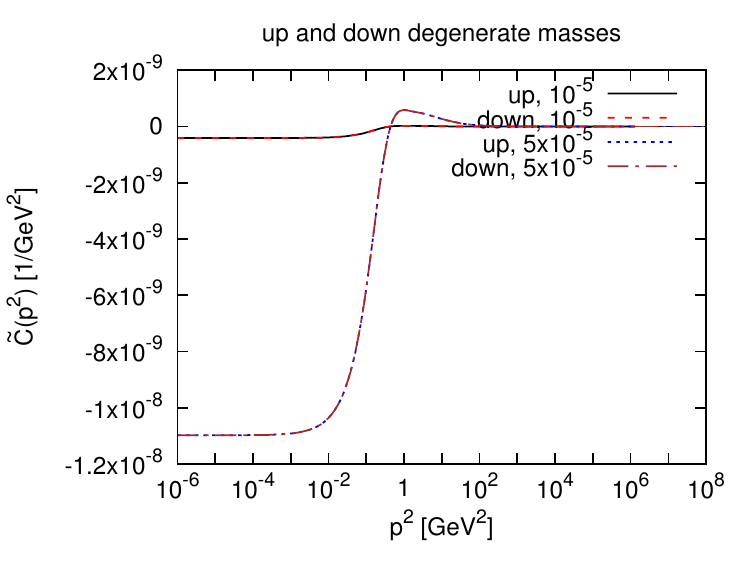}\\
 \includegraphics[width=0.5\linewidth]{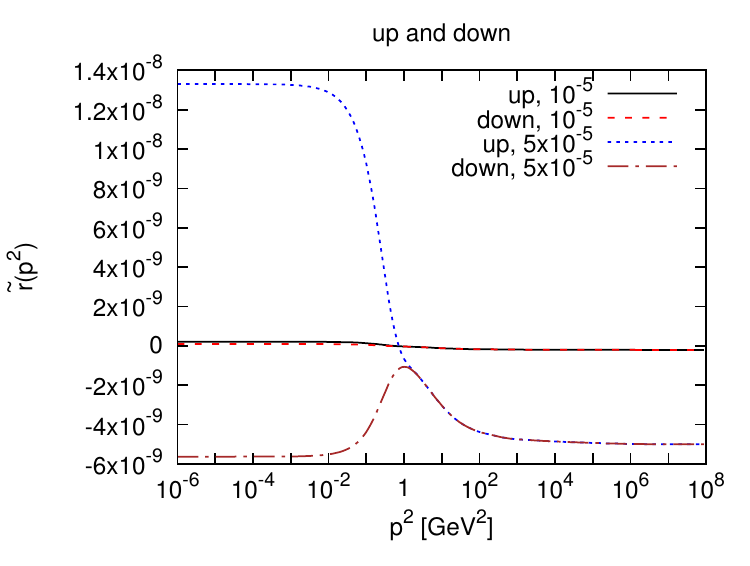}\includegraphics[width=0.5\linewidth]{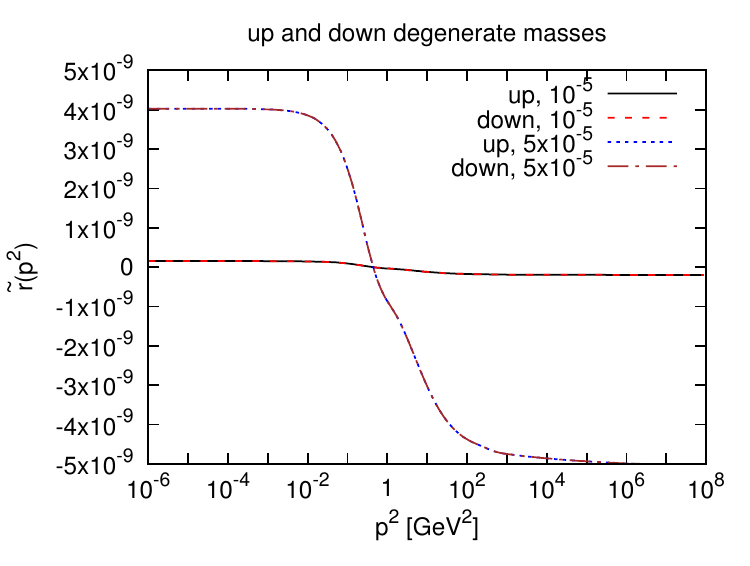}
  \caption{The axial dressing function (top panel) and the ratio (\ref{Eqn:RLR}) (bottom panel) 
  for the up quark and down quark at two different breaking strengths strength. 
  The left panels show the physical mass splitting while in the right panel both masses are degenerate.}
  \label{Pic:Cud01}
\end{figure*}

Increasing the mass, the situation for up and down quarks is shown in figure \ref{Pic:Cud01}. 
The absolute value of $\tilde{C}$ is different for up quark and down quark, 
but for $g_{\text{w}}= 10^{-5}$ the behavior for up and down quark is as in the chiral case. 
Slightly increasing $g_{\text{w}}$ to $5 \cdot 10^{-5}$ entails a drastic qualitative change. 
For the up quark $\tilde{C}$ is still positive in the UV and negative in the IR, 
but for the down quark it remains positive for all momenta. 
Therefore the up quark still flips its chirality
at long distances, but the down quark does not do so.

To understand the origin of this effect, it is helpful to study the degenerate mass case, 
also shown in figure \ref{Pic:Cud01}. 
There is no (numerically detectable) difference between up and down 
quark\footnote{Which is not trivial, as even at tree-level the flavor propagation is not symmetric.}, 
and $\tilde{C}$ changes again sign, as in the chiral limit. 
For higher values of $g_{\text{w}}$ the absolute value of $\tilde{C}$ just increases in the IR. 
This implies, that the different behavior of $\tilde{C}$ for the non-degenerate 
case is due to the mass splitting of the quarks. 
Since at tree-level only in the pseudoscalar channel a contribution proportional to the mass splitting occurs, 
this effect must have been propagated by the QCD interaction to the axial channel. 
Moreover, the effect becomes already important and the, compared to QCD, 
very small mass splitting and a very small breaking scale of $g_{\text{w}}\approx 5 \cdot 10^{-5}$. 
This can only happen if there is a strong non-linear amplification mechanism is at work. 
This implies that the QCD medium strongly affects the helicity at long ranges, 
but only for non-degenerate quark masses. That was certainly not expected.

\begin{figure}
 \includegraphics[width=\linewidth]{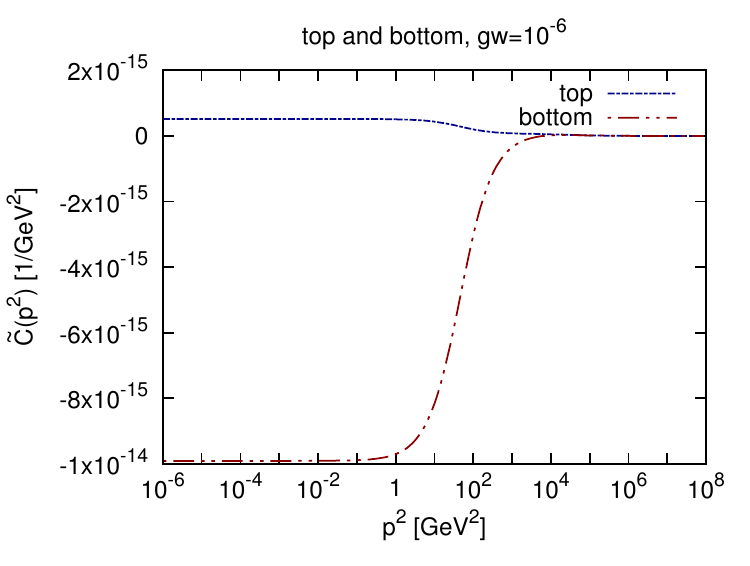}
 \includegraphics[width=\linewidth]{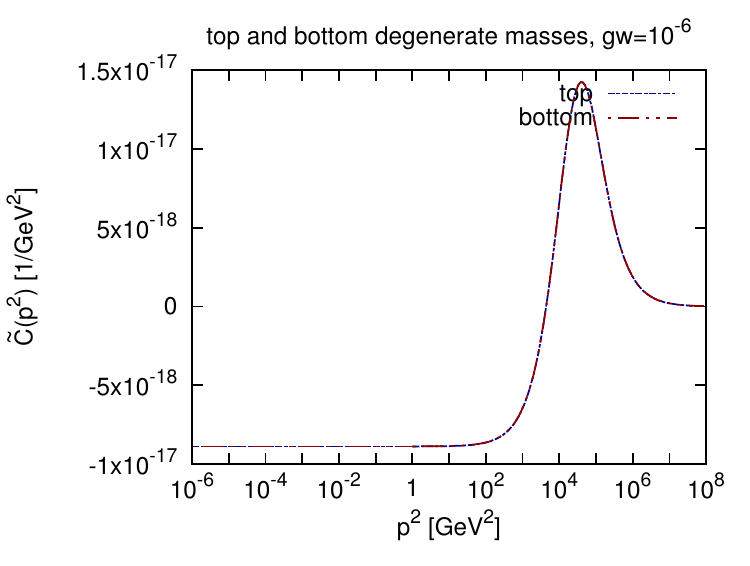}
 \caption{The axial channel for the top and bottom quark (top panel) and heavy degenerate 
 quarks (bottom panel) at $g_{\text{w}}=10^{-6}$ (top panel) .}
  \label{Pic:Ctb01}
\end{figure}

The same is true for the other quark generations, see for the second generation figure \ref{Pic:Rch01}. 
The effect is still there for the third generation, with its very large mass splitting, 
see figure \ref{Pic:Ctb01}, and there occurs already at an even smaller breaking strength of $g_{\text{w}}=10^{-6}$. 
Thus the absolute value of the involved mass scales amplifies the non-linear backcoupling, 
such that it occurs at weaker breaking strength.

The corresponding relative ratios are shown in figure \ref{Pic:RAF01} in appendix \ref{RR}. 
These graphs support the existence of a transition scale, 
where the left-handed and right-handed contribution change their relative contribution.

\begin{figure*}
 \includegraphics[width=0.5\linewidth]{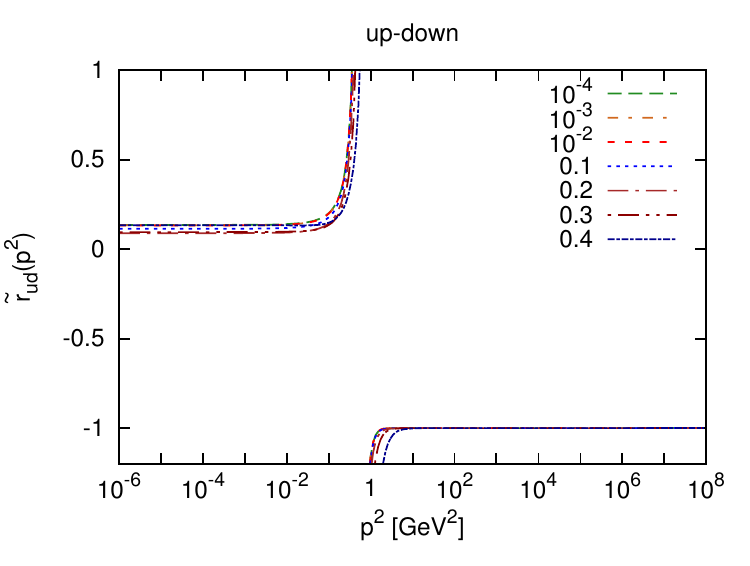}\includegraphics[width=0.5\linewidth]{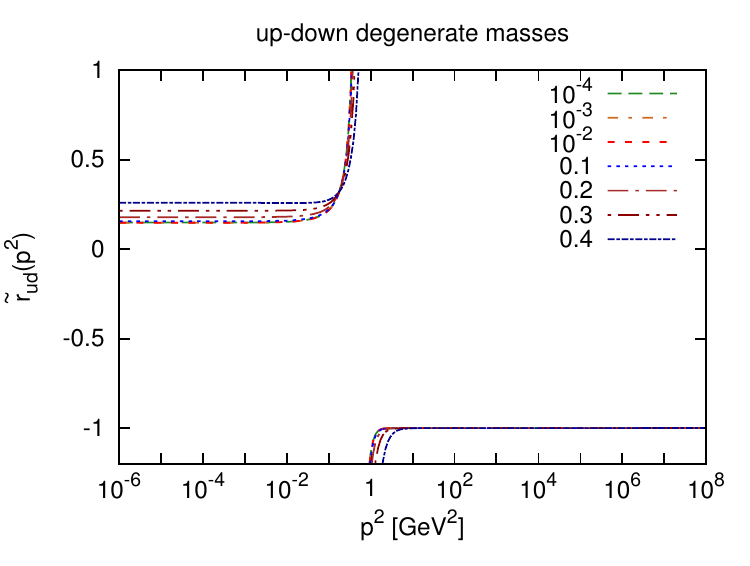}\\
 \includegraphics[width=0.5\linewidth]{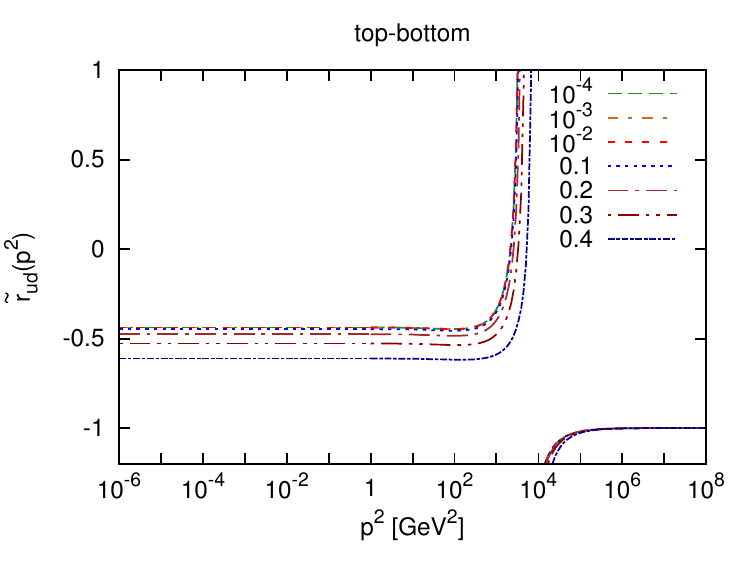}\includegraphics[width=0.5\linewidth]{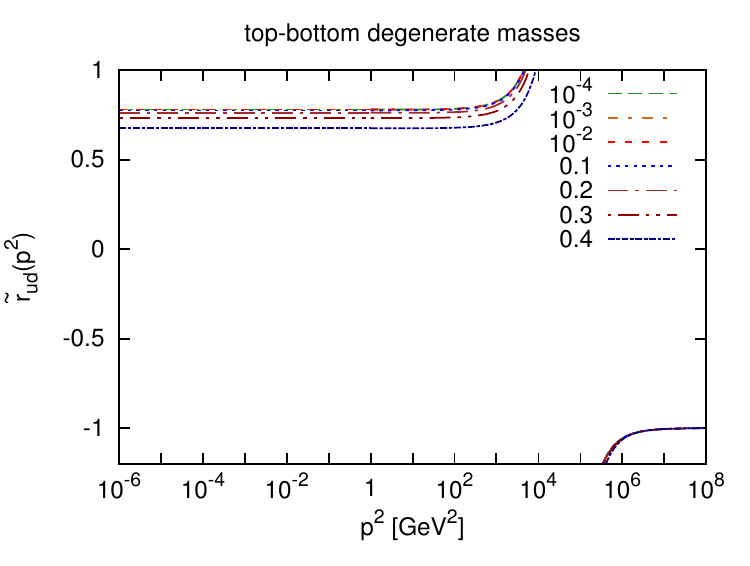}
 \caption{The flavor-off-diagonal ratio (\ref{Eqn:RLR}) for different $g_{\text{w}}$ for the first generation 
 (top panels) and third generation (bottom panels) for physical mass splittings 
 (first panel) and degenerate masses (bottom panels).}
  \label{Pic:RMF01}
\end{figure*}

For the mixed flavor case the effect is solely driven by the absolute value of the mass splitting, 
and the breaking strength plays only a minor role. 
This is shown in figure \ref{Pic:RMF01}. 
Whether there is a mass-splitting or not plays only a role if the mass splitting is large enough, 
i.\ e.\ in the third generation. 
Only then the behavior with or without mass splitting differ qualitatively in the infrared. 
In fact, already the second generation is sufficient for this, as can be seen in figure \ref{Pic:Rch01}.

%%%%%%%%%%%%%%%%%%%%%%%%%%%%%%%%%%%%%%%%%%%%%%%%%%%%%%%%%%%%%%%%%%%%%%%%%%%%%%%%%%%%%%%%%%%%%%
\subsection{Schwinger function}
\label{Masspole}

\begin{figure}
 \includegraphics[width=\linewidth]{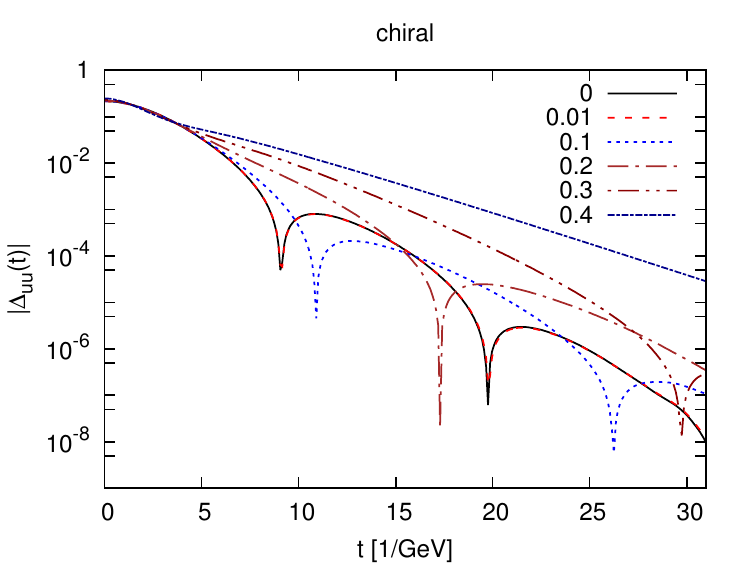}
 \includegraphics[width=\linewidth]{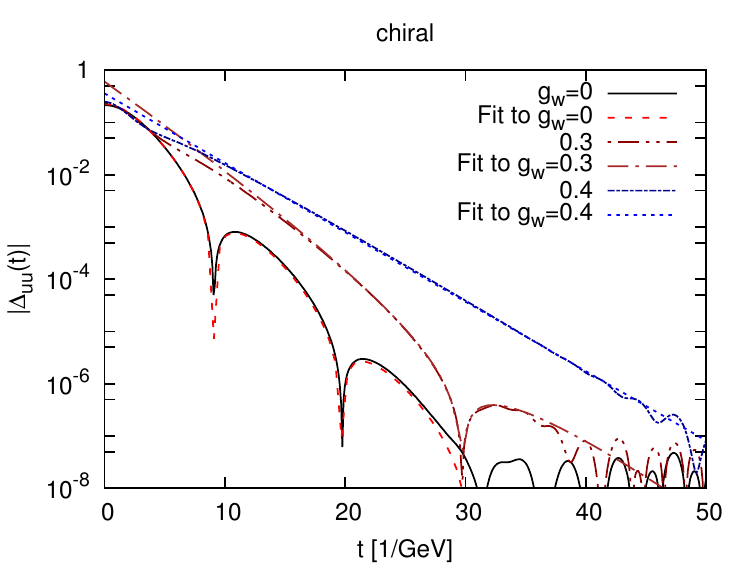}
  \caption{Flavor-diagonal Schwinger function in the chiral limit for different 
  $g_{\text{w}}$ (top panel) and with fits (bottom panel).}
  \label{Pic:SWPch}
\end{figure}

As noted, the analytic structure is accessible through the Schwinger function (\ref{Eqn:SWF}). 
Let us now consider the Schwinger function in the chiral limit for the flavor-diagonal dressing function $B$. 
The other flavor-diagonal dressing functions do not lead to qualitatively new results, 
and are even quantitatively similar, so these will be skipped here.

The results are shown in figure \ref{Pic:SWPch}. 
The Schwinger function shows an oscillatory behavior, consistent with the form (\ref{Eqn:CPS}), 
and thus complex conjugate poles, as in pure QCD for the rainbow-ladder truncation \cite{Alkofer:2003jj}. 
In fact, a fit using a more detailed ansatz, see appendix \ref{a:fpsf}, 
of this type works very well, as is also shown in figure \ref{Pic:SWPch}. 
The values of the fit parameters are listed, for completeness, in appendix \ref{a:fpsf}.

However, the oscillation period starts to substantially increase for  $g_{\text{w}}>0.01$, 
up to a point where at large $g_{\text{w}}$ the first zero 
crossing has moved to a time which we can no longer numerically resolve reliably. 
Thus, the imaginary part shrinks with increasing breaking. 
The curvature at short distances is still not quite right for a physical particle. 
A similar behavior, though with a suppression of oscillations for decreasing interaction strength, 
has already been observed for adjoint scalar particles \cite{Maas:2004se,Maas:2005rf}. 
This strongly suggests that the interaction strength plays a crucial role 
for the scale at which negative norm contributions become relevant, 
even though the coupling does not differentiate between positive-norm and negative-norm states.

At the same time the steepness decreases, making the real part smaller. 
Thus, the increase in $g_{\text{w}}$ moves in total the poles closer to the origin,
as both real and imaginary part decrease.

\begin{figure}
 \includegraphics[width=\linewidth]{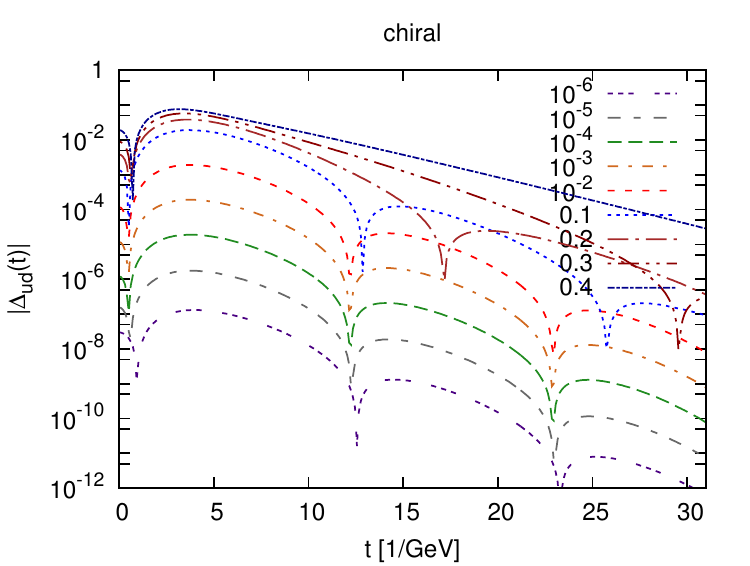}
 \includegraphics[width=\linewidth]{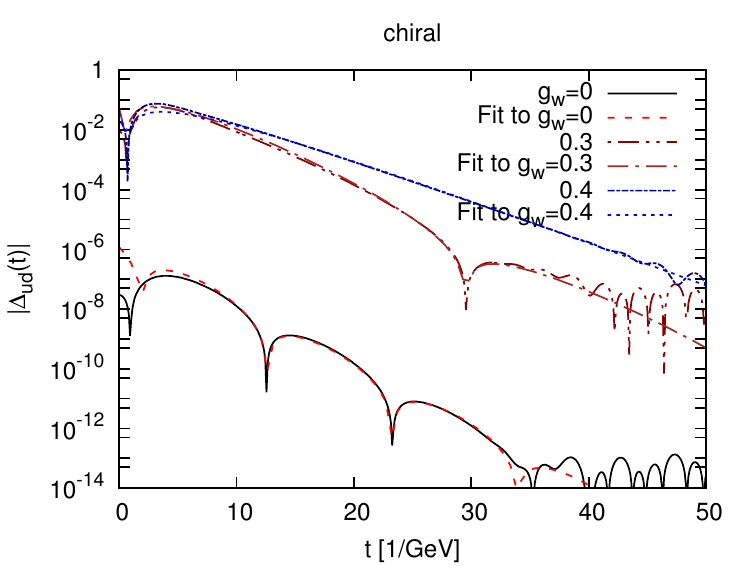}
  \caption{Flavor-off-diagonal Schwinger function in the chiral limit 
  for different $g_{\text{w}}$ (top panel) and with fits (bottom panel).}
  \label{Pic:SWMch}
\end{figure}

The flavor-off-diagonal Schwinger function, again only the scalar part as the others are very similar, 
is shown in figure \ref{Pic:SWMch}. 
In principle, it shows a very similar behavior as for the flavor-diagonal part, 
except that it always retains a first zero crossing, relatively independent of $g_{\text{w}}$, at very short times. 
It is still possible to fit it using the same fit form. 
The results for the fit are also listed in appendix \ref{a:fpsf}. 
It is found that the zero crossing at small momenta comes from the phase shift $\delta$ in (\ref{Eqn:CPS}). 
It is very close to $\pi/2$ and causes the sign change for the Schwinger function at small $t$.
Still, the position of the pole also moves towards the origin with increasing breaking strength.

The Schwinger function for the first two generations show the same behavior as in the chiral limit, 
see figure \ref{Pic:SW} in appendix \ref{SF}. 
For the third generation the fall-off was too fast, due to the large real part, 
as that any unambiguous statements could be drawn before numerical noise drowns out the signal.

It is a quite interesting result that the breaking pushes the poles closer to the origin. 
As we expect a change of physics when crossing the threshold $g_W\gtrsim 0.4$, 
this could be a first indication of a drastic change at strong breaking. 
However, this is probably not of relevance to neutron star physics. 
On the downside, the decreasing distance to the origin will create additional 
problems in any mesonic correlators in rainbow-ladder calculations 
\cite{Alkofer:2000wg,Fischer:2006ub,Roberts:2015lja}. 
In these cases more elaborate schemes will be necessary than a tree-level breaking, 
which we are currently developing.

%%%%%%%%%%%%%%%%%%%%%%%%%%%%%%%%%%%%%%%%%%%%%%%%%%%%%%%%%%%%%%%%%%%%%%%%%%%%%%%%%%%%%%
%%%%%%%%%%%%%%%%%%%%%%%%%%%%%%%%%%%%%%%%%%%%%%%%%%%%%%%%%%%%%%%%%%%%%%%%%%%%%%%%%%%%%
\section{Conclusion}
\label{Conclusion}

We have calculated the quark propagator in the presence of explicit flavor, C and P symmetry breaking.
Moreover, we took into account the non-linear back-coupling from QCD in the rainbow-ladder truncation. 
The latter lead to qualitative effects, even for relatively small explicit 
breaking strengths, at long (hadronic) distances. 
They also couple in a highly non-trivial way the various dressing functions to each other. 
This was particularly visible in the way how effects from mass splitting 
and mass averages surfaced in various dressing functions. 
The non-linear amplification also surfaced in other ways. 
This is a very important insight: External perturbations can, 
even in rainbow-ladder truncation, be substantially amplified by the strong interactions. 
This must be regarded as a warning that even small effects 
can play a non-perturbatively large role when QCD is involved.

From the point of view of physics, another interesting insight is obtained 
when considering how left-handed and right-handed particle propagation changes. 
Under particular conditions, flips between handedness can be amplified at 
long distances by the strong interaction. 
This can deplete or enlarge the available particles in some handedness. 
As the weak interactions only couple to a particular handedness, 
this can increases or decrease the reservoir of particles 
which are weakly interacting in a system. 
If this pertains to the full system, this can influence the dynamics in forming or merging 
neutron stars, as this could alter, e.\ g., the opacity for neutrinos. 
This is even more important as the typical range where this occurs is only of the size of a hadron.

Concluding, this investigations showed that weak interactions effects, even if themselves small, 
can be amplified by the strong interactions, and this backcoupling can have qualitative impact. 
Keeping this in mind will be important in the next step, when relaxing the assumption of a reservoir, 
and taking the weak interactions explicitly into account, including the emitted neutrinos and electrons. 
This will require to work on a hadronic level, which is our next aim.\\

\noindent{\bf Acknowledgements}\\

We are grateful to Helios Sanchis-Alepuz, Jordi Paris-Lopez and Adrian Lorenz Blum
for helpful discussions. W.\ M.\ has been supported by the FWF doctoral school W1203-N16.

%%%%%%%%%%%%%%%%%%%%%%%%%%%%%%%%%%%%%%%%%%%%%%%%%%%%%%%%%%%%%%%%%%%%%%%%%%%%%%%%%%%%
\appendix

\section{Structure of the Quark Propagator}
 \label{Structure of QP}

In the following a number of useful relations between the dressing functions of 
the quark propagator and its inverse will be collected. 
Combining the flavor elements in a matrix, e.\ g.\ for the vectorial channel as
\begin{align}
A &= \begin{pmatrix}
      A_{uu} & A_{ud} \\
      A_{du} & A_{dd}
     \end{pmatrix},
  \label{Eqn:SA1}
\end{align}
allows for a compact notation. Writing the propagator and its inverse in terms of these matrices yields
\begin{align}
 P(p^2) &=  \tilde{A}(p^2) \img \slashed{p} + \tilde{B}(p^2) \unit + 
 \tilde{C}(p^2) \img \slashed{p} \gamma^5 + \tilde{D}(p^2) \gamma^5,
 \nonumber \\
 P^{-1} (p^2) &= - A(p^2) \img \slashed{p} + B(p^2) \unit + 
 C(p^2) \img \slashed{p} \gamma^5 + D(p^2) \gamma^5.
  \label{Eqn:SPA1}
\end{align}
The propagator satisfies the condition
\begin{align}
  P^{-1} P &= \unit.
  \label{Eqn:PIPR}
\end{align}
yielding relations between the matrix-valued dressing functions
\begin{align}
  A \tilde{A} p^{2} + B \tilde{B} + C \tilde{C} p^{2} + D \tilde{D} &= \unit
  \nonumber \\
  - A \tilde{B} + B \tilde{A} + C \tilde{D} -D \tilde{C} &= 0
  \nonumber \\ 
  -A \tilde{D} -D \tilde{A} + B \tilde{C} + C \tilde{B} &= 0
  \label{Eqn:PIPR2} \\
  A \tilde{C} p^{2} + C \tilde{A} p^{2} + B \tilde{D} + D \tilde{B} &=0
  \nonumber
\end{align}
While this system is a system of linear equations for either the matrix elements of the propagator or its inverse, an explicit solution is of little use. The expressions become extremely lengthy, and therefore prohibitively expensive to evaluate during numerical calculations. Therefore, in our investigations we always solved such equations numerically at double precision.

%%%%%%%%%%%%%%%%%%%%%%%%%%%%%%%%%%%%%%%%%%%%%%%%%%%%%%%%%%%%%%%%%%%%%%%%%%%%%%%%%%%%%%%%%
\section{DSEs for Quark Propagators}
\label{DSE for QP}

To derive the DSEs for the different dressing functions, insert in equation 
(\ref{Eqn:DSE_QP}) the quark propagator of equation (\ref{Eqn:SQP}) 
and project out the different channels, by taking suitable traces. 
We define the following two kernels
\begin{align}
 K_{1}(p,q,k) &=
  12 \pi C_F \frac{\alpha(k^2)}{k^2},
 \nonumber \\
 K_2 (p,q,k) &=  4 \pi C_F \frac{\alpha(k^2)}{k^2 p^2} 
 \left[(p \cdot q) + 2 \frac{(p\cdot k)(q \cdot k)}{k^2} \right],
 \label{Eqn:Kernel}
\end{align}
where $C_F=(N_c^2-1)/2N_c$ and $N_c$ is the number of colors, i.\ e.\ $N_c=3$.

We further define the functional $\Pi_{i}$ for $i=1,2$ as
\begin{align}
 \Pi_{i,A}(f,p^2) &= Z_{2,A}(\mu^2,\Lambda^2) \int^{\Lambda} \frac{\dd^4 q}{(2 \pi)^4} \left\{
  K_{i}(p,q,k)f(q^2,\mu^2)\right\}.
 \label{Eqn:Pi}
\end{align}
This yields the DSEs of the the different dressing 
functions for flavor-diagonal elements $A \in \left\{ u, d\right\}$
\begin{align}
 A_{AA}(p^2,\mu^2) &= Z_{2,A}(\mu^2,\Lambda^2) \left[ 1+ \Pi_{2,A}(\tilde{A}_{AA},p^2)\right],
 \nonumber \\
 B_{AA}(p^2,\mu^2) &= Z_{2,A}(\mu^2,\Lambda^2) \left[ m_{A}+ \Pi_{1,A}(\tilde{B}_{AA},p^2)\right],
  \nonumber \\
 C_{AA}(p^2,\mu^2) &= Z_{2,A}(\mu^2,\Lambda^2) \Pi_{2,A}(\tilde{C}_{AA},p^2),
  \label{Eqn:DSE_DF1}\\
 D_{AA}(p^2,\mu^2) &= Z_{2,A}(\mu^2,\Lambda^2) \Pi_{1,A}(\tilde{D}_{AA},p^2).
 \nonumber 
\end{align}
and for mixed flavors, $A,B \in \left\{ u, d\right\}, A \neq B$,
\begin{align}
 A_{AB}(p^2,\mu^2) &= \sqrt{Z_{2,A}(\mu^2,\Lambda^2)Z_{2,B}(\mu^2,\Lambda^2)} g_{\text{w}}\nonumber\\
 &+ Z_{2,B}(\mu^2,\Lambda^2) \Pi_{2,A}(\tilde{A}_{BA},p^2),
 \nonumber \\
 B_{AB}(p^2,\mu^2) &= Z_{2,B}(\mu^2,\Lambda^2) \Pi_{1,A}(\tilde{B}_{BA},p^2),
  \nonumber \\
  C_{AB}(p^2,\mu^2) &= \sqrt{Z_{2,A}(\mu^2,\Lambda^2)Z_{2,B}(\mu^2,\Lambda^2)} g_{\text{w}}\nonumber\\
 &+ Z_{2,B}(\mu^2,\Lambda^2) \Pi_{2,A}(\tilde{C}_{BA},p^2),
 \label{Eqn:DSE_DF2} \\
 D_{AB}(p^2,\mu^2) &= Z_{2,B}(\mu^2,\Lambda^2) \Pi_{1,A}(\tilde{D}_{BA},p^2).
 \nonumber 
\end{align}
This system of equation is technically very similar to the ordinary rainbow-ladder truncation. 
Therefore a numerical solution using standard fixed-point iteration schemes is possible, and was done here. 
Only the the quark propagator was numerically inverted at every step, 
due to the involved structure, as discussed in appendix \ref{Structure of QP}.

There are, however, a few more subtle numerical issues to be mentioned.
To perform the integral for $\Pi_{i,A}$ the dressing functions $\tilde{A}$, $\tilde{B}$, $\tilde{C}$
and $\tilde{D}$ are evaluated for various momenta using interpolation.
We performed our calculations using linear and cubic interpolation.
If we use a precision of $5 \times 10^{-5}$, which is considered to be sufficient in the standard fixed-point 
iteration scheme, for $A$, $B$, $C$ and $D$
then we get different numerical solutions for linear and cubic interpolation.
By increasing the precision the solution from the linear interpolation approaches the solution 
from the cubic interpolation for all dressing functions except for $D_{AA}$ and $\tilde{D}_{AA}$.
Especially, using linear interpolation we get different solutions for $D_{AA}$ and $\tilde{D}_{AA}$
by using different start values for the iteration. 
In contrast to this, we get the same solutions for different start values using the cubic interpolation. 
Thus, we consider the solutions from the cubic interpolation as stable, and used them throughout this work.

In addition, we used a precision of $5 \times 10^{-7}$, instead of the standard value, 
for $A$, $B$, $C$ and $D$ and $2^8=256$ grid points. 
In this case the difference for the solutions from the linear and cubic interpolation
were at most in the third significant digit, and thus lead essentially to the same results.

Let us note that our precision for the iteration procedure is for the dressing functions 
of the inverse propagator ($A$, $B$, $C$ and $D$). 
The dressing functions of the propagator ($\tilde{A}$, $\tilde{B}$, $\tilde{C}$
and $\tilde{D}$) are calculated by a numerical inversion, with a precision of roughly $10^{-20}$.

%%%%%%%%%%%%%%%%%%%%%%%%%%%%%%%%%%%%%%%%%%%%%%%%%%%%%%%%%%%%%%%%%%%%%%%%%%%%%%%%%%
\section{Numerical Results}
\label{Numerical Results}

%%%%%%%%%%%%%%%%%%%%%%%%%%%%%%%%%%%%%%%%%%%%%%%%%%%%%%%%%%%%%%%%%%%%%%%%%%%%%%%%%%5
\subsection{Vector Channel}
\label{VC}

\begin{figure*}[ht]
 \includegraphics[width=0.5\linewidth]{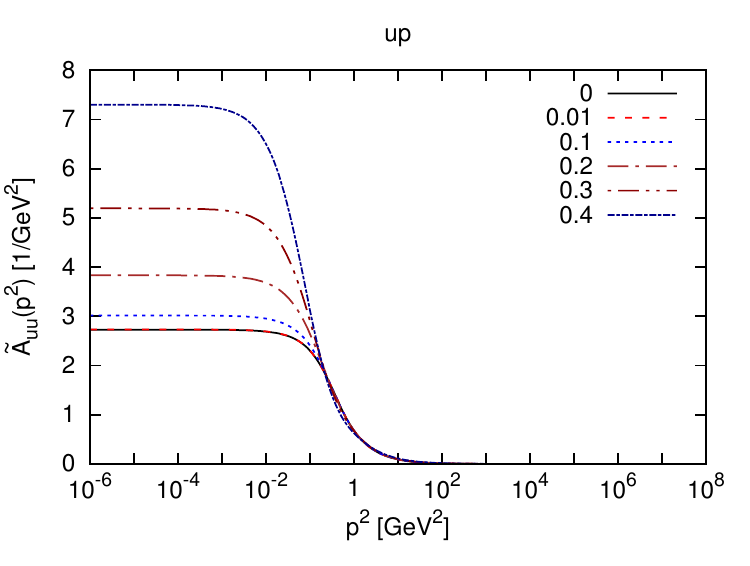}\includegraphics[width=0.5\linewidth]{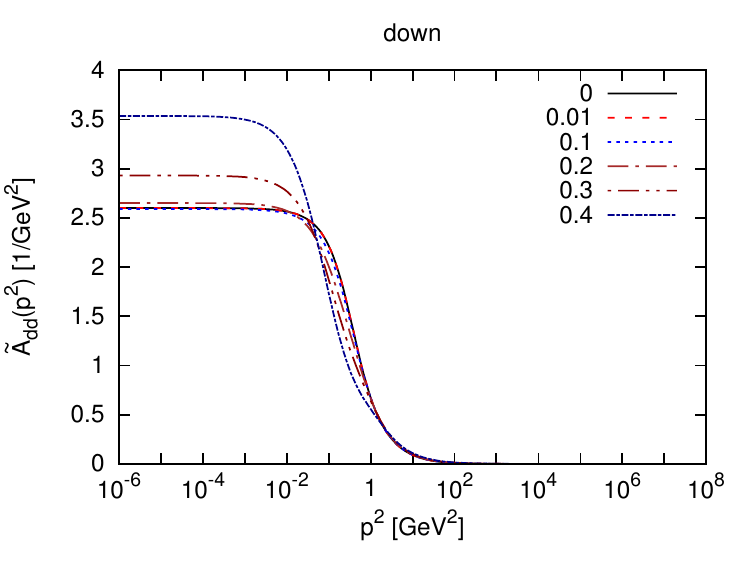}\\
 \includegraphics[width=0.5\linewidth]{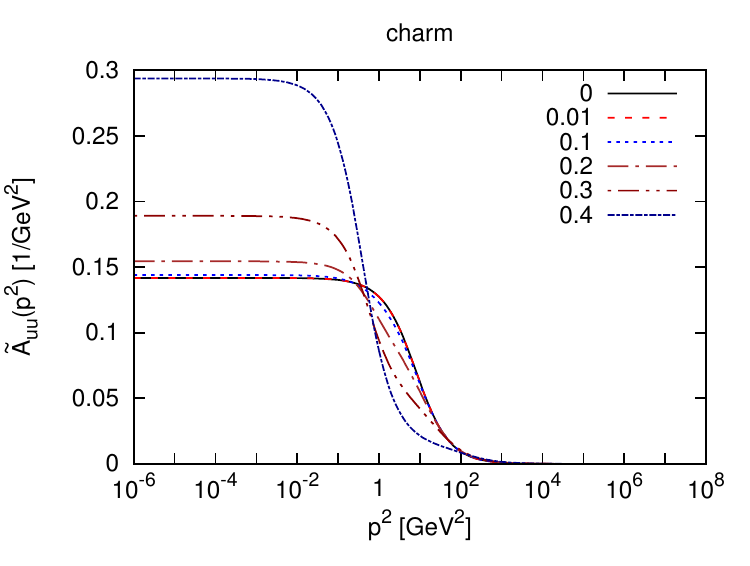}\includegraphics[width=0.5\linewidth]{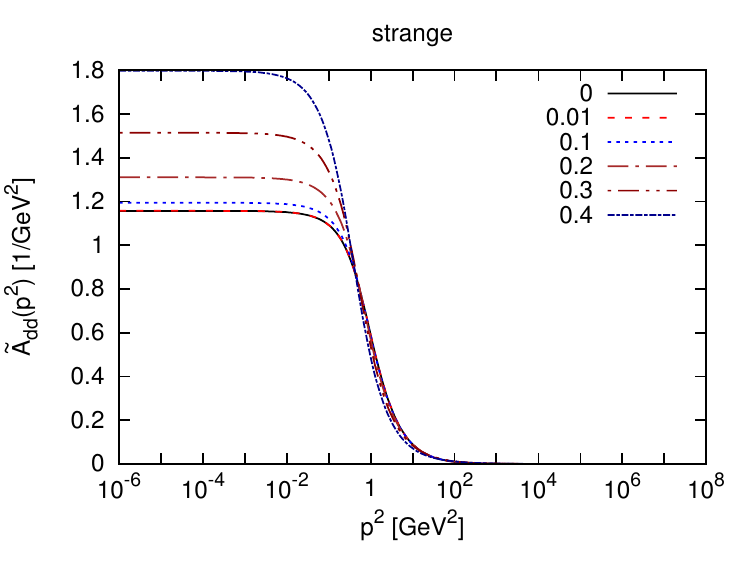}\\
 \includegraphics[width=0.5\linewidth]{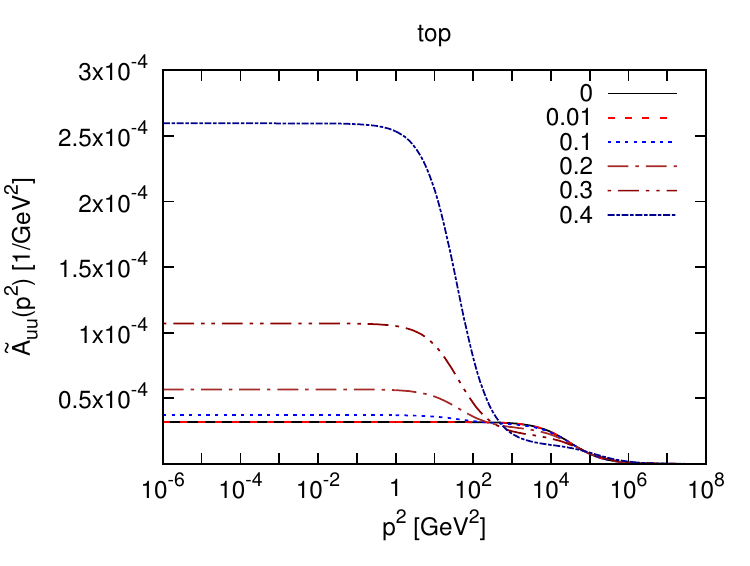}\includegraphics[width=0.5\linewidth]{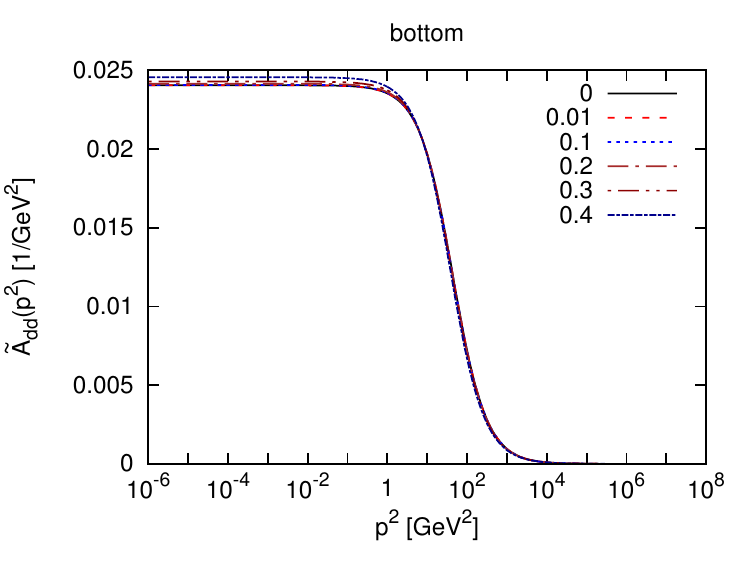}
 \caption{The flavor-diagonal vector dressing function for different values of $g_{\text{w}}$. 
 From top to bottom generations one to three are shown, and the left-hand panels show up-type quarks, 
 and the right-hand panels down-type quarks.}
  \label{Pic:APAF01}
\end{figure*}

The flavor-diagonal vector dressing functions for the different generations and values 
of the breaking strength are shown in figure \ref{Pic:APAF01}. 
A detailed discussion is given in section \ref{WFR}.

\begin{figure*}[ht]
 \includegraphics[width=0.5\linewidth]{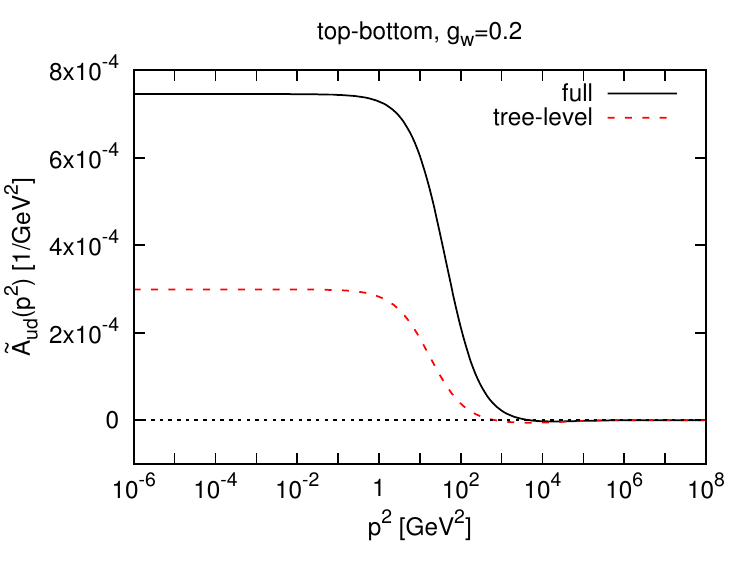}\includegraphics[width=0.5\linewidth]{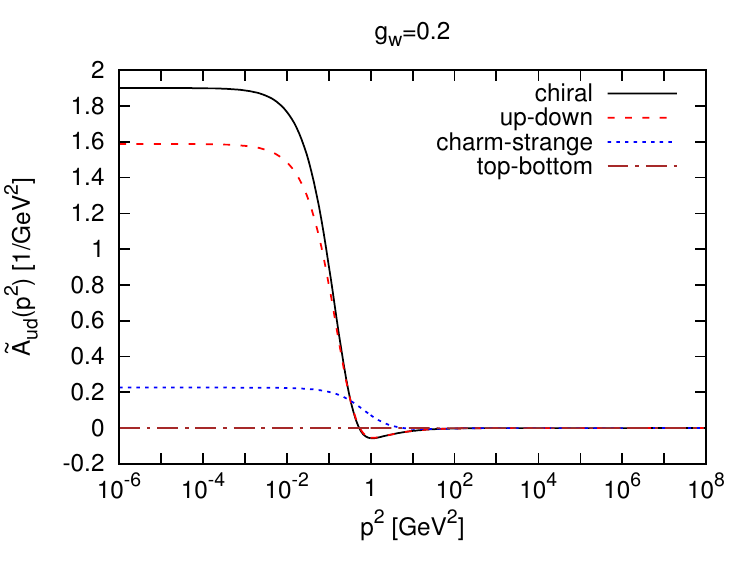}
 \caption{The flavor-off-diagonal vector dressing function for $g_{\text{w}}=0.2$ 
 in comparison to tree-level (left panel) and for different generations (right panel).}
  \label{Pic:AM01}
\end{figure*}

\begin{figure*}
 \includegraphics[width=0.5\linewidth]{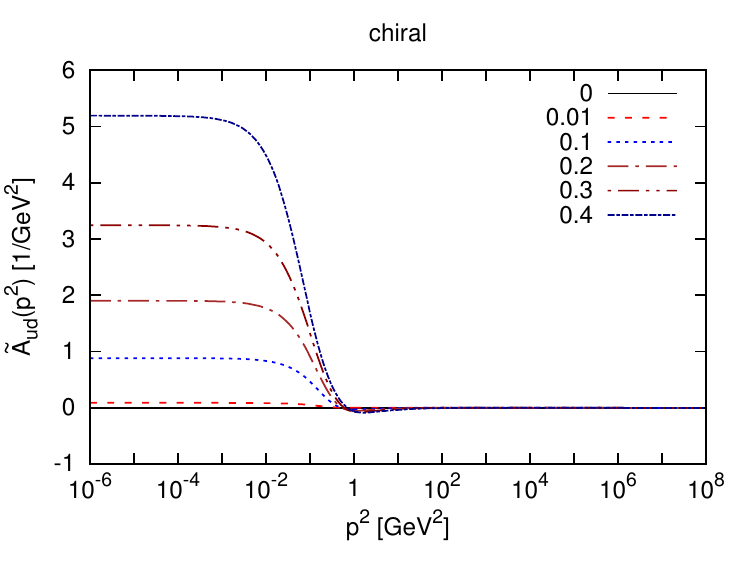}\includegraphics[width=0.5\linewidth]{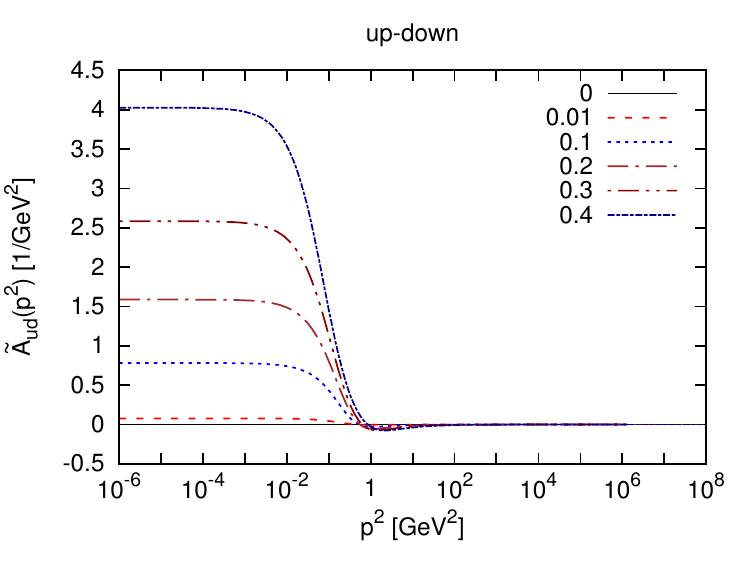}\\
 \includegraphics[width=0.5\linewidth]{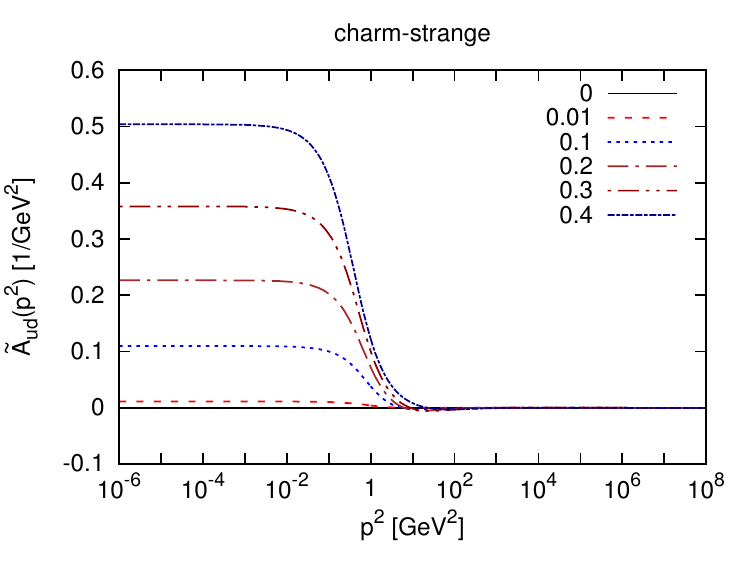}\includegraphics[width=0.5\linewidth]{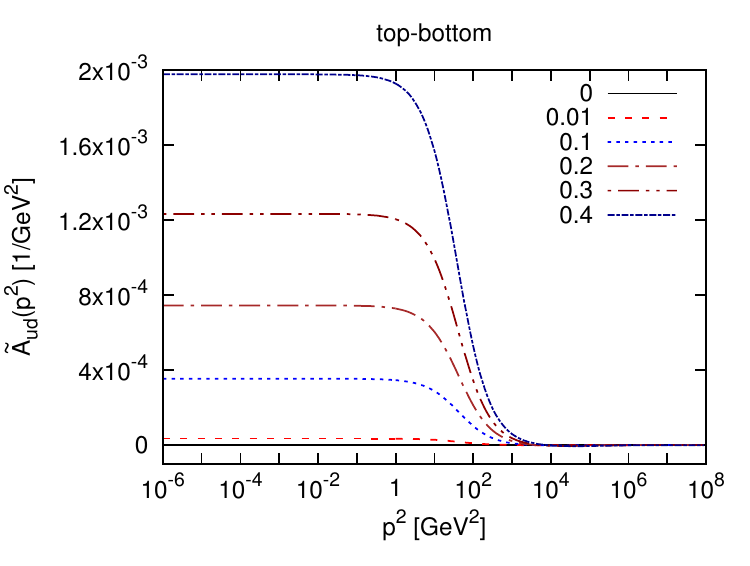}
 \caption{The flavor-off-diagonal vector dressing function for the chiral limit 
 and the different generations for different $g_{\text{w}}$.}
  \label{Pic:AMAF01}
\end{figure*}

A comparison between the different generations, the chiral limit, and tree-level for the 
flavor-off-diagonal vector dressing function is shown in figure \ref{Pic:AM01} 
for a fixed value $g_{\text{w}}=0.2$. 
The tree-level value is given by, see equation (\ref{Eqn:TLprop}),
\begin{align}
 \tilde{A}_{0,ud} &= \frac{g_{\text{w}}(m_u m_d - p^{2})}{N(p^{2})}.
 \label{Eqn:TLAM}
\end{align}
This demonstrates that at tree-level the dressing function is negative for momenta 
$p^2 \geq m_u m_d$ and positive for lower momenta. 
The full dressing function exhibits the same behavior, 
but the absolute value is substantially larger. 
Note especially that in the chiral limit the tree-level propagator is proportional 
to $-p^2$ and thus negative for all momenta. 
The full dressing function is, however, positive in the IR. 
As the masses of up and down quark are comparatively small, 
the result for them is close to the one in the chiral limit.

The dependence on the generation and $g_{\text{w}}$ is shown in figure \ref{Pic:AMAF01}. 
In every generation $\tilde{A}_{ud}$ increases with $g_{\text{w}}$. 
The zero of $\tilde{A}_{ud}$ shifts for higher momenta for heavier mass. 
This is already the case for the tree-level propagator, where the zero is determined by the condition $p^2=m_u m_d$.

%%%%%%%%%%%%%%%%%%%%%%%%%%%%%%%%%%%%%%%%%%%%%%%%%%%%%%%%%%%%%%%%%%%%%%%%%%%%%%%%%%5
\subsection{Scalar Channel}
\label{SC}

\begin{figure*}[ht]
 \includegraphics[width=0.5\linewidth]{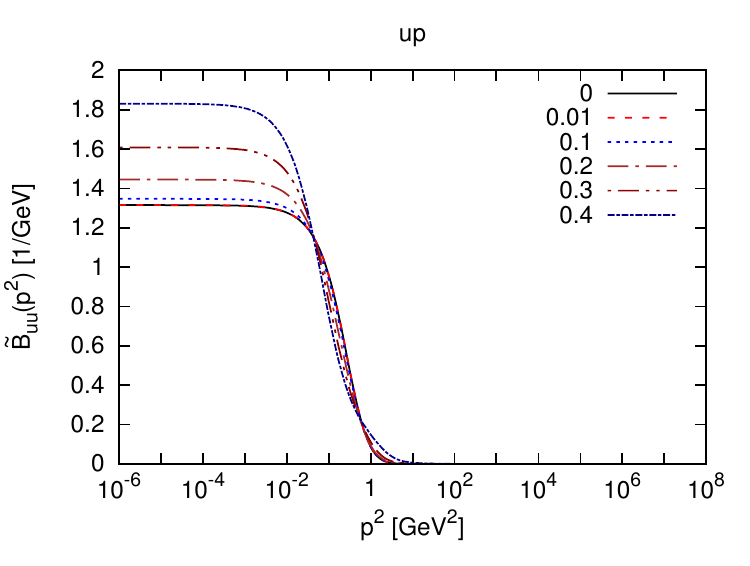}\includegraphics[width=0.5\linewidth]{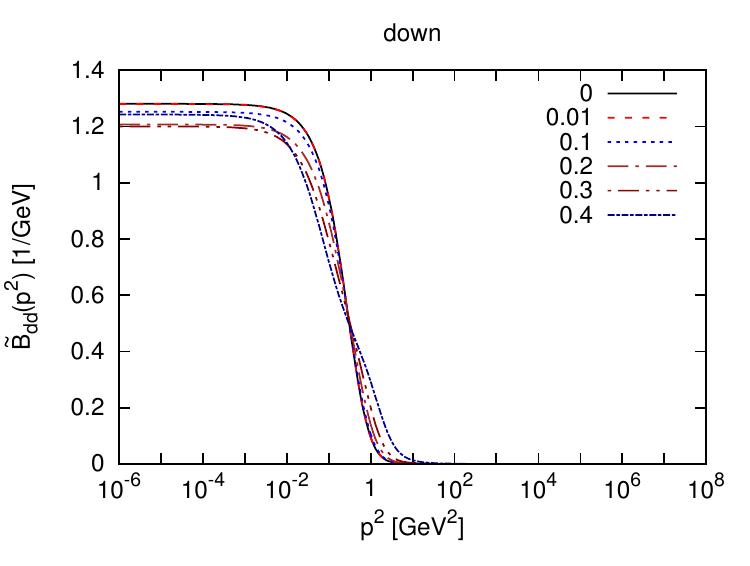}\\
 \includegraphics[width=0.5\linewidth]{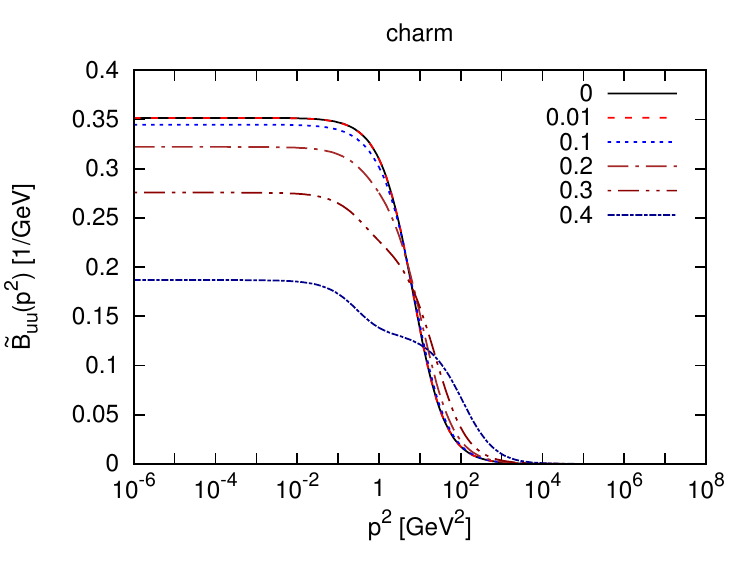}\includegraphics[width=0.5\linewidth]{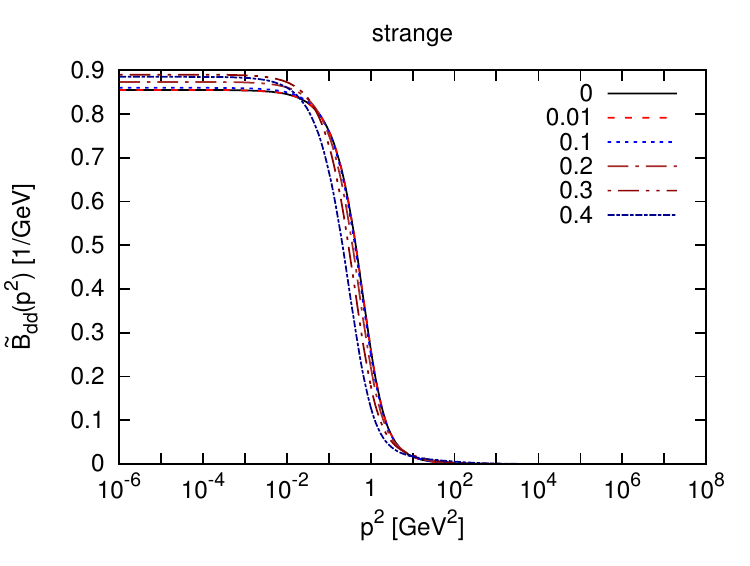}\\
 \includegraphics[width=0.5\linewidth]{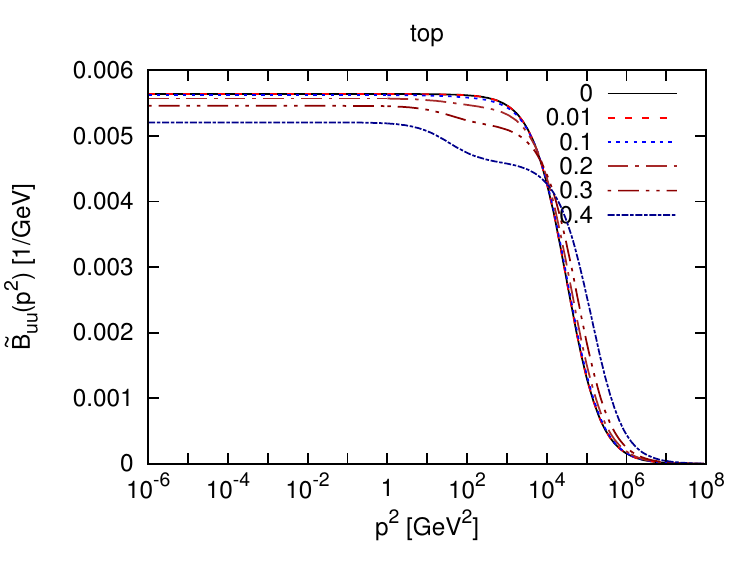}\includegraphics[width=0.5\linewidth]{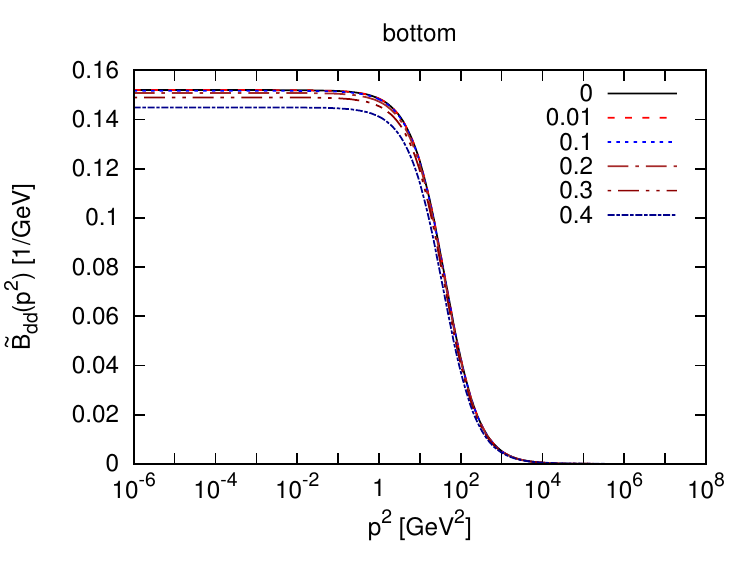}
 \caption{The flavor-diagonal scalar dressing function for different $g_{\text{w}}$ for. 
 From top to bottom generations one to three are shown, and the left-hand panels show up-type quarks,
 and the right-hand panels down-type quarks.}
  \label{Pic:BPAF01}
\end{figure*}

\begin{figure*}[ht]
 \includegraphics[width=0.5\linewidth]{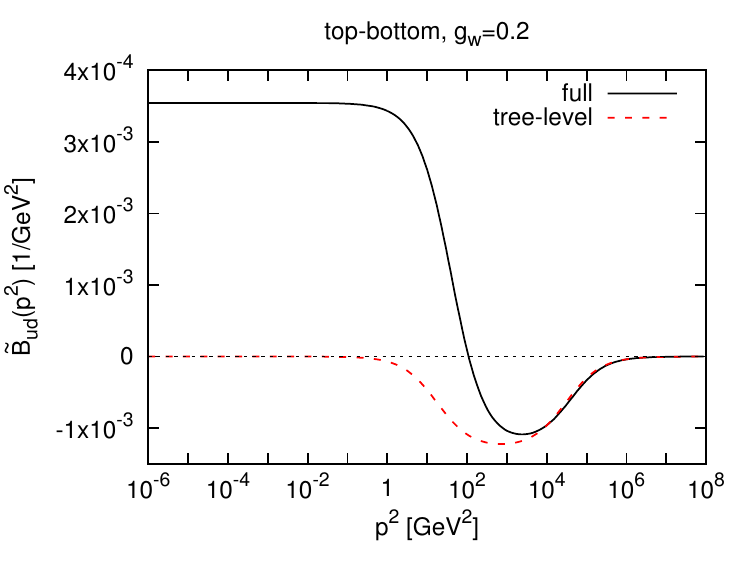}\includegraphics[width=0.5\linewidth]{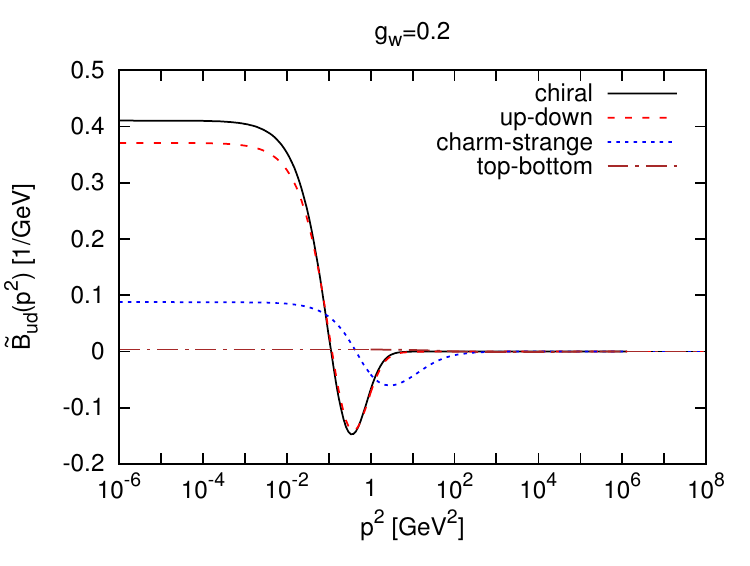}
 \caption{The flavor-off-diagonal vector dressing function for the third generation 
 at $g_{\text{w}}=0.2$ in comparison to tree-level (left panel) and the flavor-off-diagonal scalar 
 dressing function for different generations and the chiral limit at $g_{\text{w}}=0.2$ (right panel).}
  \label{Pic:BM01}
\end{figure*}

\begin{figure*}[ht]
 \includegraphics[width=0.5\linewidth]{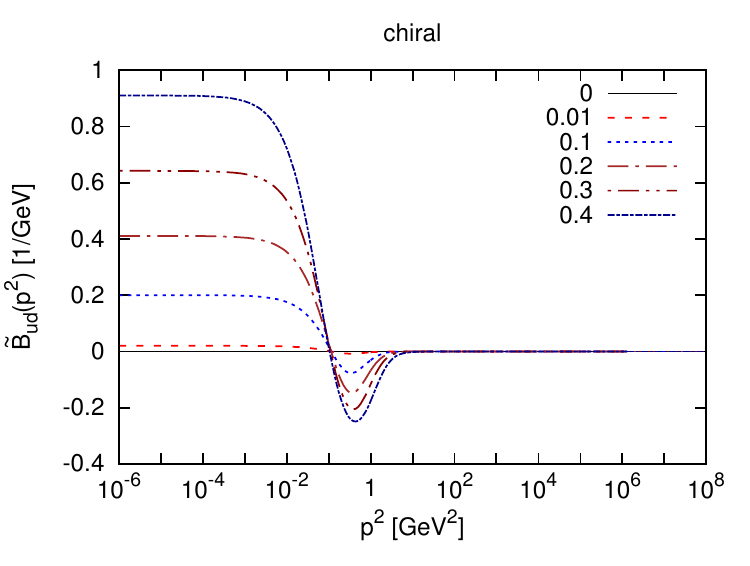}\includegraphics[width=0.5\linewidth]{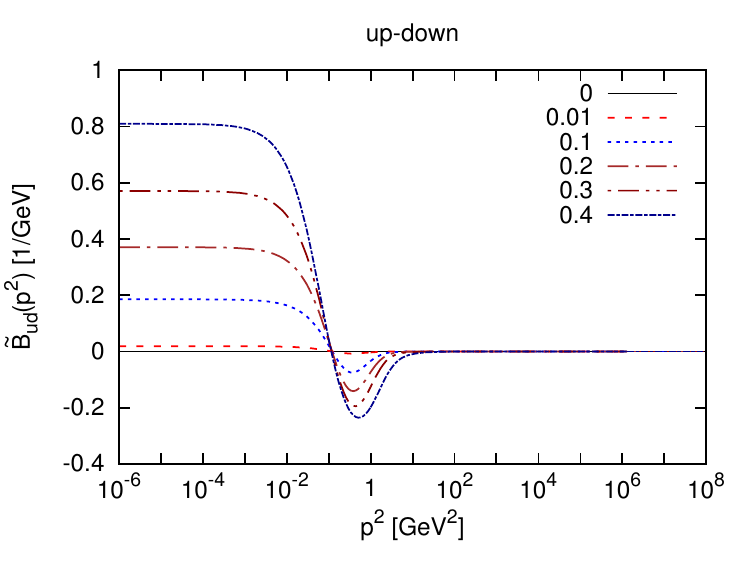}\\
 \includegraphics[width=0.5\linewidth]{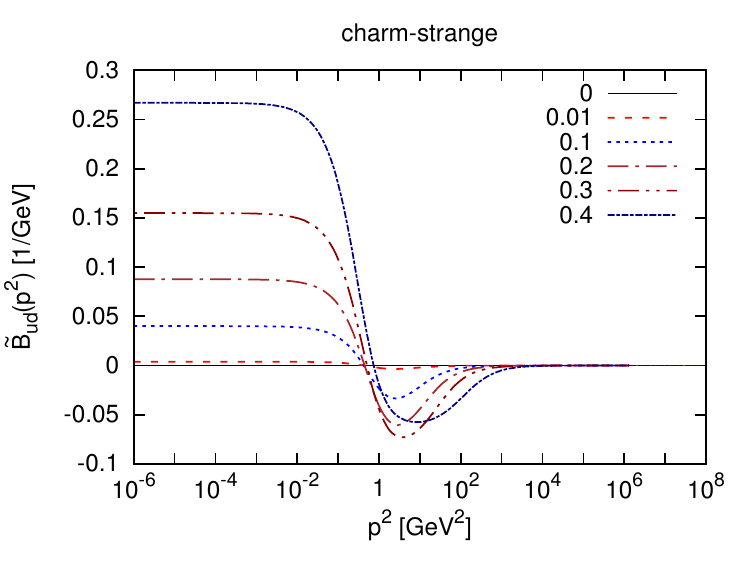}\includegraphics[width=0.5\linewidth]{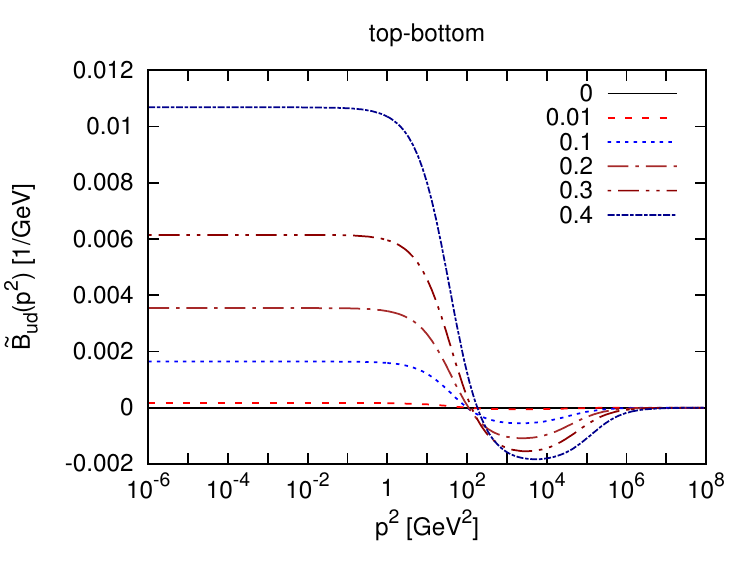}
 \caption{The flavor-off-diagonal scalar dressing function for different values $g_{\text{w}}$ 
 and the different quark generations and the chiral limit.}
  \label{Pic:BMAF01}
\end{figure*}

Complementing the results in section \ref{MF} the flavor-diagonal scalar dressing functions for 
the different quarks are shown in figure \ref{Pic:BPAF01}.

A comparison to the tree-level case of the third generation flavor-off-diagonal scalar dressing function 
is shown in figure \ref{Pic:BM01}. 
Note that the tree-level result is, see equation (\ref{Eqn:TLprop}),
\begin{align}
 \tilde{B}_{0,ud} &= -\frac{g_{\text{w}}(m_u + m_d ) p^{2}}{N(p^{2})}.
 \label{Eqn:TLBM}
\end{align}
Thus, at tree-level $\tilde{B}_{0,ud}$ is negative for all momenta, in particular in the UV. 
The latter is also seen in the full case. But it switches to a positive value in the IR. 
This qualitative behavior is also seen for the other generations, 
as is also plotted in figure \ref{Pic:BM01}, but at differing absolute values. 
This persists even in the chiral limit. 
The value of this quantity is found to also increase when increasing $g_{\text{w}}$, 
which is shown in figure \ref{Pic:BMAF01}.

%%%%%%%%%%%%%%%%%%%%%%%%%%%%%%%%%%%%%%%%%%%%%%%%%%%%%%%%%%%%%%%%%%%%%%%%%%%%%%%%%%5
\subsection{Axial Channel}
\label{AC}

\begin{figure*}[ht]
 \includegraphics[width=0.5\linewidth]{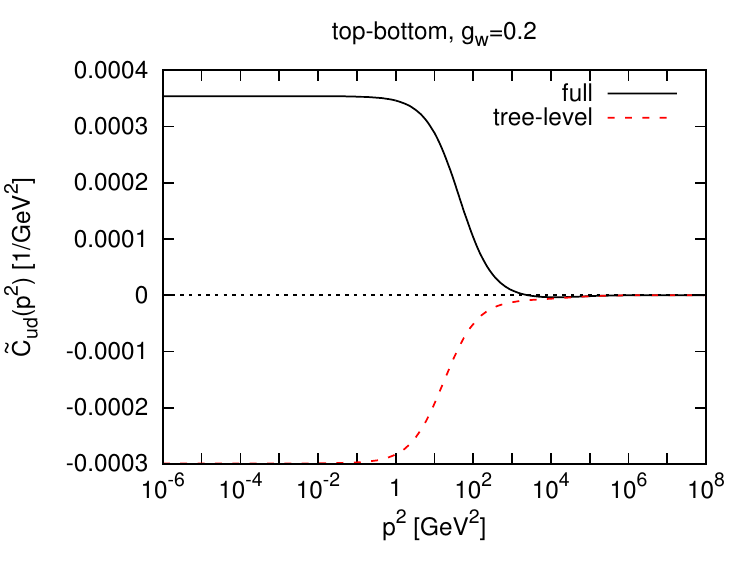}\includegraphics[width=0.5\linewidth]{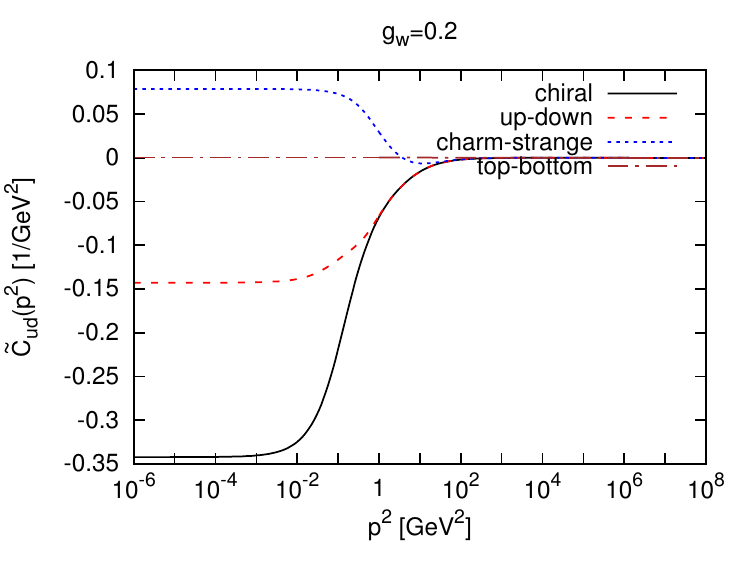}
 \caption{The flavor-off-diagonal axial dressing function at $g_{\text{w}}=0.2$ for 
 the third generation compared to tree-level (left panel) and for the different generations (right panel).}
  \label{Pic:CM01}
\end{figure*}

\begin{figure*}[ht]
 \includegraphics[width=0.5\linewidth]{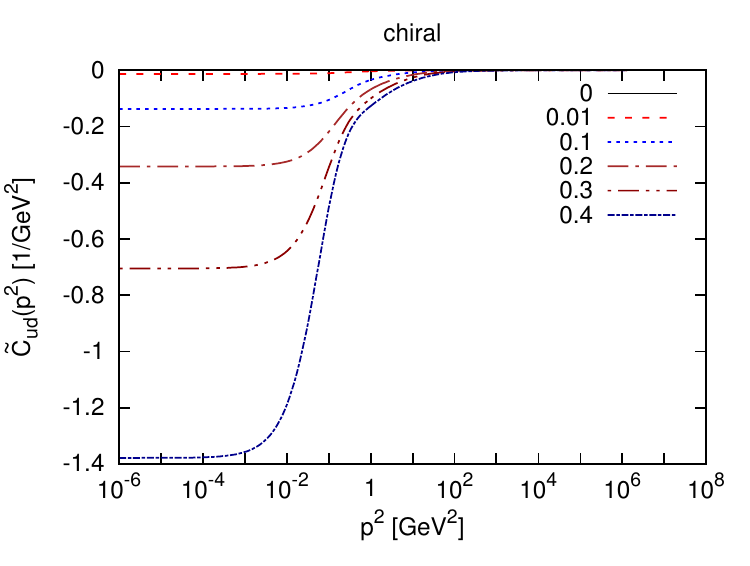}\includegraphics[width=0.5\linewidth]{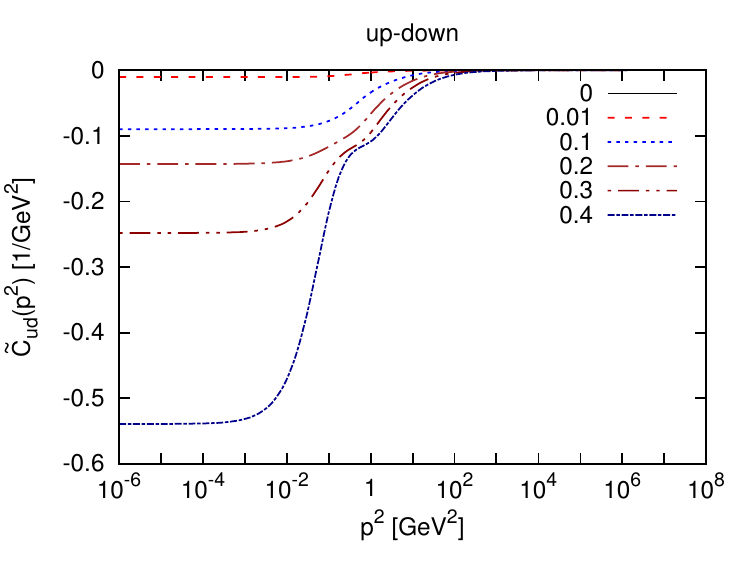}\\
 \includegraphics[width=0.5\linewidth]{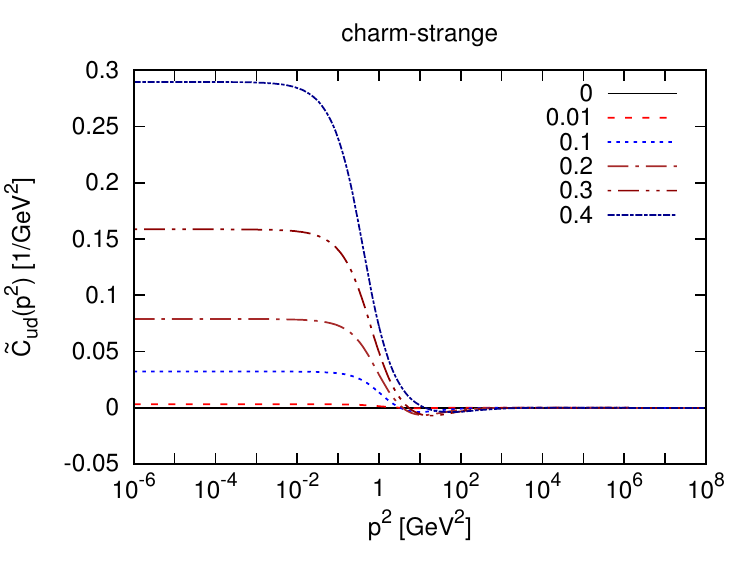}\includegraphics[width=0.5\linewidth]{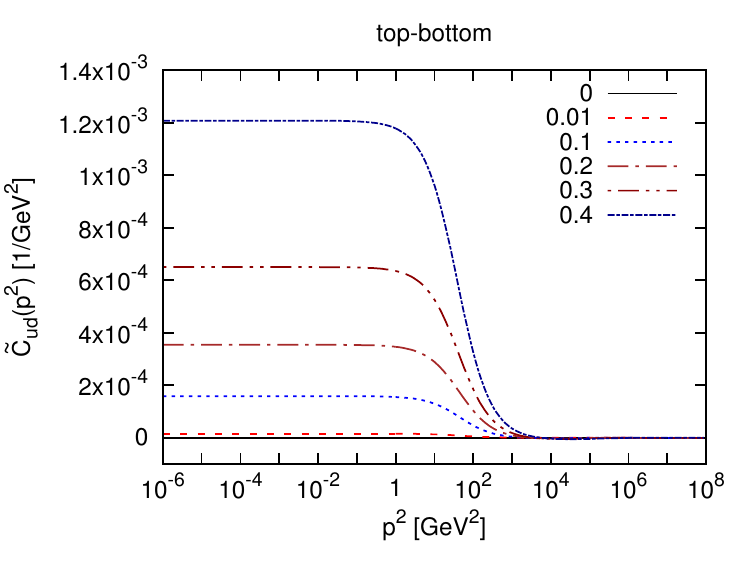}
 \caption{The dependence of the flavor-off-diagonal axial dressing function on 
 $g_{\text{w}}$ for the different generations and the chiral limit.}
  \label{Pic:CMAF01}
\end{figure*}

The flavor-off-diagonal axial dressing function is shown for the various generations and 
the chiral limit in comparison to tree-level in figure \ref{Pic:CM01} at fixed $g_{\text{w}}=0.2$. 
The tree-level propagator takes the form, see equation (\ref{Eqn:TLprop}),
\begin{align}
 \tilde{C}_{0,ud} &= -\frac{g_{\text{w}}(m_u m_d + p^{2})}{N(p^{2})}.
 \label{Eqn:TLCM}
\end{align}
Therefore the full dressing function becomes negative in the UV. 
In the IR it is positive for the third and second generation, 
as is visible from figure \ref{Pic:CM01}, but remains negative for the first generation and the chiral limit. 
The change of the relative ratio between the left-handed and right-handed contribution
is related to the sign of $\tilde{C}$, see equation (\ref{Eqn:RLR}).
For the second and third generation the contribution from the mass splitting is 
high enough to make $\tilde{C}$ positive in the IR.
A detailed discussion is given in section \ref{Parity Violation}.

In figure \ref{Pic:CMAF01} the dependence on $g_{\text{w}}$ is shown for the different generation and the chiral limit. 
The higher the value of $g_{\text{w}}$, the larger the dressing function. 
But the qualitative behavior is unchanged, and therefore entirely controlled by the masses.

%%%%%%%%%%%%%%%%%%%%%%%%%%%%%%%%%%%%%%%%%%%%%%%%%%%%%%%%%%%%%%%%%%%%%%%%%%%%%%%%%%5
\subsection{Pseudo-scalar Channel}
\label{PC}

\begin{figure*}[ht]
 \includegraphics[width=0.5\linewidth]{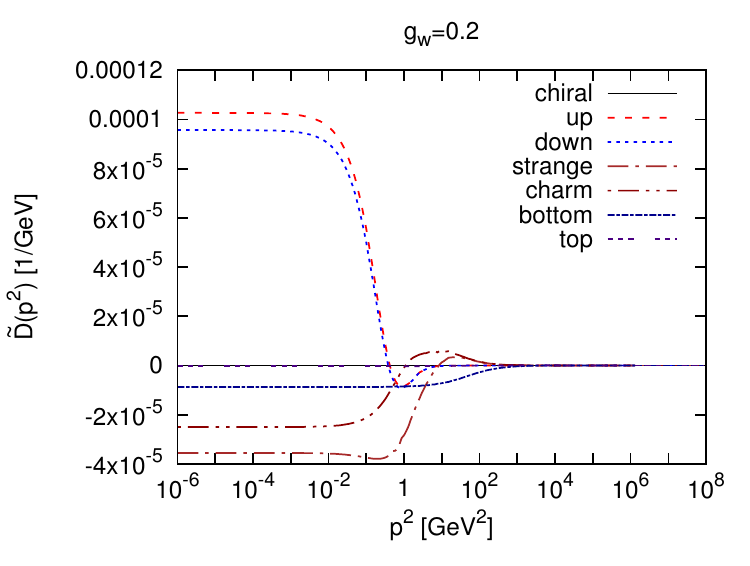}\includegraphics[width=0.5\linewidth]{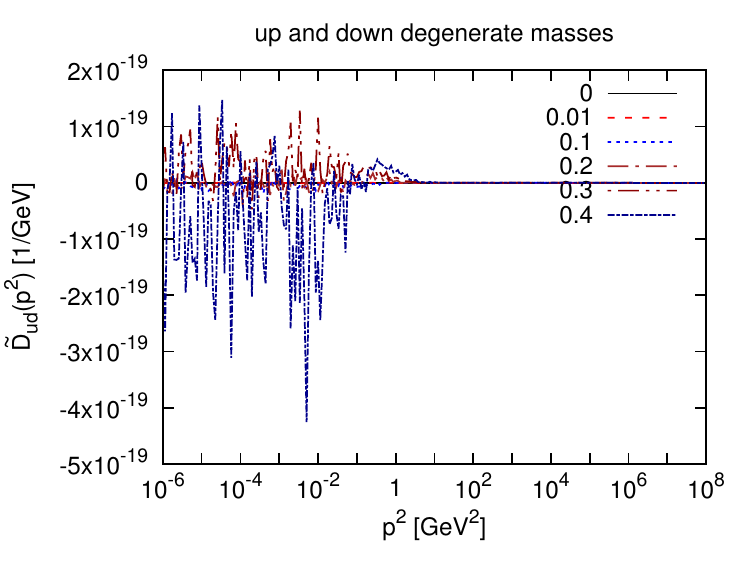}
 \caption{The left-hand panel shows the flavor-diagonal pseudo-scalar dressing function 
 for the different quarks at $g_{\text{w}}=0.2$. 
 In the right-hand panel the flavor-off-diagonal pseudo-scalar dressing function
 is shown for degenerate (here: up and down) quarks masses.}
  \label{Pic:D01}
\end{figure*}

\begin{figure*}[ht]
 \includegraphics[width=0.5\linewidth]{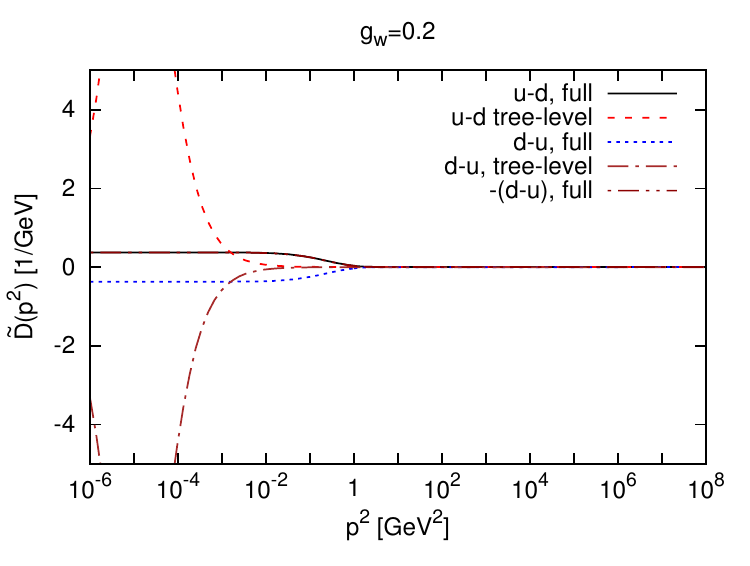}\includegraphics[width=0.5\linewidth]{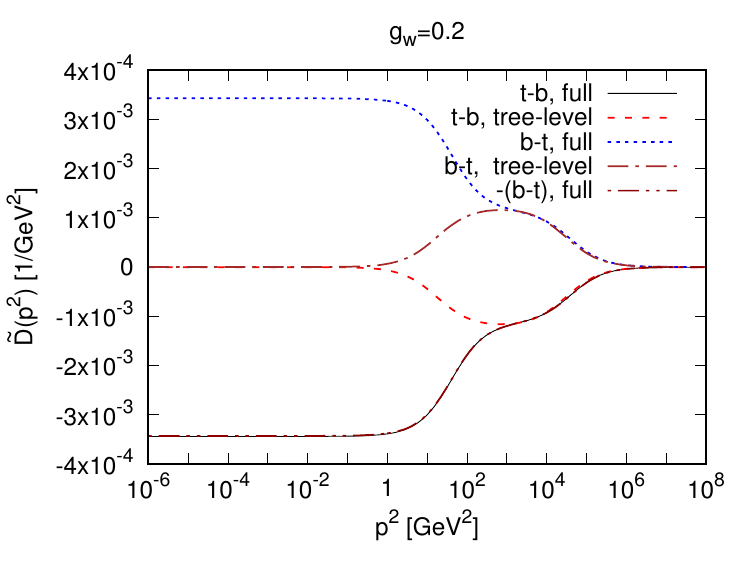}\\
 \includegraphics[width=0.5\linewidth]{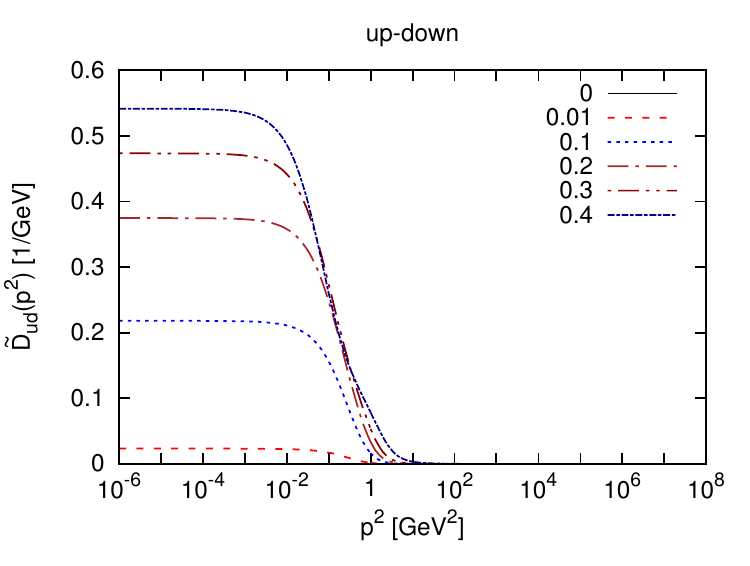}\includegraphics[width=0.5\linewidth]{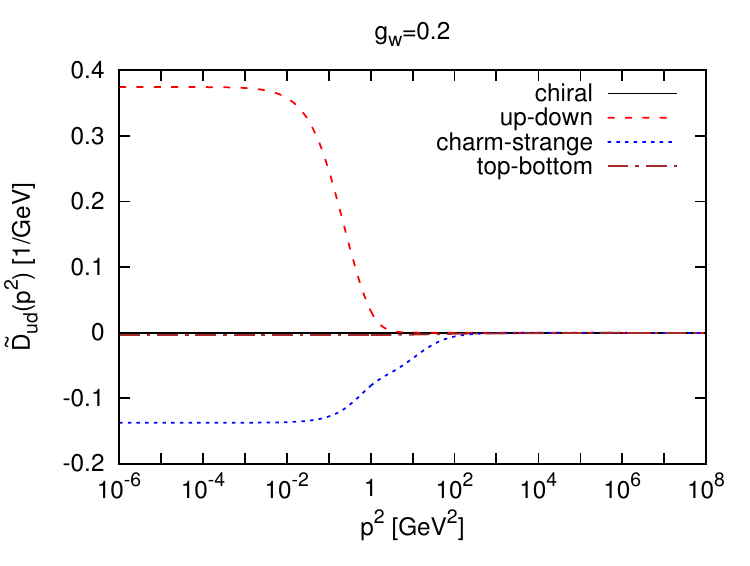}
 \caption{The flavor-off-diagonal pseudo-scalar dressing function in comparison to tree-level for the 
 first (top-left panel) and third (top-right panel) generation, as well as a function of $g_{\text{w}}$ 
 for the first generation (bottom-left panel) and for the different generations at fixed $g_{\text{w}}=0.2$ (bottom-right panel).}
  \label{Pic:DMAF01}
\end{figure*}

At tree-level the flavor-diagonal pseudo-scalar dressing function vanishes.
The full dressing function has a finite value, which is depicted
in figure \ref{Pic:D01} for all quarks at $g_{\text{w}}=0.2$.

The flavor-off-diagonal dressing function at tree-level is zero 
for degenerate quark masses, 
\begin{align}
 \tilde{D}_{0,ud} &= - \tilde{D}_{0,du} = -\frac{g_{\text{w}}(m_u - m_d ) p^{2}}{N(p^{2})},
 \label{Eqn:TLDM}
\end{align}
as can be obtained from equation (\ref{Eqn:TLprop}).
This is also true for the full case, also shown in figure \ref{Pic:D01}, 
and especially so for the chiral limit. 
For non-degenerate quark masses the dressing function does no longer vanish. 
This is in as far remarkable as non-trivial effects due to (non-)degeneracy propagate in 
other dressing functions, as discussed in the main text. 
We also find that the difference between up-to-down and down-to-up dressing function, 
as discussed in section \ref{QP}, also persists in the full case, and both differ by a sign. 
This effect does not propagate to other dressing functions, where we do not find any difference.

The dependence of the flavor-off-diagonal pseudo-scalar dressing function on generation and $g_{\text{w}}$, 
also in comparison to tree-level, is shown in figure \ref{Pic:DMAF01}.
Here also the sign switch between up-to-down and down-to-up is shown. 
Note however that the sign in the infrared is again different for the first generation and the second and third generation.
This is because of the switch of relative sign in the mass difference from the first to the other generation, 
as is already the case at tree-level in (\ref{Eqn:TLDM}). 
The size, but not the qualitative features, increase again with $g_{\text{w}}$. 
The size of the dressing function also decreases with increasing quark mass.

%%%%%%%%%%%%%%%%%%%%%%%%%%%%%%%%%%%%%%%%%%%%%%%%%%%%%%%%%%%%%%%%%%%%%%%%%%%%%%%%%%
\subsection{Relative Ratio}
\label{RR}

\begin{figure*}[ht]
 \includegraphics[width=0.5\linewidth]{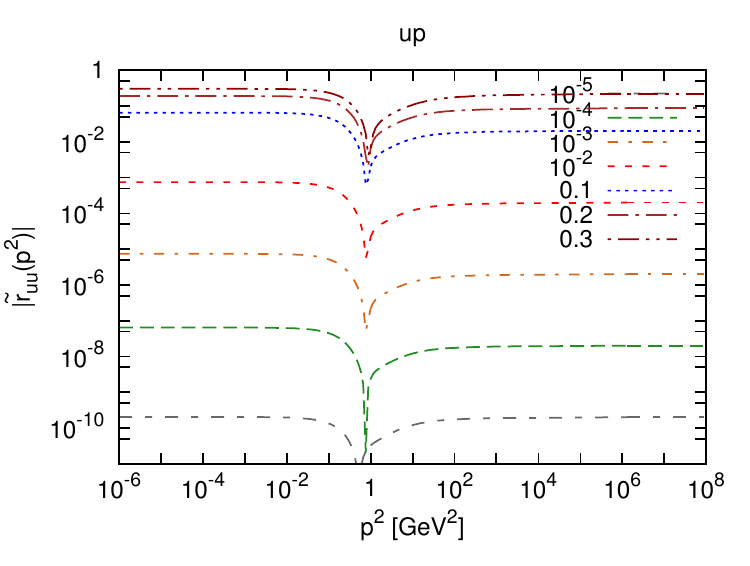}\includegraphics[width=0.5\linewidth]{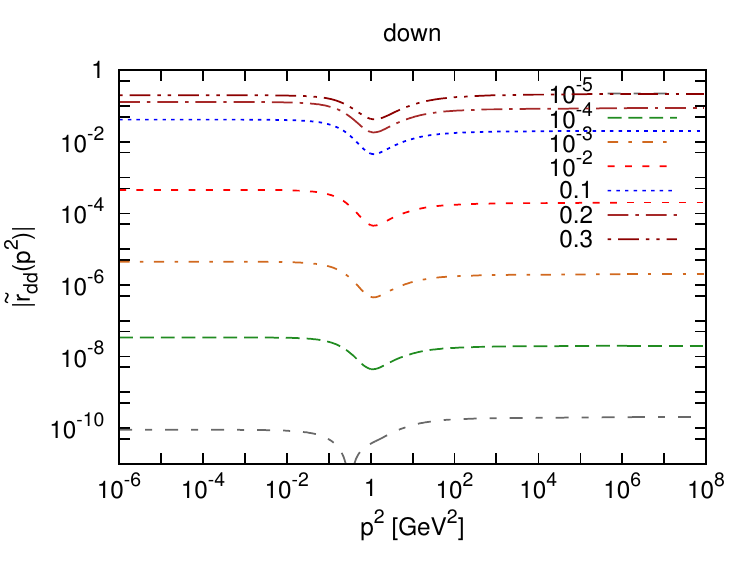}\\
 \includegraphics[width=0.5\linewidth]{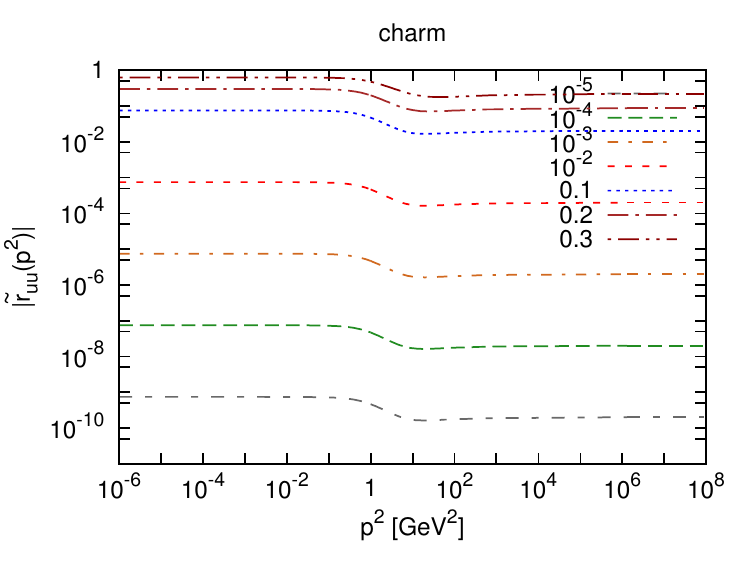}\includegraphics[width=0.5\linewidth]{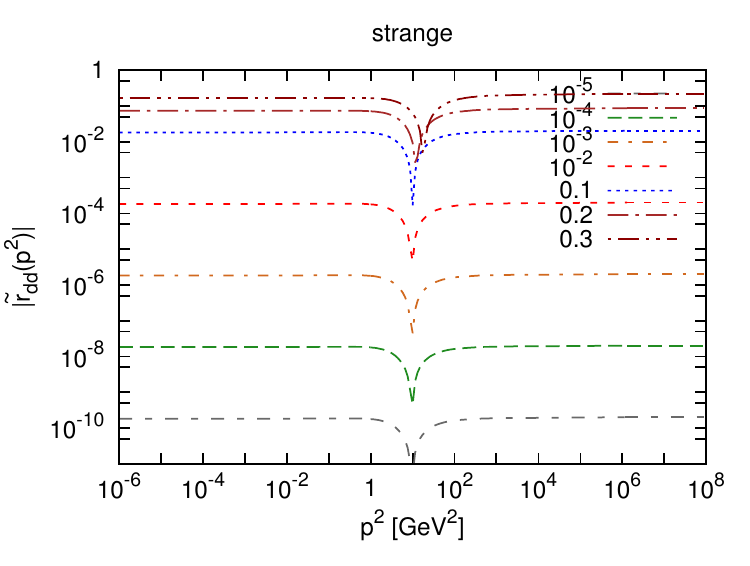}\\
 \includegraphics[width=0.5\linewidth]{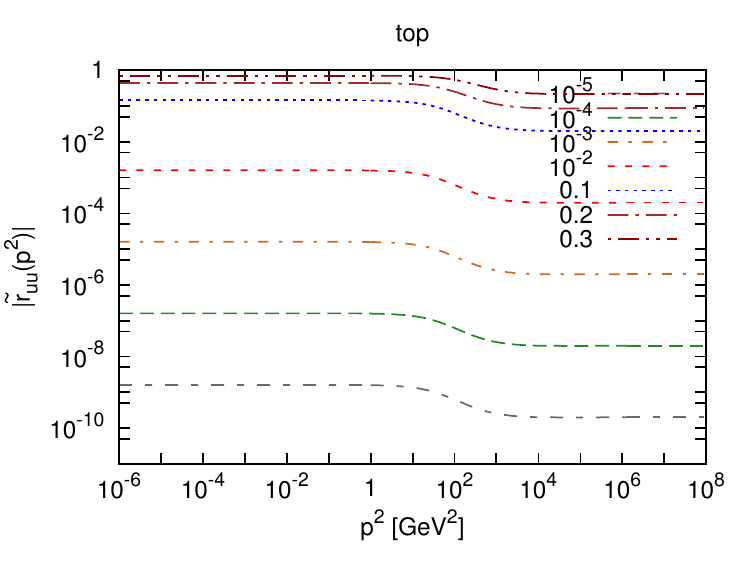}\includegraphics[width=0.5\linewidth]{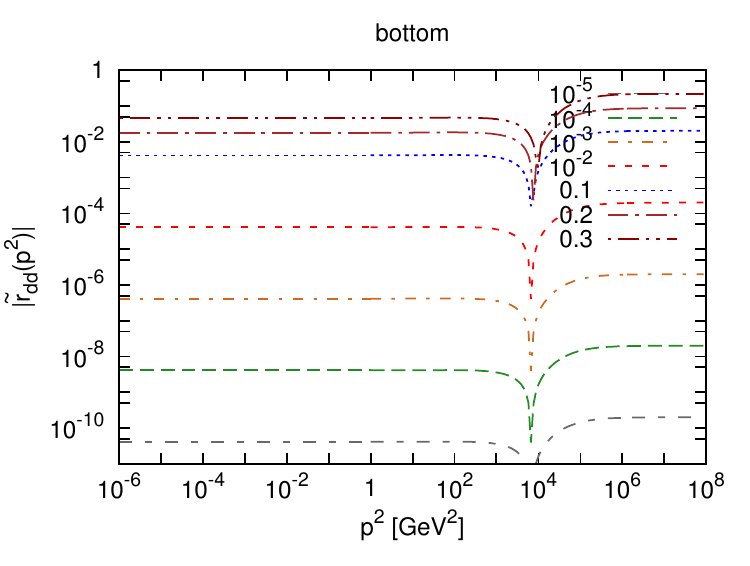}
 \caption{Ratio (\ref{Eqn:RLR}) for the left-handed and right-handed flavor-diagonal 
 quark propagator for different values of $g_{\text{w}}$. 
 From top to bottom generations one to three are shown, and the left-hand panels show up-type quarks, 
 and the right-hand panels down-type quarks.}
  \label{Pic:RAF01}
\end{figure*}

In figure \ref{Pic:RAF01} the relative ratio (\ref{Eqn:RLR}) for the flavor-diagonal elements are shown, 
see section \ref{Parity Violation} for a detailed discussion.

%%%%%%%%%%%%%%%%%%%%%%%%%%%%%%%%%%%%%%%%%%%%%%%%%%%%%%%%%%%%%%%%%%%%%%%%%%%%%%%%%%5
\subsection{Schwinger function}
\label{SF}

\begin{figure*}[ht]
 \includegraphics[width=0.5\linewidth]{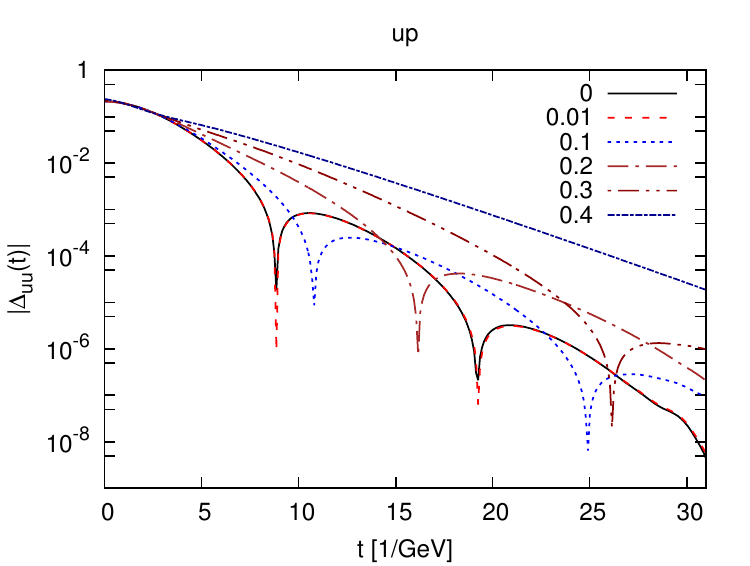}\includegraphics[width=0.5\linewidth]{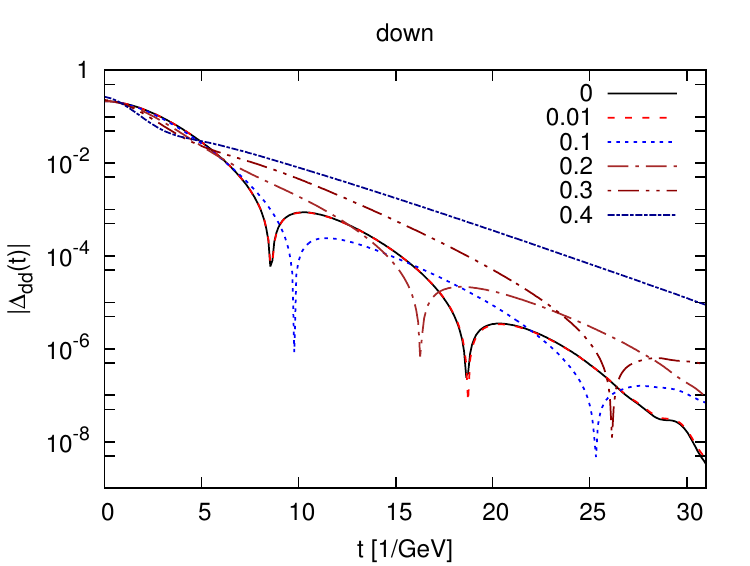}\\
 \includegraphics[width=0.5\linewidth]{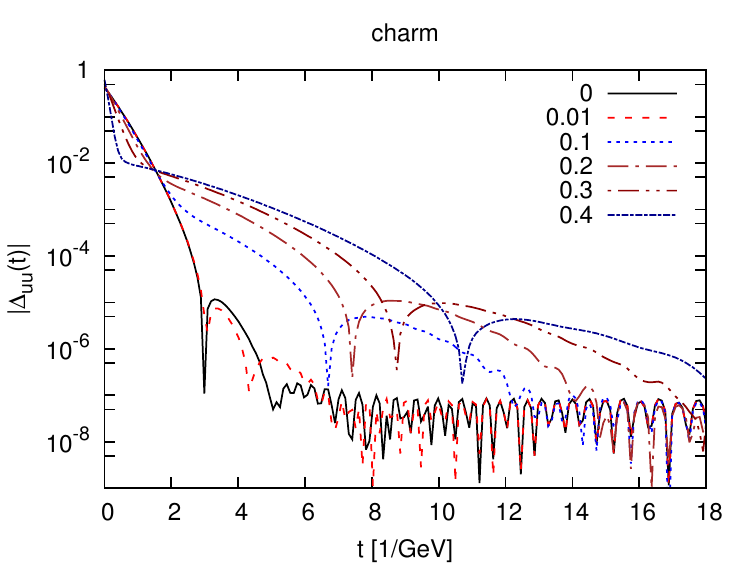}\includegraphics[width=0.5\linewidth]{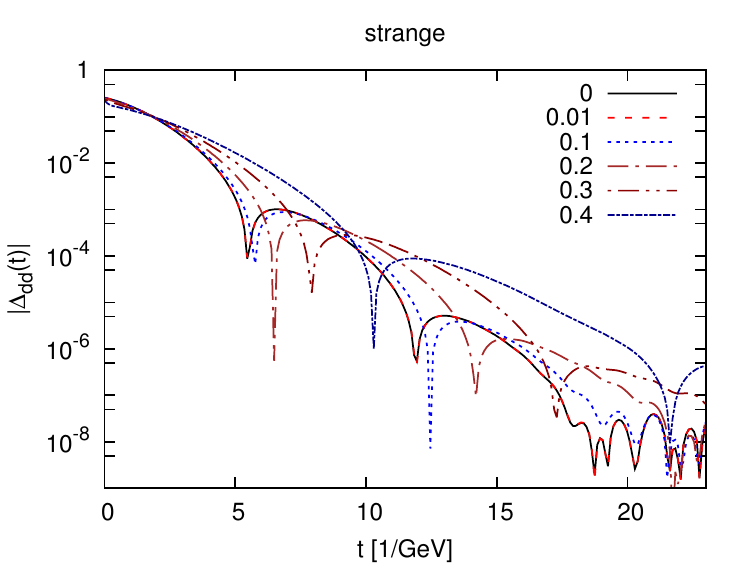}\\ 
 \includegraphics[width=0.5\linewidth]{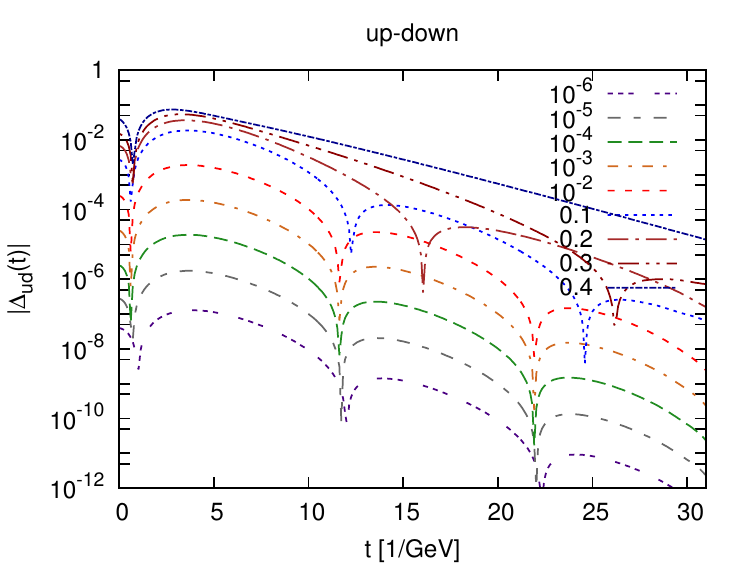}\includegraphics[width=0.5\linewidth]{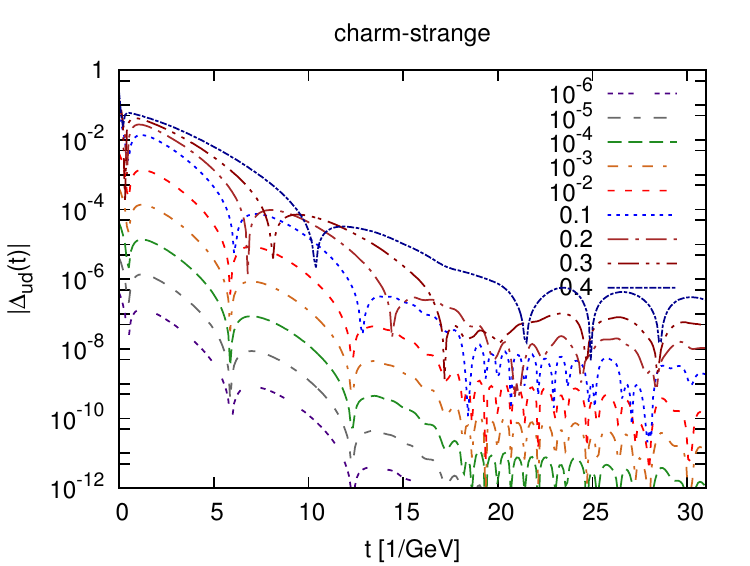}
  \caption{The Schwinger function for different values of $g_{\text{w}}$. 
  Top and middle panels show generations one and two and the left-hand panels show up-type quarks, 
  and the right-hand panels down-type quarks. 
  The bottom panels show the flavor-off-diagonal elements for the first (left panel) and second (right panel) generation.}
  \label{Pic:SW}
\end{figure*}

In figure \ref{Pic:SW} the Schwinger function for the scalar dressing function for the first 
and second generation, both for flavor-diagonal and flavor-off-diagonal elements, are shown. 
The third generation's large tree-level mass leads to a too quick drowning in numerical noise to provide
any reasonable results. 
For a detailed discussion see section section \ref{Masspole}.

\section{Fit parameters for the Schwinger function}
\label{a:fpsf}

The values for the fit-function \cite{Maas:2011se}
\begin{align}
 &\Delta(t)=\frac{e^2}{2m^3\sin\left(2\phi\right)}e^{-tm\cos\left(\phi\right)}\nonumber\\
 &\times\left(\sin\left(\phi+tm\sin\left(\phi\right)\right)+\frac{fm^2}{e^2}\sin\left(\phi-tm\sin\left(\phi\right)\right)\right)\nonumber,
\end{align}
for the Schwinger function in section \ref{Masspole} are listed for the various cases studied in tables 
\ref{tab:Pure} and \ref{tab:mix} for the flavor-diagonal and flavor-off-diagonal elements, respectively. 
The fit form (\ref{Eqn:CPS}) in section \ref{Schwinger} captures the essential features of this form. 
See \cite{Maas:2011se} for a more detailed discussion. 
The errors are an estimate of how strongly the fit can be distorted without 
introducing any substantial deviations from the data.

\begin{table*}[t]
 \begin{tabular}{l||c||c|c|c||c|c|c|c}
 \toprule
  & $g_{\text{w}}$ & $a$ & $b$ & $\delta$ & $e$ & $f$ & $m$& $\phi$ \\
 \toprule
 chiral & 
  \begin{tabular}{c}
  $0$ \\ $0.01$ \\  $0.05$ \\  $0.1$ \\  $0.2$ \\  $0.3$ \\  $0.4$ \\
  \end{tabular} &
  \begin{tabular}{c}
  $0.54 \pm 0.05$ \\ $0.54 \pm 0.05$ \\  $0.54 \pm 0.05$ \\  $0.47 \pm 0.05$ \\
  $0.42 \pm 0.05$ \\  $0.38 \pm 0.05$ \\  $0.31 \pm 0.05$ \\
  \end{tabular} &
  \begin{tabular}{c}
  $0.30 \pm 0.01$ \\ $0.30 \pm 0.01$ \\  $0.26 \pm 0.01$ \\ $0.21 \pm 0.01$ \\
  $0.17 \pm 0.01$ \\  $0.05 \pm 0.01$ \\  $10^{-7} \pm 0.01$ \\
  \end{tabular} &
  \begin{tabular}{c}
  $-1.13 \pm 0.05$ \\ $-1.13 \pm 0.05$ \\  $-0.90 \pm 0.05$ \\ $-0.66 \pm 0.05$ \\
  $-0.14 \pm 0.05$ \\  $0 \pm 0.05$ \\  $0 \pm 0.05$ \\
  \end{tabular} &
  \begin{tabular}{c}
  $0.44 \pm 0.05$ \\ $0.44 \pm 0.05$ \\  $0.48 \pm 0.05$ \\ $0.25 \pm 0.05$ \\
  $0.16 \pm 0.05$ \\  $0.26 \pm 0.05$ \\  $0.15 \pm 0.05$ \\
  \end{tabular} &
  \begin{tabular}{c}
  $0.043 \pm 0.005$ \\ $0.043 \pm 0.005$ \\  $0.14 \pm 0.05$ \\ $0.12 \pm 0.05$ \\
  $-0.04 \pm 0.05$ \\  $0.46 \pm 0.05$ \\  $0.25 \pm 0.05$ \\
  \end{tabular} &
  \begin{tabular}{c}
  $0.61 \pm 0.05$ \\ $0.61 \pm 0.05$ \\  $0.60 \pm 0.05$ \\ $0.51 \pm 0.05$ \\
  $0.45 \pm 0.05$ \\  $0.38 \pm 0.05$ \\  $0.31 \pm 0.05$ \\
  \end{tabular} &
  \begin{tabular}{c}
  $0.51 \pm 0.05$ \\ $0.51 \pm 0.05$ \\  $0.46 \pm 0.05$ \\ $0.41 \pm 0.05$ \\
  $0.38 \pm 0.05$ \\  $0.14 \pm 0.05$ \\  $10^{-9} \pm 0.05$ \\
  \end{tabular} \\
   \toprule
 up & 
  \begin{tabular}{c}
  $0$ \\ $0.01$ \\  $0.05$ \\  $0.1$ \\  $0.2$ \\  $0.3$ \\  $0.4$ \\
  \end{tabular} &
  \begin{tabular}{c}
  $0.54 \pm 0.05$ \\ $0.54 \pm 0.05$ \\  $0.54 \pm 0.05$ \\  $0.49 \pm 0.05$ \\
  $0.45 \pm 0.05$ \\  $0.39 \pm 0.05$ \\  $0.34 \pm 0.05$ \\
  \end{tabular} &
  \begin{tabular}{c}
  $0.31 \pm 0.01$ \\ $0.31 \pm 0.01$ \\  $0.27 \pm 0.01$ \\ $0.22 \pm 0.01$ \\
  $0.17 \pm 0.01$ \\  $0.06 \pm 0.01$ \\  $10^{-7} \pm 0.01$ \\
  \end{tabular} &
  \begin{tabular}{c}
  $-1.13 \pm 0.05$ \\ $-1.13 \pm 0.05$ \\  $-0.92 \pm 0.05$ \\ $-0.82 \pm 0.05$ \\
  $-1.15 \pm 0.05$ \\  $0 \pm 0.05$ \\  $0 \pm 0.05$ \\
  \end{tabular} &
  \begin{tabular}{c}
  $0.46 \pm 0.05$ \\ $0.46 \pm 0.05$ \\  $0.45 \pm 0.05$ \\ $0.30 \pm 0.05$ \\
  $0.25 \pm 0.05$ \\  $0.27 \pm 0.05$ \\  $0.22 \pm 0.05$ \\
  \end{tabular} &
  \begin{tabular}{c}
  $-0.045 \pm 0.005$ \\ $-0.045 \pm 0.005$ \\  $0.10 \pm 0.05$ \\ $0.11 \pm 0.05$ \\
  $-0.02 \pm 0.05$ \\  $0.47 \pm 0.05$ \\  $0.44 \pm 0.05$ \\
  \end{tabular} &
  \begin{tabular}{c}
  $0.63 \pm 0.05$ \\ $0.63 \pm 0.05$ \\  $0.60 \pm 0.05$ \\ $0.53 \pm 0.05$ \\
  $0.48 \pm 0.05$ \\  $0.39 \pm 0.05$ \\  $0.34 \pm 0.05$ \\
  \end{tabular} &
  \begin{tabular}{c}
  $0.51 \pm 0.05$ \\ $0.51 \pm 0.05$ \\  $0.47 \pm 0.05$ \\ $0.43 \pm 0.05$ \\
  $0.36 \pm 0.05$ \\  $0.14 \pm 0.05$ \\  $10^{-9} \pm 0.05$ \\
  \end{tabular} \\
   \toprule
 down & 
  \begin{tabular}{c}
  $0$ \\ $0.01$ \\  $0.05$ \\  $0.1$ \\  $0.2$ \\  $0.3$ \\  $0.4$ \\
  \end{tabular} &
  \begin{tabular}{c}
  $0.56 \pm 0.05$ \\ $0.56 \pm 0.05$ \\  $0.56 \pm 0.05$ \\  $0.50 \pm 0.05$ \\
  $0.46 \pm 0.05$ \\  $0.39 \pm 0.05$ \\  $0.34 \pm 0.05$ \\
  \end{tabular} &
  \begin{tabular}{c}
  $0.32 \pm 0.01$ \\ $0.32 \pm 0.01$ \\  $0.28 \pm 0.01$ \\ $0.20 \pm 0.01$ \\
  $0.19 \pm 0.01$ \\  $0.06 \pm 0.01$ \\  $10^{-7} \pm 0.01$ \\
  \end{tabular} &
  \begin{tabular}{c}
  $-1.13 \pm 0.05$ \\ $-1.13 \pm 0.05$ \\  $-0.86 \pm 0.05$ \\ $-0.40 \pm 0.05$ \\
  $-1.45 \pm 0.05$ \\  $0 \pm 0.05$ \\  $0 \pm 0.05$ \\
  \end{tabular} &
  \begin{tabular}{c}
  $0.47 \pm 0.05$ \\ $0.47 \pm 0.05$ \\  $0.46 \pm 0.05$ \\ $0.24 \pm 0.05$ \\
  $0.17 \pm 0.05$ \\  $0.22 \pm 0.05$ \\  $0.17 \pm 0.05$ \\
  \end{tabular} &
  \begin{tabular}{c}
  $-0.046 \pm 0.005$ \\ $-0.046 \pm 0.005$ \\  $0.15 \pm 0.05$ \\ $0.15 \pm 0.05$ \\
  $-0.06 \pm 0.05$ \\  $0.30 \pm 0.05$ \\  $0.25 \pm 0.05$ \\
  \end{tabular} &
  \begin{tabular}{c}
  $0.64 \pm 0.05$ \\ $0.64 \pm 0.05$ \\  $0.63 \pm 0.05$ \\ $0.54 \pm 0.05$ \\
  $0.49 \pm 0.05$ \\  $0.40 \pm 0.05$ \\  $0.34 \pm 0.05$ \\
  \end{tabular} &
  \begin{tabular}{c}
  $0.51 \pm 0.05$ \\ $0.51 \pm 0.05$ \\  $0.46 \pm 0.05$ \\ $0.39 \pm 0.05$ \\
  $0.39 \pm 0.05$ \\  $0.15 \pm 0.05$ \\  $10^{-9} \pm 0.05$ \\
  \end{tabular}
 \end{tabular}
\caption{Fit values for the flavor-diagonal Schwinger functions.}
\label{tab:Pure}
\end{table*}

\begin{table*}[t]
 \begin{tabular}{l||c||c|c|c||c|c|c|c}
 \toprule
  & $g_{\text{w}}$ & $a$ & $b$ & $\delta$ & $e$ & $f$ & $m$& $\phi$ \\
 \toprule
 chiral & 
  \begin{tabular}{c}
  $10^{-6}$ \\ $10^{-5}$ \\ $10^{-4}$ \\ $10^{-3}$ \\ $10^{-2}$ \\ $0.05$ \\
  $0.1$ \\ $0.2$ \\ $0.3$ \\ $0.4$
  \end{tabular} &
  \begin{tabular}{c}
  $0.48 \pm 0.05$ \\  $0.47 \pm 0.05$ \\  $0.47 \pm 0.05$ \\  $0.47 \pm 0.05$ \\  $0.47 \pm 0.05$ \\
  $0.48 \pm 0.05$ \\  $0.49 \pm 0.05$ \\  $0.43 \pm 0.05$ \\  $0.42 \pm 0.05$ \\  $0.35 \pm 0.05$
  \end{tabular} &
  \begin{tabular}{c}
  $0.29 \pm 0.01$ \\ $0.29 \pm 0.01$ \\  $0.29 \pm 0.01$ \\  $0.29 \pm 0.01$ \\  $0.29 \pm 0.01$ \\
  $0.28 \pm 0.01$ \\ $0.24 \pm 0.01$ \\  $0.19 \pm 0.01$ \\  $0.11 \pm 0.01$ \\ $10^{-4} \pm 0.01$
  \end{tabular} &
  \begin{tabular}{c}
  $-2.12 \pm 0.05$ \\ $-2.03 \pm 0.05$ \\  $-2.00 \pm 0.05$ \\ $-2.00 \pm 0.05$ \\ $-2.00 \pm 0.05$ \\
  $-1.93 \pm 0.05$ \\ $-1.57 \pm 0.05$ \\  $-1.68 \pm 0.05$ \\ $-1.64 \pm 0.05$ \\ $-1.57087$ 
  \end{tabular} &
  \begin{tabular}{c}
  $(3.0 \pm 0.5)\times 10^{-5}$ \\ $(9.5 \pm 0.5)\times 10^{-4}$ \\ $(3.5 \pm 0.5)\times 10^{-3}$ \\
  $(1.2 \pm 0.5)\times 10^{-2}$ \\ $(3.5 \pm 0.5)\times 10^{-2}$ \\ $0.11 \pm 0.05$ \\
  $0.25 \pm 0.05$ \\ $0.13 \pm 0.05$ \\ $0.16 \pm 0.05$ \\ $0.10 \pm 0.05$
  \end{tabular} &
  \begin{tabular}{c}
  $(-2.5 \pm 0.5) \times 10^{-6}$ \\ $(-2.5 \pm 0.5) \times 10^{-5}$ \\ $(-3.0 \pm 0.5) \times 10^{-4}$ \\
  $(-2.8 \pm 0.5) \times 10^{-3}$ \\ $(-3.0 \pm 0.5) \times 10^{-2}$ \\ $-0.15 \pm 0.05$ \\ $-0.20 \pm 0.05$ \\
  $-0.13 \pm 0.05$ \\  $-0.23 \pm 0.05$ \\  $-0.12 \pm 0.05$
  \end{tabular} &
  \begin{tabular}{c}
  $0.56 \pm 0.05$ \\ $0.55 \pm 0.05$ \\  $0.55 \pm 0.05$ \\ $0.55 \pm 0.05$ \\ $0.55 \pm 0.05$ \\
  $0.56 \pm 0.05$ \\ $0.55 \pm 0.05$ \\ $0.47 \pm 0.05$ \\  $0.43 \pm 0.05$ \\  $0.35 \pm 0.05$ 
  \end{tabular} &
  \begin{tabular}{c}
  $0.55 \pm 0.05$ \\ $0.56 \pm 0.05$ \\  $0.56 \pm 0.05$ \\ $0.56 \pm 0.05$ \\ $0.56 \pm 0.05$ \\
  $0.53 \pm 0.05$ \\ $0.46 \pm 0.05$ \\  $0.41 \pm 0.05$ \\  $0.25 \pm 0.05$ \\ $2.857 \times 10^{-4}$ 
  \end{tabular} \\
 \toprule
 up-down & 
  \begin{tabular}{c}
  $10^{-6}$ \\ $10^{-5}$ \\ $10^{-4}$ \\ $10^{-3}$ \\ $10^{-2}$ \\ $0.05$ \\
  $0.1$ \\ $0.2$ \\ $0.3$ \\ $0.4$
  \end{tabular} &
  \begin{tabular}{c}
  $0.49 \pm 0.05$ \\  $0.49 \pm 0.05$ \\  $0.49 \pm 0.05$ \\  $0.49 \pm 0.05$ \\  $0.49 \pm 0.05$ \\
  $0.49 \pm 0.05$ \\  $0.49 \pm 0.05$ \\  $0.44 \pm 0.05$ \\  $0.44 \pm 0.05$ \\  $0.38 \pm 0.05$ \\
  \end{tabular} &
  \begin{tabular}{c}
  $0.31 \pm 0.01$ \\ $0.31 \pm 0.01$ \\  $0.31 \pm 0.01$ \\  $0.31 \pm 0.01$ \\  $0.31 \pm 0.01$ \\
  $0.29 \pm 0.01$ \\ $0.26 \pm 0.01$ \\  $0.20 \pm 0.01$ \\  $0.12 \pm 0.01$ \\ $10^{-4} \pm 0.01$
  \end{tabular} &
  \begin{tabular}{c}
  $-2.10 \pm 0.05$ \\ $-2.01 \pm 0.05$ \\  $-1.99 \pm 0.05$ \\ $-1.99 \pm 0.05$ \\ $-1.99 \pm 0.05$ \\
  $-1.90 \pm 0.05$ \\ $-1.56 \pm 0.05$ \\  $-1.71 \pm 0.05$ \\ $-1.66 \pm 0.05$ \\ $-1.57087$
  \end{tabular} &
  \begin{tabular}{c}
  $(1.8 \pm 0.5)\times 10^{-4}$ \\ $(1.1 \pm 0.5)\times 10^{-3}$ \\ $(4.5 \pm 0.5)\times 10^{-3}$ \\
  $(1.4 \pm 0.5)\times 10^{-2}$ \\ $(4.0 \pm 0.5)\times 10^{-2}$ \\ $0.11 \pm 0.05$ \\
  $0.22 \pm 0.05$ \\ $0.13 \pm 0.05$ \\ $0.15 \pm 0.05$ \\ $0.10 \pm 0.05$
  \end{tabular} &
  \begin{tabular}{c}
  $(-2.7 \pm 0.5) \times 10^{-6}$ \\ $(-2.9 \pm 0.5) \times 10^{-5}$ \\ $(-3.2 \pm 0.5) \times 10^{-4}$ \\
  $(-3.5 \pm 0.5) \times 10^{-3}$ \\ $(-3.1 \pm 0.5) \times 10^{-2}$ \\ $-0.14 \pm 0.05$ \\ $-0.15 \pm 0.05$ \\
  $-0.13 \pm 0.05$ \\  $-0.23 \pm 0.05$ \\  $-0.13 \pm 0.05$
  \end{tabular} &
  \begin{tabular}{c}
  $0.58 \pm 0.05$ \\ $0.58 \pm 0.05$ \\  $0.58 \pm 0.05$ \\ $0.58 \pm 0.05$ \\ $0.58 \pm 0.05$ \\
  $0.57 \pm 0.05$ \\ $0.55 \pm 0.05$ \\ $0.49 \pm 0.05$ \\  $0.46 \pm 0.05$ \\  $0.38 \pm 0.05$
  \end{tabular} &
  \begin{tabular}{c}
  $0.56 \pm 0.05$ \\ $0.56 \pm 0.05$ \\  $0.56 \pm 0.05$ \\ $0.56 \pm 0.05$ \\ $0.56 \pm 0.05$ \\
  $0.54 \pm 0.05$ \\ $0.48 \pm 0.05$ \\  $0.43 \pm 0.05$ \\  $0.27 \pm 0.05$ \\ $2.652 \times 10^{-4}$
  \end{tabular}
 \end{tabular}
\caption{Fit values for the flavor-off-diagonal Schwinger functions.}
\label{tab:mix}
\end{table*}

\bibliography{references}

\begin{thebibliography}{27}
\expandafter\ifx\csname natexlab\endcsname\relax\def\natexlab#1{#1}\fi
\expandafter\ifx\csname bibnamefont\endcsname\relax
  \def\bibnamefont#1{#1}\fi
\expandafter\ifx\csname bibfnamefont\endcsname\relax
  \def\bibfnamefont#1{#1}\fi
\expandafter\ifx\csname citenamefont\endcsname\relax
  \def\citenamefont#1{#1}\fi
\expandafter\ifx\csname url\endcsname\relax
  \def\url#1{\texttt{#1}}\fi
\expandafter\ifx\csname urlprefix\endcsname\relax\def\urlprefix{URL }\fi
\providecommand{\bibinfo}[2]{#2}
\providecommand{\eprint}[2][]{\url{#2}}

\bibitem[{\citenamefont{Abbott et~al.}(2016)}]{Abbott:2016blz}
\bibinfo{author}{\bibfnamefont{B.~P.} \bibnamefont{Abbott}}
  \bibnamefont{et~al.} (\bibinfo{collaboration}{Virgo, LIGO Scientific}),
  \bibinfo{journal}{Phys. Rev. Lett.} \textbf{\bibinfo{volume}{116}},
  \bibinfo{pages}{061102} (\bibinfo{year}{2016}), \eprint{1602.03837}.

\bibitem[{\citenamefont{Rosswog and Liebendoerfer}(2003)}]{Rosswog:2003rv}
\bibinfo{author}{\bibfnamefont{S.}~\bibnamefont{Rosswog}} \bibnamefont{and}
  \bibinfo{author}{\bibfnamefont{M.}~\bibnamefont{Liebendoerfer}},
  \bibinfo{journal}{Mon. Not. Roy. Astron. Soc.}
  \textbf{\bibinfo{volume}{342}}, \bibinfo{pages}{673} (\bibinfo{year}{2003}),
  \eprint{astro-ph/0302301}.

\bibitem[{\citenamefont{Sekiguchi et~al.}(2011)\citenamefont{Sekiguchi, Kiuchi,
  Kyutoku, and Shibata}}]{Sekiguchi:2011zd}
\bibinfo{author}{\bibfnamefont{Y.}~\bibnamefont{Sekiguchi}},
  \bibinfo{author}{\bibfnamefont{K.}~\bibnamefont{Kiuchi}},
  \bibinfo{author}{\bibfnamefont{K.}~\bibnamefont{Kyutoku}}, \bibnamefont{and}
  \bibinfo{author}{\bibfnamefont{M.}~\bibnamefont{Shibata}},
  \bibinfo{journal}{Phys. Rev. Lett.} \textbf{\bibinfo{volume}{107}},
  \bibinfo{pages}{051102} (\bibinfo{year}{2011}), \eprint{1105.2125}.

\bibitem[{\citenamefont{Faber and Rasio}(2012)}]{Faber:2012rw}
\bibinfo{author}{\bibfnamefont{J.~A.} \bibnamefont{Faber}} \bibnamefont{and}
  \bibinfo{author}{\bibfnamefont{F.~A.} \bibnamefont{Rasio}},
  \bibinfo{journal}{Living Rev. Rel.} \textbf{\bibinfo{volume}{15}},
  \bibinfo{pages}{8} (\bibinfo{year}{2012}), \eprint{1204.3858}.

\bibitem[{\citenamefont{Neilsen et~al.}(2014)\citenamefont{Neilsen, Liebling,
  Anderson, Lehner, O'Connor, and Palenzuela}}]{Neilsen:2014hha}
\bibinfo{author}{\bibfnamefont{D.}~\bibnamefont{Neilsen}},
  \bibinfo{author}{\bibfnamefont{S.~L.} \bibnamefont{Liebling}},
  \bibinfo{author}{\bibfnamefont{M.}~\bibnamefont{Anderson}},
  \bibinfo{author}{\bibfnamefont{L.}~\bibnamefont{Lehner}},
  \bibinfo{author}{\bibfnamefont{E.}~\bibnamefont{O'Connor}}, \bibnamefont{and}
  \bibinfo{author}{\bibfnamefont{C.}~\bibnamefont{Palenzuela}},
  \bibinfo{journal}{Phys. Rev.} \textbf{\bibinfo{volume}{D89}},
  \bibinfo{pages}{104029} (\bibinfo{year}{2014}), \eprint{1403.3680}.

\bibitem[{\citenamefont{Palenzuela et~al.}(2015)\citenamefont{Palenzuela,
  Liebling, Neilsen, Lehner, Caballero, O'Connor, and
  Anderson}}]{Palenzuela:2015dqa}
\bibinfo{author}{\bibfnamefont{C.}~\bibnamefont{Palenzuela}},
  \bibinfo{author}{\bibfnamefont{S.~L.} \bibnamefont{Liebling}},
  \bibinfo{author}{\bibfnamefont{D.}~\bibnamefont{Neilsen}},
  \bibinfo{author}{\bibfnamefont{L.}~\bibnamefont{Lehner}},
  \bibinfo{author}{\bibfnamefont{O.~L.} \bibnamefont{Caballero}},
  \bibinfo{author}{\bibfnamefont{E.}~\bibnamefont{O'Connor}}, \bibnamefont{and}
  \bibinfo{author}{\bibfnamefont{M.}~\bibnamefont{Anderson}},
  \bibinfo{journal}{Phys. Rev.} \textbf{\bibinfo{volume}{D92}},
  \bibinfo{pages}{044045} (\bibinfo{year}{2015}), \eprint{1505.01607}.

\bibitem[{\citenamefont{Sekiguchi et~al.}(2015)\citenamefont{Sekiguchi, Kiuchi,
  Kyutoku, and Shibata}}]{Sekiguchi:2015dma}
\bibinfo{author}{\bibfnamefont{Y.}~\bibnamefont{Sekiguchi}},
  \bibinfo{author}{\bibfnamefont{K.}~\bibnamefont{Kiuchi}},
  \bibinfo{author}{\bibfnamefont{K.}~\bibnamefont{Kyutoku}}, \bibnamefont{and}
  \bibinfo{author}{\bibfnamefont{M.}~\bibnamefont{Shibata}},
  \bibinfo{journal}{Phys. Rev.} \textbf{\bibinfo{volume}{D91}},
  \bibinfo{pages}{064059} (\bibinfo{year}{2015}), \eprint{1502.06660}.

\bibitem[{\citenamefont{Foucart et~al.}(2016)\citenamefont{Foucart, Haas, Duez,
  O'Connor, Ott, Roberts, Kidder, Lippuner, Pfeiffer, and
  Scheel}}]{Foucart:2015gaa}
\bibinfo{author}{\bibfnamefont{F.}~\bibnamefont{Foucart}},
  \bibinfo{author}{\bibfnamefont{R.}~\bibnamefont{Haas}},
  \bibinfo{author}{\bibfnamefont{M.~D.} \bibnamefont{Duez}},
  \bibinfo{author}{\bibfnamefont{E.}~\bibnamefont{O'Connor}},
  \bibinfo{author}{\bibfnamefont{C.~D.} \bibnamefont{Ott}},
  \bibinfo{author}{\bibfnamefont{L.}~\bibnamefont{Roberts}},
  \bibinfo{author}{\bibfnamefont{L.~E.} \bibnamefont{Kidder}},
  \bibinfo{author}{\bibfnamefont{J.}~\bibnamefont{Lippuner}},
  \bibinfo{author}{\bibfnamefont{H.~P.} \bibnamefont{Pfeiffer}},
  \bibnamefont{and} \bibinfo{author}{\bibfnamefont{M.~A.}
  \bibnamefont{Scheel}}, \bibinfo{journal}{Phys. Rev.}
  \textbf{\bibinfo{volume}{D93}}, \bibinfo{pages}{044019}
  (\bibinfo{year}{2016}), \eprint{1510.06398}.

\bibitem[{\citenamefont{Caballero}(2016)}]{Caballero:2016lof}
\bibinfo{author}{\bibfnamefont{O.~L.} \bibnamefont{Caballero}}, in
  \emph{\bibinfo{booktitle}{{11th Latin American Symposium on Nuclear Physics
  and Applications Medellin, Colombia, November 30-December 4, 2015}}}
  (\bibinfo{year}{2016}), \eprint{1603.02755},
  \urlprefix\url{https://inspirehep.net/record/1426803/files/arXiv:1603.02755.pdf}.

\bibitem[{\citenamefont{Hasenfratz and von Allmen}(2008)}]{Hasenfratz:2007dp}
\bibinfo{author}{\bibfnamefont{P.}~\bibnamefont{Hasenfratz}} \bibnamefont{and}
  \bibinfo{author}{\bibfnamefont{R.}~\bibnamefont{von Allmen}},
  \bibinfo{journal}{JHEP} \textbf{\bibinfo{volume}{02}}, \bibinfo{pages}{079}
  (\bibinfo{year}{2008}), \eprint{0710.5346}.

\bibitem[{\citenamefont{Alkofer and von Smekal}(2001)}]{Alkofer:2000wg}
\bibinfo{author}{\bibfnamefont{R.}~\bibnamefont{Alkofer}} \bibnamefont{and}
  \bibinfo{author}{\bibfnamefont{L.}~\bibnamefont{von Smekal}},
  \bibinfo{journal}{Phys. Rept.} \textbf{\bibinfo{volume}{353}},
  \bibinfo{pages}{281} (\bibinfo{year}{2001}), \eprint{hep-ph/0007355}.

\bibitem[{\citenamefont{Fischer}(2006)}]{Fischer:2006ub}
\bibinfo{author}{\bibfnamefont{C.~S.} \bibnamefont{Fischer}},
  \bibinfo{journal}{J. Phys.} \textbf{\bibinfo{volume}{G32}},
  \bibinfo{pages}{R253} (\bibinfo{year}{2006}), \eprint{hep-ph/0605173}.

\bibitem[{\citenamefont{Alkofer et~al.}(2009)\citenamefont{Alkofer, Fischer,
  Llanes-Estrada, and Schwenzer}}]{Alkofer:2008tt}
\bibinfo{author}{\bibfnamefont{R.}~\bibnamefont{Alkofer}},
  \bibinfo{author}{\bibfnamefont{C.~S.} \bibnamefont{Fischer}},
  \bibinfo{author}{\bibfnamefont{F.~J.} \bibnamefont{Llanes-Estrada}},
  \bibnamefont{and}
  \bibinfo{author}{\bibfnamefont{K.}~\bibnamefont{Schwenzer}},
  \bibinfo{journal}{Annals Phys.} \textbf{\bibinfo{volume}{324}},
  \bibinfo{pages}{106} (\bibinfo{year}{2009}), \eprint{0804.3042}.

\bibitem[{\citenamefont{Roberts}(2016)}]{Roberts:2015lja}
\bibinfo{author}{\bibfnamefont{C.~D.} \bibnamefont{Roberts}},
  \bibinfo{journal}{J. Phys. Conf. Ser.} \textbf{\bibinfo{volume}{706}},
  \bibinfo{pages}{022003} (\bibinfo{year}{2016}), \eprint{1509.02925}.

\bibitem[{\citenamefont{Williams et~al.}(2016)\citenamefont{Williams, Fischer,
  and Heupel}}]{Williams:2015cvx}
\bibinfo{author}{\bibfnamefont{R.}~\bibnamefont{Williams}},
  \bibinfo{author}{\bibfnamefont{C.~S.} \bibnamefont{Fischer}},
  \bibnamefont{and} \bibinfo{author}{\bibfnamefont{W.}~\bibnamefont{Heupel}},
  \bibinfo{journal}{Phys. Rev.} \textbf{\bibinfo{volume}{D93}},
  \bibinfo{pages}{034026} (\bibinfo{year}{2016}), \eprint{1512.00455}.

\bibitem[{\citenamefont{Mian and Maas}(2016)}]{Mian:2016eel}
\bibinfo{author}{\bibfnamefont{W.~A.} \bibnamefont{Mian}} \bibnamefont{and}
  \bibinfo{author}{\bibfnamefont{A.}~\bibnamefont{Maas}}, in
  \emph{\bibinfo{booktitle}{{12th Conference on Quark Confinement and the
  Hadron Spectrum (Confinement XII) Thessaloniki, Greece, August 29-September
  2, 2016}}} (\bibinfo{year}{2016}), \eprint{1610.02936},
  \urlprefix\url{http://inspirehep.net/record/1490923/files/arXiv:1610.02936.pdf}.

\bibitem[{\citenamefont{Maris and Tandy}(1999)}]{Maris:1999nt}
\bibinfo{author}{\bibfnamefont{P.}~\bibnamefont{Maris}} \bibnamefont{and}
  \bibinfo{author}{\bibfnamefont{P.~C.} \bibnamefont{Tandy}},
  \bibinfo{journal}{Phys. Rev.} \textbf{\bibinfo{volume}{C60}},
  \bibinfo{pages}{055214} (\bibinfo{year}{1999}), \eprint{nucl-th/9905056}.

\bibitem[{\citenamefont{Goecke et~al.}(2011)\citenamefont{Goecke, Fischer, and
  Williams}}]{Goecke:2011pe}
\bibinfo{author}{\bibfnamefont{T.}~\bibnamefont{Goecke}},
  \bibinfo{author}{\bibfnamefont{C.~S.} \bibnamefont{Fischer}},
  \bibnamefont{and} \bibinfo{author}{\bibfnamefont{R.}~\bibnamefont{Williams}},
  \bibinfo{journal}{Phys. Lett.} \textbf{\bibinfo{volume}{B704}},
  \bibinfo{pages}{211} (\bibinfo{year}{2011}), \eprint{1107.2588}.

\bibitem[{\citenamefont{Sanchis-Alepuz
  et~al.}(2014)\citenamefont{Sanchis-Alepuz, Fischer, and
  Kubrak}}]{Sanchis-Alepuz:2014wea}
\bibinfo{author}{\bibfnamefont{H.}~\bibnamefont{Sanchis-Alepuz}},
  \bibinfo{author}{\bibfnamefont{C.~S.} \bibnamefont{Fischer}},
  \bibnamefont{and} \bibinfo{author}{\bibfnamefont{S.}~\bibnamefont{Kubrak}},
  \bibinfo{journal}{Phys. Lett.} \textbf{\bibinfo{volume}{B733}},
  \bibinfo{pages}{151} (\bibinfo{year}{2014}), \eprint{1401.3183}.

\bibitem[{\citenamefont{Hilger et~al.}(2015)\citenamefont{Hilger, Gomez-Rocha,
  and Krassnigg}}]{Hilger:2015ora}
\bibinfo{author}{\bibfnamefont{T.}~\bibnamefont{Hilger}},
  \bibinfo{author}{\bibfnamefont{M.}~\bibnamefont{Gomez-Rocha}},
  \bibnamefont{and} \bibinfo{author}{\bibfnamefont{A.}~\bibnamefont{Krassnigg}}
  (\bibinfo{year}{2015}), \eprint{1508.07183}.

\bibitem[{\citenamefont{Alkofer et~al.}(2004)\citenamefont{Alkofer, Detmold,
  Fischer, and Maris}}]{Alkofer:2003jj}
\bibinfo{author}{\bibfnamefont{R.}~\bibnamefont{Alkofer}},
  \bibinfo{author}{\bibfnamefont{W.}~\bibnamefont{Detmold}},
  \bibinfo{author}{\bibfnamefont{C.~S.} \bibnamefont{Fischer}},
  \bibnamefont{and} \bibinfo{author}{\bibfnamefont{P.}~\bibnamefont{Maris}},
  \bibinfo{journal}{Phys. Rev.} \textbf{\bibinfo{volume}{D70}},
  \bibinfo{pages}{014014} (\bibinfo{year}{2004}), \eprint{hep-ph/0309077}.

\bibitem[{\citenamefont{Maas}(2013)}]{Maas:2011se}
\bibinfo{author}{\bibfnamefont{A.}~\bibnamefont{Maas}}, \bibinfo{journal}{Phys.
  Rept.} \textbf{\bibinfo{volume}{524}}, \bibinfo{pages}{203}
  (\bibinfo{year}{2013}), \eprint{1106.3942}.

\bibitem[{\citenamefont{Habel et~al.}(1990{\natexlab{a}})\citenamefont{Habel,
  Konning, Reusch, Stingl, and Wigard}}]{Habel:1989aq}
\bibinfo{author}{\bibfnamefont{U.}~\bibnamefont{Habel}},
  \bibinfo{author}{\bibfnamefont{R.}~\bibnamefont{Konning}},
  \bibinfo{author}{\bibfnamefont{H.~G.} \bibnamefont{Reusch}},
  \bibinfo{author}{\bibfnamefont{M.}~\bibnamefont{Stingl}}, \bibnamefont{and}
  \bibinfo{author}{\bibfnamefont{S.}~\bibnamefont{Wigard}},
  \bibinfo{journal}{Z. Phys.} \textbf{\bibinfo{volume}{A336}},
  \bibinfo{pages}{423} (\bibinfo{year}{1990}{\natexlab{a}}).

\bibitem[{\citenamefont{Habel et~al.}(1990{\natexlab{b}})\citenamefont{Habel,
  Konning, Reusch, Stingl, and Wigard}}]{Habel:1990tw}
\bibinfo{author}{\bibfnamefont{U.}~\bibnamefont{Habel}},
  \bibinfo{author}{\bibfnamefont{R.}~\bibnamefont{Konning}},
  \bibinfo{author}{\bibfnamefont{H.~G.} \bibnamefont{Reusch}},
  \bibinfo{author}{\bibfnamefont{M.}~\bibnamefont{Stingl}}, \bibnamefont{and}
  \bibinfo{author}{\bibfnamefont{S.}~\bibnamefont{Wigard}},
  \bibinfo{journal}{Z. Phys.} \textbf{\bibinfo{volume}{A336}},
  \bibinfo{pages}{435} (\bibinfo{year}{1990}{\natexlab{b}}).

\bibitem[{\citenamefont{Stingl}(1996)}]{Stingl:1994nk}
\bibinfo{author}{\bibfnamefont{M.}~\bibnamefont{Stingl}}, \bibinfo{journal}{Z.
  Phys.} \textbf{\bibinfo{volume}{A353}}, \bibinfo{pages}{423}
  (\bibinfo{year}{1996}), \eprint{hep-th/9502157}.

\bibitem[{\citenamefont{Maas et~al.}(2004)\citenamefont{Maas, Wambach, Gruter,
  and Alkofer}}]{Maas:2004se}
\bibinfo{author}{\bibfnamefont{A.}~\bibnamefont{Maas}},
  \bibinfo{author}{\bibfnamefont{J.}~\bibnamefont{Wambach}},
  \bibinfo{author}{\bibfnamefont{B.}~\bibnamefont{Gruter}}, \bibnamefont{and}
  \bibinfo{author}{\bibfnamefont{R.}~\bibnamefont{Alkofer}},
  \bibinfo{journal}{Eur. Phys. J.} \textbf{\bibinfo{volume}{C37}},
  \bibinfo{pages}{335} (\bibinfo{year}{2004}), \eprint{hep-ph/0408074}.

\bibitem[{\citenamefont{Maas}(2005)}]{Maas:2005rf}
\bibinfo{author}{\bibfnamefont{A.}~\bibnamefont{Maas}}, Ph.D. thesis,
  \bibinfo{school}{Darmstadt, Tech. Hochsch.} (\bibinfo{year}{2005}),
  \eprint{hep-ph/0501150}.

\end{thebibliography}

\end{document}